\documentclass[journal,onecolumn,12pt]{IEEEtran}

\usepackage{amsmath,amsfonts,amssymb}
\usepackage{algorithmic}
\usepackage{algorithm}
\usepackage{array}
\usepackage[caption=false,font=footnotesize]{subfig}
\usepackage{textcomp}
\usepackage{stfloats}
\usepackage{url}
\usepackage{verbatim}
\usepackage{graphicx}
\usepackage{cite}
\usepackage{tikz}
\usetikzlibrary{fit, backgrounds, positioning, calc}
\usetikzlibrary{tikzmark}
\usepackage{booktabs}
\usepackage{supertabular}
\usepackage{enumerate}

\newtheorem{definition}{Definition}
\newtheorem{theorem}{Theorem}
\newtheorem{lemma}{Lemma}
\newtheorem{corollary}{Corollary}
\newtheorem{remark}{Remark}
\newtheorem{example}{Example}

\newcommand*{\diff}{\mathop{}\!\mathrm{d}}
\allowdisplaybreaks[1]

\begin{document}

\title{
The Deterministic Broadcast Channel Revisited:\\ A Bipartite Graph Approach
}

\author{
Yiyu~Qiu,
Wei~Chen,
and~H.~Vincent~Poor
\thanks{
Yiyu Qiu and Wei Chen are with the Department of Electronic Engineering, Tsinghua University, Beijing 100084, China. They are also with the State Key Laboratory of Space Network and Communications and the Beijing National Research Center for Information Science and Technology (BNRist), Beijing 100084, China (email: qiuyy22@mails.tsinghua.edu.cn; wchen@tsinghua.edu.cn).

H. Vincent Poor is with the Department of Electrical and Computer Engineering, Princeton University, Princeton, NJ 08544 USA (email: poor@princeton.edu).

\copyright\ 2026 IEEE.  Personal use of this material is permitted.  Permission from IEEE must be obtained for all other uses, in any current or future media, including reprinting/republishing this material for advertising or promotional purposes, creating new collective works, for resale or redistribution to servers or lists, or reuse of any copyrighted component of this work in other works.
}
}

\markboth{Qiu, Chen, and Poor: The Deterministic Broadcast Channel Revisited}%
{}

\maketitle

\begin{abstract}
The general broadcast channel (BC) has attracted considerable attention because it is a central problem in network or multi-user information theory. The deterministic BC (det-BC), in which both output symbols are functions of the input symbol, plays a vital role in understanding general BCs. In the 1970s, Marton and Pinsker investigated the det-BC independently using a random coding approach and hence quantified its capacity region as functions of the entropy of output symbols. In this paper, we present a purely deterministic and combinatorial treatment of the deterministic BC, which is in contrast to Marton and Pinsker’s probabilistic treatment. In particular, we show that each det-BC can be uniquely and completely characterized by a bipartite graph. Based on the bipartite graph model, we present the explicit capacity region of the general det-BC and a deterministic capacity-achieving coding scheme. Specifically, each point on the capacity region boundary is expressed as an explicit function of the degree sequences of the two parts, with boundary coordinates parametrized by the tangent slope. As a result, a closed-form capacity region, which was previously unknown except for special cases such as the Blackwell channel, can be analytically determined, without optimizing the probability distribution of input symbols numerically. For a bipartite graph consisting of multiple isolated subgraphs, there can exist a commonly decodable message while preserving the maximum achievable sum rate, and its achievable rate is explicitly characterized. Finally, the bipartite-graph formulation allows us to construct optimal finite-blocklength coding for the det-BC through binary linear programming.
\end{abstract}

\begin{IEEEkeywords}
Deterministic broadcast channel, bipartite graph, capacity region, deterministic coding, common message, finite-blocklength coding, multi-user information theory, graph homomorphism.
\end{IEEEkeywords}


\section{Introduction}
\label{sec:introduction}

The broadcast channel (BC), introduced by Cover in 1972 \cite{cover_broadcast_1972}, models a fundamental communication scenario in which a single transmitter sends information to multiple receivers over a shared medium. This setting captures the essence of downlink transmission in wireless networks, satellite communications, and various multi-user systems. While the general capacity characterization remains elusive after five decades, significant progress has been made for special cases and through the development of inner and outer bounds.


A deterministic broadcast channel (det-BC) is a special case where the output symbols are deterministic functions of the input: $Y = y(X)$ and $Z = z(X)$, rather than probabilistic transitions. Early work by van der Meulen \cite{van_der_meulen_random_1975} and Cover \cite{cover_achievable_1975} established achievable rate regions for general discrete memoryless broadcast channels (DM-BCs), transmitting common message at rate $R_0$ and independent private information at rates $R_1$ and $R_2$ to two receivers. These results were soon extended to specific deterministic settings by Gelfand \cite{gelfand_capacity_1977}, who derived the exact capacity region of the Blackwell channel (a canonical example of a det-BC) and later by Gelfand and Pinsker \cite{gelfand_capacity_1980}, who characterized the capacity region of the semi-deterministic broadcast channel, in which the output at one receiver is a deterministic function of the input while the other undergoes a stochastic transition.
Shortly thereafter, Marton \cite{marton_capacity_1977} and Pinsker \cite{pinsker_capacity_1978} independently characterized the capacity region for general det-BCs. Their fundamental result expresses achievable rates in terms of the entropies $H(Y)$, $H(Z)$, and $H(Y,Z)$, evaluated over all possible input distributions. However, this characterization is \emph{implicit}: determining the capacity region for a specific det-BC requires optimizing over the space of input distributions $P_X$, which generally must be performed numerically.


The capacity region of general DM-BCs remains one of the most significant open problems in network information theory. The strongest known inner bound is Marton's region \cite{marton_coding_1979}, while the tightest outer bound was proposed by Gohari and Nair \cite{gohari_outer_2022}, improving upon earlier bounds by Nair and El Gamal \cite{nair_outer_2007}. The gap between these bounds has been closed for certain special cases, such as degraded BCs, less noisy BCs \cite{nair_capacity_2010}, and Gaussian vector BCs \cite{geng_capacity_2014}, but remains open in general. For example, \cite{geng_information_2013} explicitly evaluated the bounds for the binary skew-symmetric BC and demonstrated a gap between Marton's inner bound and the UV outer bound, and \cite{geng_martons_2014} demonstrated that the Nair--El Gamal outer bound is not tight by analyzing the product channel.

Recent methodological advances have provided new tools for analyzing BC capacity. Gohari and Anantharam \cite{gohari_evaluation_2012} introduced perturbation techniques to impose cardinality constraints on Marton's region. Several works have evaluated capacity regions by computing weighted sum rates, thereby characterizing the supporting hyperplanes of the achievable region \cite{gohari_martons_2014,anantharam_evaluation_2019}. Gohari and Nair \cite{gohari_outer_2022} proposed the auxiliary channel framework, which strictly improved the UV outer bound by introducing auxiliary receivers.
On the computational front, numerical algorithms have been developed to evaluate capacity bounds. The Blahut--Arimoto algorithm \cite{blahut_computation_1972,arimoto_algorithm_1972}, originally designed for point-to-point channels, has been extended to broadcast channels despite challenges arising from non-convexity \cite{calvo_computation_2008}. Yasui and Matsushima \cite{yasui_toward_2010} proposed algorithms for degraded BCs, which were later generalized to compute both inner and outer bounds for general DM-BCs \cite{liu_blahut-arimoto_2022,liu_blahut-arimoto_2023}.

On the coding-theoretic side, Goela, Abbe, and Gastpar~\cite{goela_polar_2015} introduced polar codes for $m$-user deterministic broadcast channels with binary output alphabets and showed that rates on the private-message capacity region boundary are achievable using polarization techniques. Mondelli, Hassani, Sason, and Urbanke~\cite{mondelli_achieving_2015} subsequently extended these ideas to general two-user DM-BCs and showed that polar codes can achieve Marton's region. For deterministic broadcast channels, their construction specializes to the full private-message capacity region. These works focus on explicit coding constructions, whereas our emphasis is on obtaining an explicit characterization of the capacity region.


A common limitation of existing capacity characterizations is that they either express the capacity region implicitly through entropy inequalities, or require numerical optimization over input and auxiliary distributions. For a specific channel, determining whether a given rate pair is achievable generally involves optimizing over the space of input distributions, with no closed-form expression available. This lack of explicit formulas limits both theoretical insight and practical applicability.

Det-BCs are particularly amenable to a graph-theoretic treatment; since the channel outputs are deterministic functions of the input, the input-output mapping corresponds naturally to edges in a graph. This observation motivates a purely combinatorial reformulation of the capacity problem, in which probabilistic notions such as entropy and mutual information are replaced by graph-theoretic quantities such as degree sequences and vertex contractions. Graph-theoretic approaches in information theory date back to Shannon's seminal work on zero-error capacity~\cite{shannon_zero_1956}, which represents the confusability structure of a channel by a graph. Since then, graph-theoretic methods have proven effective in a wide range of communication problems, providing combinatorial perspectives that complement classical information-theoretic analysis~\cite{jukna_extremal_2011,bai_max-matching_2010,bai_diversity-multiplexing_2011,bai_unified_2012}. The use of a deterministic channel model to simplify analysis also parallels the approach of Avestimehr et al.~\cite{avestimehr_wireless_2011}, who approximated Gaussian relay networks by deterministic channels to obtain capacity characterizations.


This paper revisits the det-BC capacity problem from a purely deterministic and combinatorial perspective, in contrast to the traditional probabilistic treatment. While the capacity region can in principle be obtained by solving an entropy optimization problem over input distributions, we show that each det-BC is uniquely characterized by a bipartite graph whose degree sequences alone determine the capacity region, yielding the first explicit, closed-form characterization for arbitrary det-BCs. Our main contributions are as follows:
\begin{itemize}
    \item \textbf{Graph-theoretic framework:} We show that each det-BC is uniquely characterized by a bipartite graph, and prove that a rate pair $(R_1, R_2)$ is achievable if and only if the complete bipartite graph $K_{M_1,M_2}$ of corresponding size can be obtained by contracting the $n$-fold product channel graph (Theorem~\ref{thm:bipartite_blocklength}), thereby reducing the capacity problem to a purely combinatorial problem.
    \item \textbf{Explicit capacity region:} We derive a parametric characterization of the private-message capacity region boundary depending solely on the degree sequences of the received symbols (Theorem~\ref{thm:main_capacity}), with both achievability and converse rigorously established. This yields a closed-form capacity region for general det-BCs without requiring any numerical optimization over input distributions.
    \item \textbf{Extension to common messages:} We extend the graph-theoretic framework to characterize achievable rate triples $(R_0, R_1, R_2)$ with a common message (Theorem~\ref{thm:achievability_with_common}). The connected components of the bipartite graph carry common message naturally: when multiple components exist, the transmitter can embed a publicly decodable message while simultaneously achieving the private-message capacity region boundary. From a security perspective, this also reveals an inherent information leakage phenomenon in capacity-achieving codes.
    \item \textbf{Finite-blocklength analysis:} We formulate the finite-blocklength achievable rate problem as an integer linear program (ILP) that directly yields encoding and decoding schemes, enabling numerical computation of achievable rate pairs at moderate blocklengths. For degraded det-BCs, the ILP reduces to a bin covering problem.
\end{itemize}

The remainder of this paper is organized as follows. Section~\ref{sec:preliminaries} introduces the det-BC model and its bipartite graph representation, presenting our main capacity characterization result. Section~\ref{sec:graph} develops the graph-theoretic framework, establishing degree properties of product channels and the fundamental equivalence between achievability and graph contraction. Section~\ref{sec:derivation} provides a rigorous proof of the private-message capacity region. Section~\ref{sec:common} extends the analysis in two directions: capacity regions with common messages and finite-blocklength regimes. Section~\ref{sec:numerical} presents numerical results that validate our theoretical characterizations and illustrate geometric properties of the capacity region. Finally, Section~\ref{sec:conclusion} concludes with a summary and discussion of open problems.


\section{Preliminaries and Main Results}
\label{sec:preliminaries}


\subsection{Bipartite Graph Model for Det-BCs}
\label{subsec:bipartite}

\begin{figure}[t]
    \centering
    \subfloat[]{\includegraphics[width=.3\linewidth]{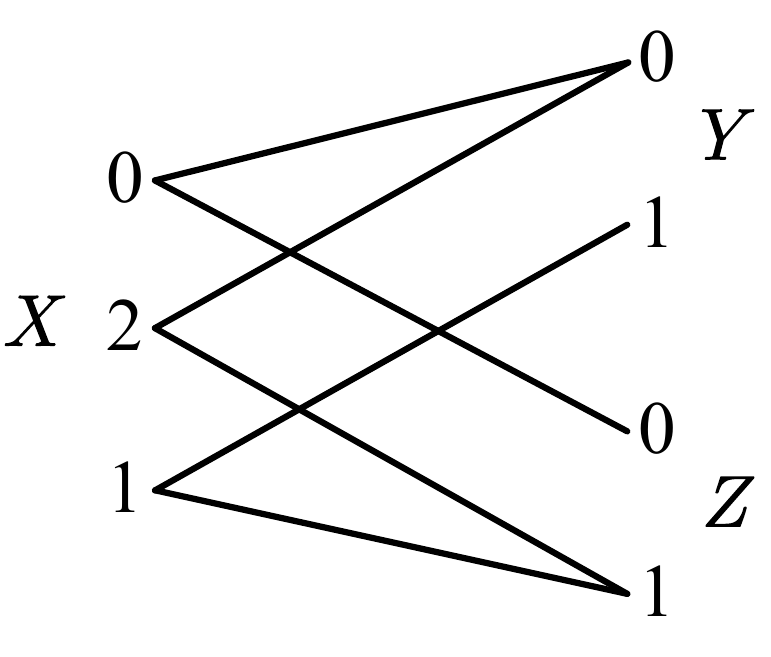}
    \label{fig:blackwell, model}}
    \hfil
    \subfloat[]{\includegraphics[width=.3\linewidth]{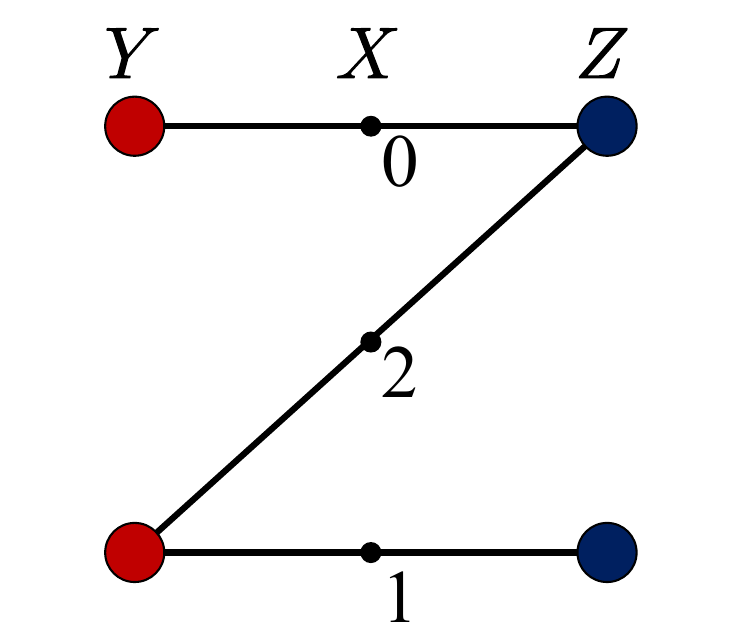}
    \label{fig:blackwell, bipartite graph}}
    \hfil
    \subfloat[]{\includegraphics[width=.3\linewidth]{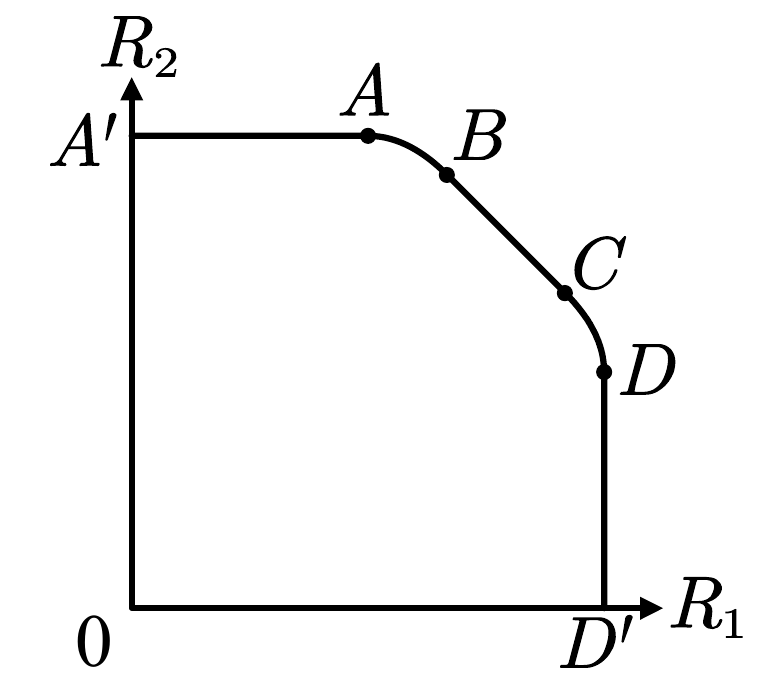}
    \label{fig:blackwell, capacity region}}
    \caption{The Blackwell channel. (a) Channel model. (b) Bipartite graph model. (c) Capacity region.}
    \label{fig:blackwell channel}
\end{figure}

Consider a two-receiver discrete memoryless broadcast channel characterized by $\mathcal{B} = (\mathcal{X}, p(y,z|x), \mathcal{Y} \times \mathcal{Z})$, where $p(y,z|x)$ denotes the channel transition probability, and $X$, $Y$, $Z$ represent the channel input and outputs taking values from finite alphabets $\mathcal{X}$, $\mathcal{Y}$, $\mathcal{Z}$, respectively. We specifically consider the class of deterministic broadcast channels (det-BCs), where both $Y$ and $Z$ are deterministic functions of $X$, i.e., $Y = y(X)$ and $Z = z(X)$. Equivalently, the transition probability satisfies $p(y,z|x) \in \{0,1\}$ for all $(x,y,z) \in \mathcal{X} \times \mathcal{Y} \times \mathcal{Z}$. Our focus is on characterizing the private-message capacity region, defined as the closure of all achievable rate pairs $(R_1, R_2)$.

Fig.~\ref{fig:blackwell channel} illustrates the classic Blackwell channel, originally introduced by Blackwell in 1963 as an example of a broadcast channel, together with its bipartite graph representation. Any det-BC can be equivalently represented as a bipartite graph $G(\mathcal{B}) = (\mathcal{Y}, \mathcal{Z}, E)$, where the edge set is given by
\begin{align}
    E = \{(y(x), z(x)) : x \in \mathcal{X}\}.
\end{align}
The two parts of the bipartite graph correspond to the two receivers. Each vertex in one part represents a received symbol for the corresponding receiver, while each edge represents a transmitted symbol. Specifically, vertices $y^{(i)} \in \mathcal{Y}$ and $z^{(j)} \in \mathcal{Z}$ are connected by an edge labeled $x^{(k)}$ if and only if $p\left(y^{(i)}, z^{(j)} | x^{(k)}\right) = 1$.

Two transmitted symbols $x^{(i)}$ and $x^{(j)}$ are said to be equivalent if $p\left(y,z|x^{(i)}\right)=p\left(y,z|x^{(j)}\right)$ for all $y,z\in\mathcal{Y}\times\mathcal{Z}$. Removing equivalent symbols does not affect the capacity region. Without loss of generality, we assume throughout this paper that all transmitted symbols are pairwise non-equivalent. Consequently, our bipartite graphs contain no multiple edges, i.e., they are simple graphs.

We define the degree of a received symbol as the degree of the corresponding vertex in the bipartite graph $G(\mathcal{B})$. Let $d_{Y,y}$ denote the degree of vertex $y$, and let $d_{Z,z}$ denote the degree of vertex $z$. The degree sequences are multisets defined as
\begin{subequations}
\begin{align}
    D_Y(\mathcal{B}) = \{d_{Y,y} \mid y \in \mathcal{Y}\},\\
    D_Z(\mathcal{B}) = \{d_{Z,z} \mid z \in \mathcal{Z}\}.
\end{align}
\end{subequations}
We can equivalently characterize the degree sequences through their multiplicity functions, which we call the \emph{degree spectrum}. Let $m_Y(d)$ denote the multiplicity function of multiset $D_Y(\mathcal{B})$, i.e., the number of vertices in receiver~1's part with degree $d$, where $d \in \{1, 2, \ldots, |\mathcal{X}|\}$. Similarly, let $m_Z(d)$ denote the multiplicity function for receiver~2. The degree spectrum provides a compact representation when the degree sequence contains repeated values.

\begin{remark}
The degree spectrum should not be confused with the graph spectrum in spectral graph theory, which refers to the set of eigenvalues of the adjacency matrix. Our degree spectrum is a purely combinatorial object counting vertex degrees.
\end{remark}


\subsection{Explicit Capacity Region of Det-BCs}
\label{subsec:explicit_capacity}

In this section, we present the main result of this paper: a characterization of the capacity region $\mathcal{C}(\mathcal{B})$ for det-BC $\mathcal{B}$. Unless otherwise specified, all rates in this paper are measured in nats.
We define the region $\tilde{\mathcal{C}}(\mathcal{B})$ as follows:

\begin{definition}
\label{def:explicit_capacity_region}
For a det-BC $\mathcal{B}$ with degree sequences $D_Y(\mathcal{B})$ and $D_Z(\mathcal{B})$, the region $\tilde{\mathcal{C}}(\mathcal{B})$ is defined as the set of rate pairs $(R_1, R_2)$ whose boundary consists of five segments, as illustrated in Fig.~\ref{fig:blackwell, capacity region}:

\textbf{Curved segment $A$--$B$:} Characterized by the parametric equations with parameter $t \in [0,1]$:
\begin{subequations}
\label{eq:parametric}
\begin{align}
    \label{eq:parametric_R1}
    R_1 &= \frac{\sum_{z\in\mathcal{Z}}d_{Z,z}^t\ln d_{Z,z}}{\sum_{z\in\mathcal{Z}}d_{Z,z}^t},\\
    \label{eq:parametric_R2}
    R_2 &= \ln\sum_{z\in\mathcal{Z}}d_{Z,z}^t - \frac{\sum_{z\in\mathcal{Z}}d_{Z,z}^t\ln d_{Z,z}^t}{\sum_{z\in\mathcal{Z}}d_{Z,z}^t}.
\end{align}
\end{subequations}
Define the probability distribution $q_Z = \{q_Z(z)\}_{z \in \mathcal{Z}}$ by
\begin{equation}
\begin{aligned}
    \label{eq:type q}
    q_Z(z)
    &= q_Z(z,D_Z,t)\\
    &= \frac{d_{Z,z}^t}{\sum_{z'\in\mathcal{Z}}d_{Z,z'}^t}, \quad z \in \mathcal{Z}.
\end{aligned}
\end{equation}
The parametric equations can then be written equivalently as
\begin{subequations}
\label{eq:parametric_simplified}
\begin{align}
    \label{eq:parametric_R1_simplified}
    R_1 &= \mathbb{E}_{q_Z}\left[\ln d_{Z,z}\right],\\
    \label{eq:parametric_R2_simplified}
    R_2 &= H(q_Z).
\end{align}
\end{subequations}
The endpoints are obtained by taking $t \to 0$ and $t = 1$:
\begin{align}
    \label{eq:endpoint_A}
    (R_{1}^A,R_{2}^A) &= \left(\frac{1}{|\mathcal{Z}|}\sum_{z\in\mathcal{Z}}\ln d_{Z,z}, \ln|\mathcal{Z}|\right),\\
    \label{eq:endpoint_B}
    (R_{1}^B,R_{2}^B) &= \left(\ln|\mathcal{X}| - H(q_Z(z,D_Z,1)), H(q_Z(z,D_Z,1))\right).
\end{align}

\textbf{Curved segment $C$--$D$:} By symmetry, this segment is characterized by parametric equations analogous to \eqref{eq:parametric}--\eqref{eq:parametric_simplified}, with $D_Z$ replaced by $D_Y$.

\textbf{Line segments:} The slopes of segments $A$--$A'$, $B$--$C$, and $D$--$D'$ are $0$, $-1$, and $\infty$, respectively.

The boundary of $\tilde{\mathcal{C}}(\mathcal{B})$ also includes two segments along the coordinate axes, namely $O$--$A'$ (along the $R_2$-axis) and $O$--$D'$ (along the $R_1$-axis), which together with the five segments above form the complete boundary. Since the region is the downward-left closure of the five segments defined above, the segments on the coordinate axes are determined implicitly and will not be referred to explicitly in the remainder of this paper.
\end{definition}

\begin{remark}
Alternatively, curved segment $A$--$B$ can be expressed using the degree spectrum $m_Z(d)$ instead of the degree sequence $D_Z$. The parametric equations become
\begin{subequations}
\label{eq:parametric_spectrum}
\begin{align}
    \label{eq:parametric_spectrum_R1}
    R_1 &= \frac{\sum_{d=1}^{|\mathcal{X}|}m_Z(d)\,d^t\ln d}{\sum_{d=1}^{|\mathcal{X}|}m_Z(d)\,d^t},\\
    \label{eq:parametric_spectrum_R2}
    R_2 &= \ln\sum_{d=1}^{|\mathcal{X}|}m_Z(d)\,d^t - \frac{\sum_{d=1}^{|\mathcal{X}|}m_Z(d)\,d^t\ln d^t}{\sum_{d=1}^{|\mathcal{X}|}m_Z(d)\,d^t}.
\end{align}
\end{subequations}
The degree spectrum representation is particularly useful when the degree sequence contains repeated values, as it provides a more compact characterization.
\end{remark}

Our main result establishes that this explicitly defined region is indeed the capacity region:

\begin{theorem}
\label{thm:main_capacity}
The capacity region of det-BC $\mathcal{B}$ is given by
\begin{align}
    \mathcal{C}(\mathcal{B}) = \tilde{\mathcal{C}}(\mathcal{B}).
\end{align}
\end{theorem}

The proof of Theorem~\ref{thm:main_capacity} is provided in Section~\ref{sec:derivation}.

\begin{remark}[Intuitive interpretation of the five boundary segments]
\label{rem:five_segments_intuition}
The five boundary segments of $\tilde{\mathcal C}(\mathcal B)$ correspond to three qualitatively different operating regimes, which can be understood intuitively as follows.

\emph{Segments $A'$--$A$ and $D$--$D'$.}
On segment~$A'$--$A$, receiver~2 operates at its single-user maximum rate $\ln\left|\mathcal{Z}\right|$, while receiver~1 simultaneously receives information at a positive rate. This is possible due to residual distinguishability among the inputs sharing the same receiver-2 output. Segment~$D$--$D'$ is the symmetric counterpart, with receiver~1 at its maximum rate $\ln\left|\mathcal{Y}\right|$.

\emph{Segments $A$--$B$ and $C$--$D$.}
Along these segments, neither receiver operates at its single-user optimum. Increasing the rate of one receiver necessarily reduces the rate available to the other, creating a tradeoff between the two users. As the operating point moves along the boundary, the boundary-achieving input distribution changes continuously, reallocating probability mass to favor one receiver or the other. Because the resulting rates are nonlinear functions of the input distribution, the resulting boundary is generally curved rather than linear.

\emph{Segment $B$--$C$.}
On segment~$B$--$C$, both receivers together already achieve the maximum total rate the channel permits, $R_1+R_2 = \ln|\mathcal{X}|$, which is attained under the uniform input distribution. In contrast to segments~$A$--$B$ and $C$--$D$, the entire segment~$B$--$C$ is achieved by the same uniform input distribution; only the allocation of the total rate between the two receivers varies. Any increase in one user's rate must therefore be compensated by an equal decrease in the other's rate, yielding the slope-$(-1)$ linear boundary.
\end{remark}

\subsection{Reduction to Known Special Cases}
\label{sec:special_cases}

We verify that Theorem~\ref{thm:main_capacity} reproduces classical closed-form results for two well-studied classes of det-BCs, thereby confirming the consistency of our graph-theoretic characterization with the existing literature.


\begin{example}[Blackwell Channel]
\label{ex:blackwell}
The Blackwell channel~\cite{gelfand_capacity_1977} has input alphabet $\mathcal{X} = \left\{0, 1, 2\right\}$ and output alphabets $\mathcal{Y} = \mathcal{Z} = \left\{0, 1\right\}$, with channel mappings $y\left(0\right) = y\left(2\right) = 0$, $y\left(1\right) = 1$, $z\left(0\right) = 0$, $z\left(1\right) = z\left(2\right) = 1$. The corresponding bipartite graph has degree sequences
\begin{equation}
    D_Y\left(\mathcal{B}\right)
    =D_Z\left(\mathcal{B}\right)
    =\left\{2, 1\right\},
    \label{eq:blackwell_degrees}
\end{equation}
where vertex $y_0$ has degree~$2$, vertex $y_1$ degree~$1$, vertex $z_0$ degree~$1$, and vertex $z_1$ degree~$2$.

We apply Theorem~\ref{thm:main_capacity} to compute the curved segment~$A$--$B$. The distribution $q_Z$ of~\eqref{eq:type q} is
\begin{equation}
    q_Z\left(z_0\right) = \frac{1}{1 + 2^{t}},
    \qquad
    q_Z\left(z_1\right) = \frac{2^{t}}{1 + 2^{t}}.
    \label{eq:blackwell_qZ}
\end{equation}
Introducing the substitution
\begin{equation}
    \alpha = q_Z\left(z_0\right) = \frac{1}{1 + 2^{t}},
    \label{eq:blackwell_alpha}
\end{equation}
where $\alpha\in \left[1/3, 1/2\right]$, with $\alpha = 1/2$ at $t = 0$ and $\alpha = 1/3$ at $t = 1$, the parametric equations~\eqref{eq:parametric}--\eqref{eq:parametric_simplified} reduce to
\begin{align}
    R_1
    &= \sum_{z} q_Z(z) \ln d_{Z,z} \notag\\
    &= (1-\alpha)\ln 2,
    \label{eq:blackwell_R1_nats}\\
    R_2
    &= H\left(q_Z\right) \notag\\
    &= -\alpha \ln \alpha - (1-\alpha)\ln(1-\alpha)\notag\\
    &= H_{\mathrm{b}}\left(\alpha\right),
    \label{eq:blackwell_R2_nats}
\end{align}
where $H_{\mathrm{b}}$ denotes the binary entropy function evaluated with the natural logarithm. By symmetry, the segment~$C$--$D$ parametrizes $R_1 = H_{\mathrm{b}}(\alpha)$, $R_2 = (1-\alpha)\ln 2$, reproducing the second curved piece of the classical result. Converting from natural logarithms to base-2 logarithms, this is precisely the curved boundary segment of the Blackwell channel capacity region derived by Gelfand~\cite{gelfand_capacity_1977}.
\end{example}


\begin{remark}[Biregular Det-BCs]
\label{rem:biregular}
Another special case is the \emph{biregular} det-BC, in which all $\mathcal{Y}$-vertices have degree $d_Y$ and all $\mathcal{Z}$-vertices have degree $d_Z$, so that $D_Y\left(\mathcal{B}\right) = \left\{d_Y, \ldots, d_Y\right\}$ and $D_Z\left(\mathcal{B}\right) = \left\{d_Z, \ldots, d_Z\right\}$. The distribution $q_Z$ of~\eqref{eq:type q} is uniform over $\mathcal{Z}$ for every $t$, so the parametric equations~\eqref{eq:parametric}--\eqref{eq:parametric_simplified} yield constant values:
\begin{subequations}
\label{eq:biregular_degenerate}
\begin{align}
R_1(t) &= \ln d_Z, \\
R_2(t) &= \ln |\mathcal{Z}|
\end{align}
\end{subequations}
for all $t\in\left[0, 1\right]$. Hence segment~$A$--$B$ degenerates to the single point $A=B=\left(\ln d_Z, \ln\left|\mathcal{Z}\right|\right)$, and by the symmetric argument segment~$C$--$D$ also degenerates to $C=D=\left(\ln\left|\mathcal{Y}\right|, \ln d_Y\right)$. Using $\left|\mathcal{Z}\right| d_Z = \left|\mathcal{X}\right|$, the full boundary of $\tilde{\mathcal{C}}\left(\mathcal{B}\right)$ reduces to the convex polygon
\begin{equation}
    \label{eq:biregular_capacity}
    \mathcal{C}\left(\mathcal{B}\right)
    =
    \left\{
    \left(R_1,R_2\right)\in\mathbb{R}_{\geq 0}^2
    \;\middle|\;
    \begin{aligned}
    R_1 &\leq \ln |\mathcal{Y}|,\\
    R_2 &\leq \ln |\mathcal{Z}|,\\
    R_1+R_2 &\leq \ln |\mathcal{X}|
    \end{aligned}
    \right\}.
\end{equation}
The resulting region coincides with the Marton--Pinsker characterization~\cite{marton_capacity_1977,pinsker_capacity_1978} specialized to the uniform input distribution $P_X = 1/|\mathcal{X}|$, under which $H(Y) = \ln\left|\mathcal{Y}\right|$, $H(Z) = \ln\left|\mathcal{Z}\right|$, and $H(Y,Z) = \ln\left|\mathcal{X}\right|$ all achieve their maximum values simultaneously. For more analyses of special cases, see Section~\ref{subsec:region w/o common}.
\end{remark}

\subsection{Simple Consequences of the Capacity Region Characterization}
\label{subsec:simple_consequences}

In this subsection, we present some simple corollaries regarding the capacity region $\tilde{\mathcal{C}}(\mathcal{B})$.

\begin{corollary}[Tangent Slope]
\label{cor:tangent_slope}
The slope of the line tangent to the curved segment~$A$--$B$
at the point $\left( R_1(t), R_2(t) \right)$ is $-t$ for
$t \in \left[ 0, 1 \right]$.
By symmetry, the tangent slope to segment~$C$--$D$ is $-1/t$.
\end{corollary}

\begin{IEEEproof}
From the parametric equations~\eqref{eq:parametric}--\eqref{eq:parametric_simplified}, the identity
\begin{equation}
    tR_1(t) + R_2(t)
    = \ln \sum_{z \in \mathcal{Z}} d_{Z,z}^{t}
    \label{eq:supporting_identity_Z}
\end{equation}
holds for all $t \in \left[ 0, 1 \right]$. Differentiating both sides with respect to $t$ yields
\begin{equation}
\begin{aligned}
  R_1(t) + t\frac{\diff R_1}{\diff t} + \frac{\diff R_2}{\diff t}
  &=\frac{\sum_{z \in \mathcal{Z}} d_{Z,z}^{t} \ln d_{Z,z}}
  {\sum_{z \in \mathcal{Z}} d_{Z,z}^{t}}\\
  &= R_1(t),
\end{aligned}
\end{equation}
where the second equality follows from~\eqref{eq:parametric_R1}. Hence,
\begin{equation*}
t\frac{\diff R_1}{\diff t} + \frac{\diff R_2}{\diff t}=0.
\end{equation*}
Therefore, wherever $\frac{\diff R_1}{\diff t}\neq 0$,
\begin{equation*}
    \frac{\mathrm{d}R_2}{\mathrm{d}R_1}=-t.
\end{equation*}
The result for segment~$C$--$D$ follows by exchanging the roles of the two receivers.
\end{IEEEproof}

\begin{corollary}[Degree Sequence Sufficiency]
\label{cor:degree_sufficiency}
The capacity region of a det-BC $\mathcal{B}$ depends on the channel transition probability $p(y,z|x)$ only through the degree sequences $D_Y(\mathcal{B})$ and $D_Z(\mathcal{B})$.
\end{corollary}

\begin{IEEEproof}
By Theorem~\ref{thm:main_capacity}, $\mathcal{C}(\mathcal{B}) = \tilde{\mathcal{C}}(\mathcal{B})$, and the latter is expressed entirely in terms of $D_Y(\mathcal{B})$ and $D_Z(\mathcal{B})$ by Definition~\ref{def:explicit_capacity_region}. 
See Remark~\ref{rem:degree_seq_sufficiency} in Section~\ref{subsec:general_proof} for an intuitive explanation of why the optimization over input distributions causes the detailed edge arrangement of $G\left(\mathcal{B}\right)$ to disappear, leaving only the degree sequences as the relevant channel parameters.
\end{IEEEproof}

\begin{corollary}[Length of Segment $B$--$C$]
\label{cor:BC_length}
The length of segment $B$--$C$ on the capacity region is given by
\begin{align}
    |BC| = \sqrt{2}\, I(Y;Z),
\end{align}
where $I(Y;Z)$ is evaluated under the uniform input distribution $P_X(x) = 1/|\mathcal{X}|$ for all $x \in \mathcal{X}$. Equivalently, the horizontal and vertical projections of segment $B$--$C$ both equal $I(Y;Z)$.
\end{corollary}

\begin{IEEEproof}
From \eqref{eq:endpoint_B} and by symmetry, the coordinates of points $B$ and $C$ are given by
\begin{align}
    (R_1^B,R_2^B) &= (\ln|\mathcal{X}| - H(q_Z(D_Z,1)), H(q_Z(D_Z,1))), \\
    (R_1^C,R_2^C) &= (H(q_Y(D_Y,1)), \ln|\mathcal{X}| - H(q_Y(D_Y,1))),
\end{align}
where $q_Z(D_Z,1)$ and $q_Y(D_Y,1)$ are defined by \eqref{eq:type q} with $t=1$. Under uniform input $P_X(x) = 1/|\mathcal{X}|$, the degree sequences are proportional to the output distributions: $q_Y(D_Y,1) = P_Y$ and $q_Z(D_Z,1) = P_Z$. Therefore,
\begin{align}
    R_1^C - R_1^B = H(Y) + H(Z) - \ln|\mathcal{X}|.
\end{align}
Since each input produces a unique output pair and the bipartite graph has no multiple edges, we have $H(X) = H(Y,Z)$. Thus,
\begin{equation}
\begin{aligned}
    R_1^C-R_1^B
    &= H(Y)+H(Z)-H(Y,Z) \\
    &= I(Y;Z).
\end{aligned}
\end{equation}
By the slope $-1$ of segment $B$--$C$, the vertical projection equals the horizontal projection, yielding $|BC| = \sqrt{2} I(Y;Z)$.
\end{IEEEproof}

We now describe a special case where the capacity region takes a simplified form. If one of the degree sequences consists entirely of ones, the det-BC becomes degraded. Without loss of generality, assume $D_Y(\mathcal{B}) = \{1, 1, \ldots, 1\}$. In this case, each output symbol $y$ at receiver~1 corresponds to a unique input symbol $x$, which in turn determines a unique output symbol $z$ at receiver~2. This implies that $Y$ uniquely determines $Z$, i.e., $Z = z(Y)$, forming a Markov chain $X \to Y \to Z$. Therefore, such channels are physically degraded broadcast channels in the sense of \cite{cover_broadcast_1972}, where receiver~2 is degraded with respect to receiver~1.

For degraded det-BCs with $D_Y(\mathcal{B}) = \{1, 1, \ldots, 1\}$, the region $\tilde{\mathcal{C}}(\mathcal{B})$ simplifies to consist only of the boundary segments $A'$--$A$--$B$--$C$. The curved segment $A$--$B$ retains the same parametric characterization as in \eqref{eq:parametric}--\eqref{eq:parametric_simplified}. The line segment $B$--$C$ has slope $-1$ and extends from point $B$ to point $C = (\ln|\mathcal{X}|, 0)$, while the vertical segment $A$--$A'$ connects point $A$ to $A' = (0, \ln|\mathcal{Z}|)$. By symmetry, if $D_Z(\mathcal{B}) = \{1, 1, \ldots, 1\}$ instead (i.e., receiver~1 is degraded with respect to receiver~2), the capacity region boundary consists of segments $D'$--$D$--$C$--$B$.


\section{Graph-Theoretic Framework}
\label{sec:graph}

In this section, we establish the graph-theoretic tools necessary for analyzing det-BC capacity region. We develop three complementary perspectives on det-BCs: the contraction order on channels and graphs (Section~\ref{subsec:contraction}), the degree properties of product channels via Kronecker products (Section~\ref{subsec:degree}), and the biadjacency matrix representation (Section~\ref{subsec:matrix}). These tools provide the foundation for our subsequent capacity region derivation. A key result is the characterization of achievable rate pairs through complete bipartite subgraphs under the contraction order (Theorem~\ref{thm:bipartite_blocklength}), which we reformulate in matrix-theoretic terms (Corollary~\ref{cor:matrix_achievability}). The degree properties of product channels (Theorem~\ref{thm:degree_kronecker}) enable us to analyze finite-blocklength codes through multiset Kronecker products. Together, these results translate the capacity analysis problem into a combinatorial question about graph/matrix contractions and covering properties.

\begin{figure}[t]
    \centering
    \subfloat[]{\includegraphics[width=.25\linewidth]{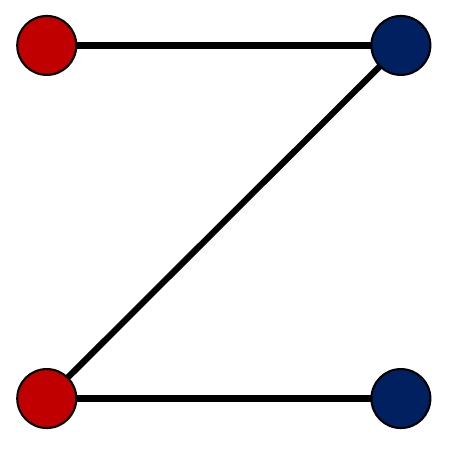}
    \label{fig:contraction, blackwell}}
    \hfil
    \subfloat[]{\includegraphics[width=.25\linewidth]{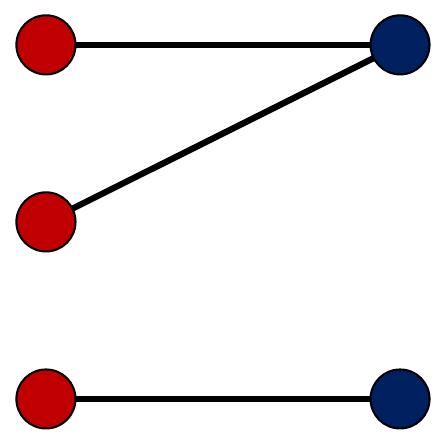}
    \label{fig:contraction, Y2Z}}
    \hfil
    \subfloat[]{\includegraphics[width=.25\linewidth]{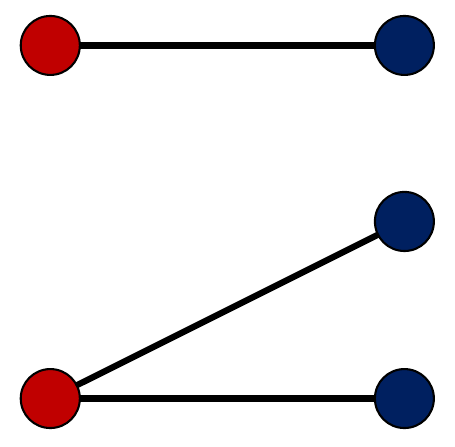}
    \label{fig:contraction, Z2Y}}
    \caption{Examples of vertex contraction and splitting. (a) Blackwell channel. (b) Degraded channel $\mathcal{B}_{Y\to Z}$ obtained by splitting receiver~1 vertices. (c) Degraded channel $\mathcal{B}_{Z\to Y}$ obtained by splitting receiver~2 vertices.}
    \label{fig:vertex contraction}
\end{figure}


\subsection{Equivalent Capacity Characterization via Contraction Order}
\label{subsec:contraction}

We introduce the concept of symbol contraction based on vertex contraction in graph theory. Vertex contraction is an operation that merges two nodes $v_1$ and $v_2$ into a single node $v$ such that $v$ is adjacent to the union of the neighbors of $v_1$ and $v_2$. As illustrated in Fig.~\ref{fig:vertex contraction}, graph (a) can be obtained from either (b) or (c) through vertex contraction. Unlike the standard graph-theoretic definition, for symbol contractions in broadcast channels, we only permit contractions between symbols at the same receiver, i.e., between vertices within the same part of the bipartite graph.

\begin{definition}[Channel Contraction Order]
\label{def:contraction_order}
For broadcast channels $\mathcal{B}_1$ and $\mathcal{B}_2$, we write $\mathcal{B}_1 \prec_c \mathcal{B}_2$ if $\mathcal{B}_1$ can be obtained from $\mathcal{B}_2$ through a finite sequence of output symbol contractions, where each contraction merges two output symbols belonging to the \emph{same} receiver. Contractions between output symbols of different receivers are not permitted. This relation induces a partial order on the space of broadcast channels, referred to as the \emph{channel contraction order}.
\end{definition}

The inverse operation of vertex contraction is vertex splitting. A vertex $v$ with degree $d \geq 2$ can be split into two vertices $v_1$ and $v_2$ by partitioning its incident edges between them. More precisely, if $v$ is adjacent to vertices $\{u_1, u_2, \ldots, u_d\}$ in the opposite part, we can split it into two new vertices $v_1$ and $v_2$ such that $v_1$ is adjacent to a subset $\{u_{i_1}, \ldots, u_{i_k}\}$ and $v_2$ is adjacent to the complementary set $\{u_{j_1}, \ldots, u_{j_{d-k}}\}$, where $\{i_1, \ldots, i_k\} \cup \{j_1, \ldots, j_{d-k}\} = \{1, \ldots, d\}$ and $k \geq 1$, $d-k \geq 1$. Since splitting is the inverse of contraction, $\mathcal{B}_1 \prec_c \mathcal{B}_2$ can be equivalently stated as: $\mathcal{B}_2$ can be obtained from $\mathcal{B}_1$ through a finite sequence of same-side output symbol splittings.

The channel contraction order is closely related to, but strictly stronger than, graph homomorphism. If $\mathcal{B}_1 \prec_c \mathcal{B}_2$, then the bipartite graph $G(\mathcal{B}_1)$ is homomorphic to $G(\mathcal{B}_2)$, i.e., there exists a graph homomorphism $G(\mathcal{B}_1) \to G(\mathcal{B}_2)$. However, the converse does not hold in general: graph homomorphism allows edges to map between arbitrary vertex pairs, whereas the contraction order requires that each contraction merges vertices within the same part of the bipartite graph. Consequently, a graph homomorphism $G(\mathcal{B}_1) \to G(\mathcal{B}_2)$ does not necessarily imply $\mathcal{B}_1 \prec_c \mathcal{B}_2$.

The contraction order has important implications for capacity regions, as established by the following lemma.

\begin{lemma}
\label{lem:contraction_capacity}
If broadcast channel $\mathcal{B}_1$ can be obtained from channel $\mathcal{B}_2$ through symbol contraction, then the capacity region of $\mathcal{B}_1$ is a proper subset of that of $\mathcal{B}_2$, i.e.,
\begin{align}
    \mathcal{C}(\mathcal{B}_1) \subset \mathcal{C}(\mathcal{B}_2) \quad \text{if} \quad \mathcal{B}_1 \prec_c \mathcal{B}_2.
\end{align}
\end{lemma}

\begin{IEEEproof}
Since the contracted received symbols are degraded versions of the original symbols, every rate pair achievable in $\mathcal{B}_1$ is achievable in $\mathcal{B}_2$, so $C(\mathcal{B}_1)\subseteq C(\mathcal{B}_2)$; it remains to show the inclusion is strict. Without loss of generality, assume that contraction is applied at receiver~1, resulting in $|\mathcal{Y}_1| < |\mathcal{Y}_2|$, where $\mathcal{Y}_1$ and $\mathcal{Y}_2$ denote the output alphabets of $\mathcal{B}_1$ and $\mathcal{B}_2$ at receiver~1, respectively. The maximum achievable rate for receiver~1 when receiver~2 receives no information is $R_{1,\max}(\mathcal{B}_i) = \ln|\mathcal{Y}_i|$ for $i \in \{1,2\}$. Therefore, $R_{1,\max}(\mathcal{B}_1) = \ln|\mathcal{Y}_1| < \ln|\mathcal{Y}_2| = R_{1,\max}(\mathcal{B}_2)$, which implies $\mathcal{C}(\mathcal{B}_1) \neq \mathcal{C}(\mathcal{B}_2)$. Since every rate pair achievable in $\mathcal{B}_1$ is also achievable in $\mathcal{B}_2$ (as $\mathcal{B}_1$ is a degraded version of $\mathcal{B}_2$), we conclude that $\mathcal{C}(\mathcal{B}_1) \subset \mathcal{C}(\mathcal{B}_2)$.
\end{IEEEproof}

Next, we establish the connection between achievable rate pairs and complete bipartite graphs. We first consider the single channel use case, then extend to the multiple-use case.

\begin{lemma}
\label{lem:bipartite_single_use}
For a single use ($n=1$) of a det-BC $\mathcal{B}$, a rate pair $(R_1, R_2)$ (in nats) is achievable if and only if there exists a complete bipartite graph $K_{M_1, M_2}$ such that $K_{M_1, M_2} \prec_c G(\mathcal{B})$, where
\begin{subequations}
\begin{align}
    R_1 &= \ln M_1,\\
    R_2 &= \ln M_2.
\end{align}
\end{subequations}
\end{lemma}

\begin{IEEEproof}
Consider a $(1, M_1, M_2)$ code that transmits one of $M_1 M_2$ messages in a single channel use. The message set can be partitioned as $\{1, 2, \ldots, M_1\} \times \{1, 2, \ldots, M_2\}$, where the first component is the message for receiver~1 and the second for receiver~2.
For each message pair $(i, j)$, the encoder selects a channel input symbol $x_{i,j} \in \mathcal{X}$, which produces outputs $y_{i,j} = y(x_{i,j}) \in \mathcal{Y}$ at receiver~1 and $z_{i,j} = z(x_{i,j}) \in \mathcal{Z}$ at receiver~2.
For reliable decoding, receiver~1 must distinguish between all messages with different first components. This requires that for all $i \neq i'$ and any $j, j'$, we have $y_{i,j} \neq y_{i',j'}$. Similarly, receiver~2 requires $z_{i,j} \neq z_{i',j'}$ for all $j \neq j'$ and any $i, i'$.

\textit{Achievability:}
Any complete bipartite graph $K_{M_1, M_2}$ such that $K_{M_1, M_2} \prec_c G(\mathcal{B})$ yields a valid $(1, M_1, M_2)$ code. The $M_1$ vertices in one part of $K_{M_1, M_2}$ correspond to distinct output symbols $\{y_1, y_2, \ldots, y_{M_1}\}$ at receiver~1, and the $M_2$ vertices in the other part correspond to distinct output symbols $\{z_1, z_2, \ldots, z_{M_2}\}$ at receiver~2. Each edge, labeled by an input symbol $x_{i,j}$, connects the corresponding output symbols. Since $K_{M_1, M_2} \prec_c G(\mathcal{B})$, there exists a mapping from the vertices and edges of $G(\mathcal{B})$ to those of $K_{M_1, M_2}$ induced by the contraction sequence. We construct the code by assigning each message pair $(i,j)$ to an edge in $K_{M_1, M_2}$, and transmitting the corresponding pre-image input symbol in $G(\mathcal{B})$.

\textit{Converse:}
Consider any $(1, M_1, M_2)$ code for $\mathcal{B}$. If two output symbols $y_{i,j}$ and $y_{i,j'}$ at receiver~1 correspond to messages with the same first component $i$, contracting them does not affect receiver~1's ability to decode its intended message, and thus does not reduce the achievable $R_1$. Similarly, contracting output symbols $z_{i,j}$ and $z_{i',j}$ at receiver~2 that correspond to messages with the same second component $j$ does not reduce $R_2$. 
By performing all such contractions, we obtain a contracted channel with exactly $M_1$ distinct output symbols at receiver~1 and $M_2$ distinct output symbols at receiver~2. Furthermore, since reliable communication of $M_1 M_2$ messages requires at least $M_1 M_2$ distinct input symbols (edges), the contracted graph must be a complete bipartite graph $K_{M_1, M_2}$. Therefore, $K_{M_1, M_2} \prec_c G(\mathcal{B})$.
\end{IEEEproof}

Using a channel $n$ times is equivalent to finite-blocklength coding with blocklength $n$. The achievable rates in the finite-blocklength regime are characterized by the following theorem.

\begin{theorem}
\label{thm:bipartite_blocklength}
For a det-BC $\mathcal{B}$ with blocklength $n$, a rate pair $(R_1, R_2)$ (in nats) is achievable if and only if there exists a complete bipartite graph $K_{M_1, M_2}$ such that $K_{M_1, M_2} \prec_c G(\mathcal{B}^n)$, where $G(\mathcal{B}^n)$ denotes the bipartite graph of the $n$-fold product channel $\mathcal{B}^n$ and
\begin{subequations}
\begin{align}
    R_1 &= \frac{1}{n}\ln M_1,\\
    R_2 &= \frac{1}{n}\ln M_2.
\end{align}
\end{subequations}
\end{theorem}

\begin{IEEEproof}
The proof follows directly from Lemma~\ref{lem:bipartite_single_use} by noting that $n$ independent uses of the det-BC $\mathcal{B}$ correspond to a single use of the product channel $\mathcal{B}^n$ with bipartite graph representation $G(\mathcal{B}^n)$.
\end{IEEEproof}

\begin{remark}
Lemma~\ref{lem:bipartite_single_use} establishes the fundamental graph-theoretic characterization for single channel use, while Theorem~\ref{thm:bipartite_blocklength} extends it to finite-blocklength coding through the product graph construction. This framework reduces capacity analysis to the combinatorial problem of finding maximum complete bipartite subgraphs under the contraction order.
\end{remark}


\subsection{Degree Properties of Product Channels}
\label{subsec:degree}

We have defined the degree sequences $D_Y(\mathcal{B})$ and $D_Z(\mathcal{B})$ as multisets in Section~\ref{subsec:bipartite}. To establish the degree properties of product channels, we introduce the Kronecker product operation on multisets.

\begin{definition}[Kronecker Product of Multisets]
\label{def:multiset_kronecker}
Let $S_1$ and $S_2$ be multisets with multiplicity functions $m_{S_1}(\cdot)$ and $m_{S_2}(\cdot)$, respectively. Their Kronecker product $S_1 \otimes S_2$ is the multiset whose multiplicity function is
\begin{align}
    \label{eq:kronecker_multiset}
    m_{S_1 \otimes S_2}(t) = \sum_{s_1 s_2 = t} m_{S_1}(s_1) m_{S_2}(s_2)
\end{align}
for all $t \geq 0$. The $n$-fold Kronecker product is defined recursively as
\begin{align}
    S^{\otimes n} = \underbrace{S \otimes S \otimes \cdots \otimes S}_{n \text{ times}}.
\end{align}
\end{definition}

\begin{remark}
The Kronecker product of multisets can be viewed as a generalization of the vector Kronecker product where element ordering is disregarded. Equivalently, it has an intuitive interpretation in terms of pairwise products: if $S_1$ and $S_2$ are multisets, then $S_1 \otimes S_2$ contains all pairwise products $s_1 \cdot s_2$ where $s_1 \in S_1$ and $s_2 \in S_2$, with multiplicities determined by \eqref{eq:kronecker_multiset}. Formally,
\begin{align}
    S_1 \otimes S_2 = \{s_1 \cdot s_2 : s_1 \in S_1, s_2 \in S_2\},
\end{align}
where the braces denote a multiset. For example, if $S_1 = \{2, 3\}$ and $S_2 = \{5, 5\}$, then $S_1 \otimes S_2 = \{10, 10, 15, 15\}$.
\end{remark}

\begin{remark}
An alternative interpretation comes from polynomial expansion under the distributive law, but without combining like terms. Viewing multiset elements as polynomial terms where multiplicities track repetitions, the Kronecker product corresponds to distributing products without subsequent addition. For instance, with $S_1 = \{2, 3\}$ and $S_2 = \{5, 5\}$, we can write symbolically $(2 + 3) \cdot (5 + 5) = 10 + 10 + 15 + 15$. This perspective clarifies why the operation is well-defined and provides computational intuition.
\end{remark}

We now apply these results to the degree sequences of det-BCs.

\begin{lemma}
\label{lem:degree_kronecker}
The degree sequences of the product channel $\mathcal{B}_1 \times \mathcal{B}_2$ are given by the Kronecker products of the degree sequences of the original channels $\mathcal{B}_1$ and $\mathcal{B}_2$:
\begin{subequations}
\label{eq:degree_product}
\begin{align}
    D_Y(\mathcal{B}_1 \times \mathcal{B}_2) &= D_Y(\mathcal{B}_1) \otimes D_Y(\mathcal{B}_2),\\
    D_Z(\mathcal{B}_1 \times \mathcal{B}_2) &= D_Z(\mathcal{B}_1) \otimes D_Z(\mathcal{B}_2).
\end{align}
\end{subequations}
\end{lemma}

\begin{IEEEproof}
Consider a vertex $\mathbf{y} = (y_1, y_2)$ in the receiver~1 part of $G(\mathcal{B}_1 \times \mathcal{B}_2)$, where $y_1 \in \mathcal{Y}_1$ and $y_2 \in \mathcal{Y}_2$. By the definition of the product channel, when input $(x_1, x_2)$ is transmitted, the output at receiver~1 is $(y_1(x_1), y_2(x_2))$. The degree of vertex $\mathbf{y}$ equals the number of input pairs $(x_1, x_2)$ that produce this output, which is the product of the number of inputs producing $y_1$ in $\mathcal{B}_1$ and the number producing $y_2$ in $\mathcal{B}_2$:
\begin{align}
    d_{\mathbf{y}} = d_{Y,y_1}(\mathcal{B}_1) \cdot d_{Y,y_2}(\mathcal{B}_2).
\end{align}
Since every combination of outputs from $\mathcal{Y}_1$ and $\mathcal{Y}_2$ appears in the product channel, the multiset of all vertex degrees is precisely $D_Y(\mathcal{B}_1) \otimes D_Y(\mathcal{B}_2)$ by Definition~\ref{def:multiset_kronecker}. The result for receiver~2 follows by the same argument.
\end{IEEEproof}

\begin{theorem}
\label{thm:degree_kronecker}
The degree sequences of the $n$-fold product channel $\mathcal{B}^n$ are given by the $n$-fold Kronecker products of the degree sequences of the original channel $\mathcal{B}$:
\begin{subequations}
\label{eq:degree_nfold}
\begin{align}
    D_Y(\mathcal{B}^n) &= D_Y^{\otimes n}(\mathcal{B}),\\
    D_Z(\mathcal{B}^n) &= D_Z^{\otimes n}(\mathcal{B}).
\end{align}
\end{subequations}
\end{theorem}

\begin{IEEEproof}
The result follows by applying Lemma~\ref{lem:degree_kronecker} recursively $n-1$ times.
\end{IEEEproof}


\subsection{Biadjacency Matrix Representation}
\label{subsec:matrix}

In this subsection, we introduce an equivalent matrix representation of det-BCs based on the biadjacency matrix of the bipartite graph. This algebraic framework provides a convenient alternative to the graph-theoretic perspective and facilitates the analysis of product channels and contraction operations. All results derived for bipartite graphs have natural counterparts in the matrix representation.

\begin{definition}[Biadjacency Matrix]
\label{def:biadjacency}
For a bipartite graph $G(\mathcal{B}) = (\mathcal{Y}, \mathcal{Z}, E)$, the biadjacency matrix $A(\mathcal{B})$ is the $|\mathcal{Y}| \times |\mathcal{Z}|$ binary matrix defined by
\begin{align}
    A_{ij} = \begin{cases}
        1, & \text{if } (y^{(i)}, z^{(j)}) \in E,\\
        0, & \text{otherwise}.
    \end{cases}
\end{align}
\end{definition}

The product operation of det-BCs corresponds to the Kronecker product of their biadjacency matrices.

\begin{lemma}
\label{lem:matrix_product}
For any det-BC $\mathcal{B}$ with blocklength $n$,
\begin{align}
    A(\mathcal{B}^n) = A^{\otimes n}(\mathcal{B}).
\end{align}
\end{lemma}

\begin{IEEEproof}
This follows directly from the definition of the product channel and the Kronecker product of matrices: an edge exists in $G(\mathcal{B}^n)$ between $(y^{(i_1)}, \ldots, y^{(i_n)})$ and $(z^{(j_1)}, \ldots, z^{(j_n)})$ if and only if $A_{i_k,j_k} = 1$ for all $k = 1, \ldots, n$, which is precisely the condition for $(A^{\otimes n})_{(i_1,\ldots,i_n),(j_1,\ldots,j_n)} = 1$.
\end{IEEEproof}

The contraction operations on bipartite graphs have natural matrix interpretations. Element-wise OR operations on rows of the biadjacency matrix $A(\mathcal{B})$ correspond to contracting vertices on the $\mathcal{Y}$ side of $G(\mathcal{B})$; analogously, OR operations on columns correspond to contractions on the $\mathcal{Z}$ side. This motivates the following matrix-based definition of the contraction order.

\begin{definition}[Matrix Contraction Order]
\label{def:matrix_contraction}
Let $A_1$ and $A_2$ be biadjacency matrices. We write $A_1 \prec_c A_2$ if $A_1$ can be obtained from $A_2$ through a finite sequence of the following operations:
\begin{itemize}
    \item \emph{Row contractions}: replacing a set of rows by their element-wise OR;
    \item \emph{Column contractions}: replacing a set of columns by their element-wise OR.
\end{itemize}
\end{definition}

The matrix contraction order is equivalent to the graph contraction order defined in Section~\ref{sec:graph}.

\begin{lemma}
\label{lem:matrix_graph_equivalence}
Let $\mathcal{B}_1$ and $\mathcal{B}_2$ be det-BCs. Then 
\begin{align}
    A(\mathcal{B}_1) \prec_c A(\mathcal{B}_2) \quad \Longleftrightarrow \quad G(\mathcal{B}_1) \prec_c G(\mathcal{B}_2).
\end{align}
\end{lemma}

\begin{IEEEproof}
The equivalence follows by construction: each row (resp. column) contraction in the matrix corresponds to a vertex contraction on the $\mathcal{Y}$ side (resp. $\mathcal{Z}$ side) of the bipartite graph, and vice versa. The element-wise OR operation on rows (columns) precisely captures the union of edge neighborhoods required in the definition of vertex contraction.
\end{IEEEproof}

The equivalence between bipartite graphs and their biadjacency matrices allows us to restate Theorem~\ref{thm:bipartite_blocklength} in matrix-theoretic terms.

\begin{corollary}[Matrix Characterization of Achievability]
\label{cor:matrix_achievability}
A rate pair $(R_1, R_2)$ is achievable for det-BC $\mathcal{B}$ with blocklength $n$ if and only if there exists an all-ones matrix $J_{M_1, M_2}$ such that
\begin{align}
    J_{M_1, M_2} \prec_c A^{\otimes n}(\mathcal{B}),
\end{align}
where $R_1 = \frac{1}{n}\ln M_1$ and $R_2 = \frac{1}{n}\ln M_2$, and $J_{M_1, M_2}$ denotes the $M_1 \times M_2$ all-ones matrix.
\end{corollary}

\begin{IEEEproof}
This is an immediate consequence of Theorem~\ref{thm:bipartite_blocklength} and Lemma~\ref{lem:matrix_graph_equivalence}. A complete bipartite graph $K_{M_1, M_2}$ corresponds to the all-ones matrix $J_{M_1, M_2}$, and the contraction orders on graphs and matrices are equivalent.
\end{IEEEproof}


\section{Derivation of the Explicit Capacity Region}
\label{sec:derivation}

In this section, we prove Theorem~\ref{thm:main_capacity} by establishing that $\mathcal{C}(\mathcal{B}) = \tilde{\mathcal{C}}(\mathcal{B})$ for all det-BCs. Our proof strategy consists of three components. Section~\ref{subsec:type} develops the method of types for product channels, introducing type-induced submatrices that enable asymptotic analysis of finite-blocklength codes. Section~\ref{subsec:degraded_proof} establishes the result for degraded det-BCs, where the forest structure of the bipartite graph permits a more direct analysis. Section~\ref{subsec:general_proof} then extends to general det-BCs by demonstrating that any det-BC's capacity region can be characterized as the intersection of two degraded channel capacity regions obtained through vertex splitting operations.


\subsection{The Method of Types for Product Channels}
\label{subsec:type}

To analyze the Kronecker powers of the biadjacency matrix, we adapt the method of types to the bipartite graph representation. This subsection develops the combinatorial machinery required for the achievability proof by characterizing the properties of specific submatrices of $A^{\otimes n}(\mathcal{B})$.

Recall that the biadjacency matrix $A^{\otimes n}(\mathcal{B})$ has dimensions $|\mathcal{Y}|^n \times |\mathcal{Z}|^n$. Each entry $A^{\otimes n}_{(\mathbf{y}, \mathbf{z})}$ corresponds to whether there exists an input sequence $\mathbf{x} \in \mathcal{X}^n$ that produces output sequences $\mathbf{y} \in \mathcal{Y}^n$ at receiver~1 and $\mathbf{z} \in \mathcal{Z}^n$ at receiver~2. Each row is indexed by a sequence $\mathbf{y} = (y_{k_1}, \ldots, y_{k_n})$, and each column by a sequence $\mathbf{z} = (z_{\ell_1}, \ldots, z_{\ell_n})$.

Following the standard method of types \cite{csiszar_information_2015}, we characterize sequences by their empirical distributions. For the det-BC, the input alphabet $\mathcal{X}$ can be identified with the set of edges in the bipartite graph $G(\mathcal{B})$, or equivalently, with pairs $(y^{(i)}, z^{(j)})$ such that $A_{ij} = 1$. We denote such input symbols by $x^{(i,j)}$ to emphasize the correspondence.

\begin{definition}[Input Type]
\label{def:input_type}
For an input sequence $\mathbf{x} = (x_1, \ldots, x_n) \in \mathcal{X}^n$, let $n_X(i,j)$ denote the number of occurrences of symbol $x^{(i,j)}$ in $\mathbf{x}$, where $x^{(i,j)}$ corresponds to the input that produces output $(y^{(i)}, z^{(j)})$. The \emph{type} of $\mathbf{x}$ is the empirical distribution $P_{\mathbf{x}}$ on $\mathcal{X}$ defined by
\begin{align}
    P_{\mathbf{x}}\left(x^{(i,j)}\right) = \frac{n_X(i,j)}{n}, \quad \forall (i,j): A_{ij} = 1.
\end{align}
For any distribution $P_\mathbf{x}$ on $\mathcal{X}$, the set of sequences of type $P_\mathbf{x}$ is called the \emph{type class} of $P_\mathbf{x}$, denoted by $T(P_\mathbf{x})$.
\end{definition}

\begin{definition}[Output Types]
\label{def:output_type}
For sequences $\mathbf{y} \in \mathcal{Y}^n$ and $\mathbf{z} \in \mathcal{Z}^n$, let $n_Y(i)$ denote the number of occurrences of symbol $y^{(i)}$ in $\mathbf{y}$, and let $n_Z(j)$ denote the number of occurrences of symbol $z^{(j)}$ in $\mathbf{z}$. The marginal types are
\begin{align}
    P_{\mathbf{y}}\left(y^{(i)}\right) &= \frac{n_Y(i)}{n}, \quad i = 1, \ldots, |\mathcal{Y}|,\\
    P_{\mathbf{z}}\left(z^{(j)}\right) &= \frac{n_Z(j)}{n}, \quad j = 1, \ldots, |\mathcal{Z}|.
\end{align}
Let $T(P_\mathbf{y})$ and $T(P_\mathbf{z})$ denote the set of sequences $\mathbf{y}$ of type $P_\mathbf{y}$ and the set of sequences $\mathbf{z}$ of type $P_\mathbf{z}$. The \emph{joint type} $P_{\mathbf{y},\mathbf{z}}$ is defined by
\begin{align}
    P_{\mathbf{y},\mathbf{z}}\left(y^{(i)}, z^{(j)}\right) = \frac{n_{Y,Z}(i,j)}{n},
\end{align}
where $n_{Y,Z}(i,j)$ counts the number of positions $k \in \{1, \ldots, n\}$ where $y_k = y^{(i)}$ and $z_k = z^{(j)}$ simultaneously.
\end{definition}

\begin{remark}
For deterministic broadcast channels, the input type and the joint output type are equivalent: $n_X(i,j) = n_{Y,Z}(i,j)$ for all positions where $A_{ij} = 1$, since each input uniquely determines the output pair.
\end{remark}

\begin{definition}[Type-Induced Subgraph]
\label{def:type_subgraph}
For a det-BC $\mathcal{B}$ and a fixed input type $P_{\mathbf{x}}$ on $\mathcal{X}$, the \emph{type-induced subgraph} $G(\mathcal{B}^n, P_{\mathbf{x}})$ is the edge-induced subgraph of $G(\mathcal{B}^n)$ whose edge set is the type class $T(P_{\mathbf{x}})$ and whose vertex sets consist of all endpoints of edges in $T(P_{\mathbf{x}})$.
\end{definition}

\begin{remark}[Typical Vertex Sets]
The type-induced subgraph $G(\mathcal{B}^n, P_{\mathbf{x}})$ is a bipartite graph with vertex parts $T(P_{\mathbf{y}})$ and $T(P_{\mathbf{z}})$, where $P_{\mathbf{y}}$ and $P_{\mathbf{z}}$ are the marginal distributions induced by $P_{\mathbf{x}}$ through the deterministic channel mapping.
\end{remark}

\begin{remark}[Relation to Biadjacency Matrix]
The biadjacency matrix $A(\mathcal{B}^n, P_{\mathbf{x}})$ of the type-induced subgraph can be viewed as a type-constrained submatrix of $A^{\otimes n}(\mathcal{B})$. It is obtained by selecting rows and columns indexed by $T(P_{\mathbf{y}})$ and $T(P_{\mathbf{z}})$, respectively, and retaining only those entries corresponding to edges in $T(P_{\mathbf{x}})$. Entries whose row and column indices lie in the typical sets but whose underlying input sequence does not belong to $T(P_{\mathbf{x}})$ are set to zero.
\end{remark}

\begin{lemma}[Asymptotic Properties of Type-Induced Subgraphs]
\label{lem:submatrix_asymptotics}
Let $P_{\mathbf{x}}$ be a fixed input type with induced joint output type $P_{Y,Z}$. For sufficiently large $n$, the type-induced subgraph $G(\mathcal{B}^n, P_{\mathbf{x}})$ is a biregular bipartite graph with vertex set cardinalities
\begin{subequations}
\label{eq:typical_set_sizes}
\begin{equation}
\begin{aligned}
    |T(P_{\mathbf{y}})|
    &= \frac{n!}{\prod_{i=1}^{|\mathcal{Y}|} n_Y(i)!} \\
    &= \exp\left[ n H(Y) + O(\ln n) \right],
\end{aligned}
\end{equation}
\begin{equation}
\begin{aligned}
    |T(P_{\mathbf{z}})|
    &= \frac{n!}{\prod_{j=1}^{|\mathcal{Z}|} n_Z(j)!}\\
    &= \exp\left[ n H(Z) + O(\ln n) \right],
\end{aligned}
\end{equation}
\end{subequations}
and vertex degrees
\begin{subequations}
\label{eq:degree_growth}
    \begin{equation}
    \begin{aligned}
    d_Y(P_{\mathbf{x}})
    &= \prod_{i=1}^{|\mathcal{Y}|}
    \frac{n_Y(i)!}
    {\prod_{j=1}^{|\mathcal{Z}|} n_{Y,Z}(i,j)!} \\
    &= \exp\left[
    n H(Z|Y) + O(\ln n)
    \right],
    \end{aligned}
    \end{equation}
    \begin{equation}
    \begin{aligned}
    d_Z(P_{\mathbf{x}})
    &= \prod_{j=1}^{|\mathcal{Z}|}
    \frac{n_Z(j)!}
    {\prod_{i=1}^{|\mathcal{Y}|} n_{Y,Z}(i,j)!} \\
    &= \exp\left[
    n H(Y|Z) + O(\ln n)
    \right].
    \end{aligned}
    \end{equation}
\end{subequations}
where all entropies are evaluated with respect to $P_{\mathbf{y},\mathbf{z}}$.
\end{lemma}

\begin{IEEEproof}
We first establish biregularity. For any two sequences $\mathbf{y}, \mathbf{y}' \in T(P_{\mathbf{y}})$, there exists a permutation $\pi$ of $\{1, \ldots, n\}$ such that $\mathbf{y}' = \pi(\mathbf{y})$. Due to the memoryless structure of the product channel, this permutation induces a bijection between edges incident to $\mathbf{y}$ and edges incident to $\mathbf{y}'$ within $T(P_{\mathbf{x}})$. Therefore, all vertices in $T(P_{\mathbf{y}})$ have the same degree $d_Y(P_{\mathbf{x}})$. By symmetry, all vertices in $T(P_{\mathbf{z}})$ have degree $d_Z(P_{\mathbf{x}})$, confirming biregularity.

For the asymptotic cardinalities, we apply Stirling's approximation $\ln n! = n \ln n - n + O(\ln n)$ to the multinomial coefficients. For $|T(P_{\mathbf{y}})|$:
\begin{align}
    &\ln |T(P_{\mathbf{y}})| = \ln n! - \sum_{i} \ln n_Y(i)! \notag\\
    &= n \ln n - n - \sum_{i} (n_Y(i) \ln n_Y(i) - n_Y(i)) + O(\ln n) \notag\\
    &= -n \sum_{i} P_Y(i) \ln P_Y(i) + O(\ln n) \notag\\
    &= n H(Y) + O(\ln n).
\end{align}

For the degree $d_Z(P_{\mathbf{x}})$, we have
\begin{align}
    &\ln d_Z(P_{\mathbf{x}}) = \sum_{j} \left( \ln n_Z(j)! - \sum_{i} \ln n_{Y,Z}(i,j)! \right) \notag\\
    &= \sum_{j} \left[ n_Z(j) \ln n_Z(j) - \sum_{i} n_{Y,Z}(i,j) \ln n_{Y,Z}(i,j) \right] \notag\\
    &\quad\ + O(\ln n) \notag\\
    &= -n \sum_{j} P_Z(j) \sum_{i} P_{Y|Z}(i|j) \ln P_{Y|Z}(i|j) + O(\ln n) \notag\\
    &= n \sum_{j} P_Z(j) H(Y|Z=z^{(j)}) + O(\ln n) \notag\\
    &= n H(Y|Z) + O(\ln n).
\end{align}
The derivation for $d_Y(P_{\mathbf{x}})$ and $|T(P_{\mathbf{z}})|$ follows by symmetry.
\end{IEEEproof}

\begin{remark}[Role of Lemma~7]
\label{rem:lemma7_role}
Lemma~\ref{lem:subgraph_covering} is a technical tool used in the achievability proof. Building on the biregularity and asymptotic vertex counts established in Lemma~\ref{lem:submatrix_asymptotics}, it shows via a random binning argument that the corner point $\left( H(Y|Z), H(Z) \right)$ is achievable (Corollary~\ref{cor:corner_points}). This corner point is in turn the building block for the achievability of the entire boundary of $\tilde{\mathcal{C}}\left( \mathcal{B} \right)$ established in Lemma~\ref{lem:degraded_achievability} and Lemma~\ref{lem:general_achievability}. The converse direction, developed in Section~\ref{sec:derivation} through the degraded capacity region converse (Lemma~\ref{lem:degraded_converse}), the supporting-hyperplane representation (Lemma~\ref{lem:SHR}), the intersection of degraded capacity regions (Lemma~\ref{lem:intersection_identity}), and the general converse (Lemma~\ref{lem:general_converse}), does not rely on Lemma~\ref{lem:subgraph_covering} and may be read independently of it.
\end{remark}

\begin{lemma}
\label{lem:subgraph_covering}
For any det-BC $\mathcal{B}$ and any input distribution $P_X$, the type-induced subgraph $G(\mathcal{B}^n, P_{\mathbf{x}})$ satisfies
\begin{align}
    \label{eq:subgraph_contraction_prob}
    \lim_{n \to \infty} \Pr\left\{ K_{M_1, M_2} \prec_c G(\mathcal{B}^n, P_{\mathbf{x}}) \right\} = 1,
\end{align}
if for any $\epsilon > 0$, the message sizes satisfy
\begin{align}
    \frac{1}{n} \ln M_1 &= H(Y|Z) - \epsilon, \\
    \frac{1}{n} \ln M_2 &= H(Z) + O\left(\frac{\ln n}{n}\right),
\end{align}
where the entropies are evaluated with respect to the joint distribution $P_{Y,Z}$ induced by $P_X$.
\end{lemma}

\begin{IEEEproof}
\textit{Intuition.}
By Lemma~\ref{lem:submatrix_asymptotics}, $G\left( \mathcal{B}^n, P_\mathbf{x} \right)$ is a biregular bipartite graph in which every row vertex (sequence in $\mathcal{T}\left( P_\mathbf{y} \right)$) has degree $d_Y\left( P_\mathbf{x} \right)$ and every column vertex (sequence in $\mathcal{T}\left( P_\mathbf{z} \right)$) has degree $d_Z\left( P_\mathbf{x} \right)$. By Definition~\ref{def:contraction_order}, establishing $K_{M_1,M_2} \prec_c G\left( \mathcal{B}^n, P_\mathbf{x} \right)$ amounts to partitioning the row vertices into $M_1$ bins such that, after row contraction (element-wise OR), every bin is connected to every column vertex. We construct such a partition by assigning row vertices to the $M_1$ bins uniformly at random. Each column vertex $\mathbf{z}_j$ has $d_Z\left( P_\mathbf{x} \right) = \exp\left( nH(Y|Z) + O(\ln n) \right)$ neighbors among the $\left| \mathcal{T}\left( P_\mathbf{y} \right) \right|= \exp\left( nH(Y) + O(\ln n) \right)$ row vertices. With $M_1 = \left\lfloor \exp\left( n\left( H(Y|Z) - \epsilon \right) \right) \right\rfloor$, each bin contains roughly $\exp\left( n\left( I(Y;Z) + \epsilon \right) + O(\ln n) \right)$ row vertices, an exponentially larger sample than the fraction $1/M_1$ of $\left| \mathcal{T}\left( P_\mathbf{y} \right) \right|$ needed in expectation to cover the $d_Z\left( P_\mathbf{x} \right)$ neighbors of $\mathbf{z}_j$. A standard hypergeometric tail bound then shows that the probability that a given bin misses $\mathbf{z}_j$'s neighborhood decays doubly exponentially in $n$. The union bound over all $M_1 M_2$ bin-column pairs in~\eqref{eq:double exponential dominates} shows that this failure probability vanishes as $n \to \infty$, even though $M_1 M_2$ itself grows only exponentially in $n$, as discussed following~\eqref{eq:double exponential dominates}. We now make this argument precise.

To establish the contraction $K_{M_1, M_2} \prec_c G(\mathcal{B}^n, P_{\mathbf{x}})$, or equivalently $J_{M_1, M_2} \prec_c A(\mathcal{B}^n, P_{\mathbf{x}})$ in the matrix representation, we employ a random binning argument. Partition the row vertex set $T(P_{\mathbf{y}})$ into $M_1$ disjoint bins $\{B_1, \dots, B_{M_1}\}$ of equal size, while keeping all column vertices in $T(P_{\mathbf{z}})$ distinct (i.e., $M_2 = |T(P_{\mathbf{z}})|$). By Definition~\ref{def:matrix_contraction}, a contraction to $J_{M_1, M_2}$ succeeds if and only if every bin $B_i$ has at least one edge connecting to every column vertex $\mathbf{z}_j \in T(P_{\mathbf{z}})$.

Set $M_1 = \lfloor \exp(n(H(Y|Z)-\epsilon)) \rfloor$ for some $\epsilon > 0$. Each bin then has size
\begin{align}
    |B_i| = \left\lfloor \frac{|T(P_{\mathbf{y}})|}{M_1} \right\rfloor.
\end{align}
The floor function introduces an $O(1)$ additive term, which is absorbed into the $O(\ln n)$ term when taking logarithms. By Lemma~\ref{lem:submatrix_asymptotics},
\begin{align}
    \ln |B_i| &= \ln |T(P_{\mathbf{y}})| - \ln M_1 + O(\ln n) \notag\\
    &= n H(Y) - n(H(Y|Z) - \epsilon) + O(\ln n) \notag\\
    &= n(I(Y;Z) + \epsilon) + O(\ln n).
\end{align}

Let $E_{i,j}$ denote the failure event that bin $B_i$ is not connected to column vertex $\mathbf{z}_j$. By the biregularity established in Lemma~\ref{lem:submatrix_asymptotics}, each column vertex has degree $d_Z = \exp\left[n H(Y|Z) + O(\ln n)\right]$. When randomly selecting $|B_i|$ rows from the $|T(P_{\mathbf{y}})|$ available rows to form bin $B_i$, the probability of avoiding all $d_Z$ neighbors of $\mathbf{z}_j$ follows a hypergeometric distribution:
\begin{align}
    \Pr\{E_{i,j}\}
    &= \frac{\binom{|T(P_{\mathbf{y}})| - d_Z}{|B_i|}}{\binom{|T(P_{\mathbf{y}})|}{|B_i|}} \notag \\
    &= \prod_{k=0}^{|B_i|-1} \left( 1 - \frac{d_Z}{|T(P_{\mathbf{y}})| - k} \right) \notag \\
    &\leq \left( 1 - \frac{d_Z}{|T(P_{\mathbf{y}})|} \right)^{|B_i|}.
\end{align}
Taking the logarithm and applying the inequality $\ln(1-a) \leq -a$ for $0 < a < 1$, we obtain
\begin{equation}
\begin{aligned}
    \ln \Pr\{E_{i,j}\}
    &\leq |B_i| \ln \left( 1 - \frac{d_Z}{|T(P_{\mathbf{y}})|} \right) \\
    &\leq - |B_i| \frac{d_Z}{|T(P_{\mathbf{y}})|} \\
    &= -\exp\left[ n(I(Y;Z) + \epsilon)+O(\ln n)\right] \frac{\exp\left[nH(Y|Z)+O(\ln n)\right]}{\exp\left[nH(Y)+O(\ln n)\right]} \\
    &= -\exp\left[n \epsilon + O(\ln n) \right].
\end{aligned}
\end{equation}

Applying the union bound over all $M_1 \times M_2$ bin-column pairs, the failure probability satisfies
\begin{equation}
\begin{aligned}
    &\quad \Pr\left\{K_{M_1,M_2} \nprec_c G(\mathcal{B}^n,P_{\mathbf{x}})\right\}\\
    &\leq\sum_{i=1}^{M_1}\sum_{j=1}^{M_2}\Pr\{E_{i,j}\}\\
    &\leq M_1M_2\,\exp\left\{-\exp\left[n\epsilon+O(\ln n)\right]\right\}\\
    &=\exp\left[n(H(Y,Z)-\epsilon)+O(\ln n)\right] \exp\left\{-\exp\left[n\epsilon+O(\ln n)\right]\right\}.
    \label{eq:probability, subgraph covering}
\end{aligned}
\end{equation}
Since $M_1 M_2 = \exp\left[n(H(Y,Z)-\epsilon) + O(\ln n)\right]$ grows \emph{exponentially} in $n$, while the factor $\exp\left\{-\exp\left[n\epsilon + O(\ln n)\right]\right\}$ decays \emph{doubly exponentially} in $n$, the product on the right-hand side of~\eqref{eq:probability, subgraph covering} vanishes as $n\to\infty$. To see this, note that for any fixed constants $a>0$ and $b>0$,
\begin{equation}
\label{eq:double exponential dominates}
\begin{aligned}
    &\quad\,\exp\left[an+O(\ln n)\right]\,\exp\left[-e^{\,bn+O(\ln n)}\right] \\
    &=\exp\left[an+O(\ln n)-e^{\,bn+O(\ln n)}\right]\to 0
    \quad\text{as }n\to\infty,
\end{aligned}
\end{equation}
because the exponential term $e^{bn}$ grows strictly faster than the linear term $an$ for any $b>0$. Applying this with $a=H(Y,Z)-\epsilon$ and $b=\epsilon$ establishes \eqref{eq:subgraph_contraction_prob} and completes the proof.
\end{IEEEproof}

\begin{corollary}
\label{cor:corner_points}
For any det-BC $\mathcal{B}$ and any input distribution $P_X$, the rate pair 
\begin{align}
    (R_1, R_2) = (H(Y|Z), H(Z))
\end{align}
is achievable, where the entropies are evaluated with respect to the joint distribution $P_{Y,Z}$ induced by $P_X$. By symmetry, the rate pair $(H(Y), H(Z|Y))$ is also achievable.
\end{corollary}

\begin{IEEEproof}
Fix $\epsilon > 0$ and set $M_1 = \lfloor \exp(n(H(Y|Z) - \epsilon)) \rfloor$ and $M_2 = |T(P_{\mathbf{z}})|$. By Lemma~\ref{lem:subgraph_covering}, we have
\begin{align}
    \Pr\left\{ K_{M_1, M_2} \prec_c G(\mathcal{B}^n, P_{\mathbf{x}}) \right\} = 1 - o(1)
\end{align}
as $n \to \infty$. Since $G(\mathcal{B}^n, P_{\mathbf{x}})$ is a subgraph of $G(\mathcal{B}^n)$ by Definition~\ref{def:type_subgraph}, and the contraction operation is monotonic with respect to edge inclusion (Definition~\ref{def:matrix_contraction}), i.e., adding edges can only facilitate contraction into a complete bipartite graph, we have $K_{M_1, M_2} \prec_c G(\mathcal{B}^n)$ with probability $1 - o(1)$. 

By Theorem~\ref{thm:bipartite_blocklength}, this establishes the achievability of
\begin{equation}
\begin{aligned}
    (R_1, R_2)
    &= \lim_{n\to\infty}\left( \frac{1}{n} \ln M_1, \frac{1}{n} \ln M_2 \right) \\
    &= \left(H(Y|Z) - \epsilon, H(Z)\right).
\end{aligned}
\end{equation}
Since $\epsilon$ can be chosen arbitrarily small and the capacity region is the closure of achievable rate pairs, the boundary point $(H(Y|Z), H(Z))$ is achievable. The symmetric result $(H(Y), H(Z|Y))$ follows by exchanging the roles of receivers~1 and~2.
\end{IEEEproof}

\begin{remark}
The rate pair $(H(Y|Z), H(Z))$ coincides with the corner point of the capacity region previously established via probabilistic arguments, derivable from the entropy inequalities. But we recover this result through a purely combinatorial argument, demonstrating the power of graph-theoretic framework.
\end{remark}


\subsection{Degraded Det-BCs}
\label{subsec:degraded_proof}

In this subsection, we establish that the capacity region $\tilde{\mathcal{C}}(\mathcal{B})$ characterized in Section~\ref{subsec:explicit_capacity} equals the true capacity region $\mathcal{C}(\mathcal{B})$ for degraded det-BCs. Without loss of generality, we consider the physically degraded case where $X \to Y \to Z$ forms a Markov chain, i.e., $Z = z(Y)$. The bipartite graph representations of such channels have a special structural property: they are forests (collections of disjoint trees). This forest structure arises because each vertex $y$ at receiver~1 has degree 1, connecting to a unique input symbol, which in turn determines a unique output at receiver~2. Consequently, the degree sequence $D_Z(\mathcal{B})$ uniquely determines the channel structure up to vertex relabeling. We prove the result by establishing achievability in Lemma~\ref{lem:degraded_achievability} and the converse in Lemma~\ref{lem:degraded_converse}, which together yield Theorem~\ref{thm:degraded_capacity}.

\begin{lemma}[Achievability]
\label{lem:degraded_achievability}
Any rate pair $(R_1, R_2) \in \tilde{\mathcal{C}}(\mathcal{B})$ is achievable for the degraded det-BC $\mathcal{B}$.
\end{lemma}

\begin{IEEEproof}
The proof follows directly from the universal achievability of corner points established in Corollary~\ref{cor:corner_points}. We construct a specific input distribution to achieve any point on the curved boundary segment $A$--$B$. For $t \in [0,1]$, define the probability distribution $q_Z$ on $\mathcal{Z}$ by
\begin{align}
    q_Z(z) = \frac{d_{Z,z}^t}{\sum_{z' \in \mathcal{Z}} d_{Z,z'}^t},\quad z\in\mathcal{Z},
\end{align}
as in Eq.~\eqref{eq:type q}. Since each $y \in \mathcal{Y}$ maps to a unique $z \in \mathcal{Z}$ under the degraded channel, we define the conditional distribution
\begin{align}
    \label{eq:uniform_conditional}
    P_{Y|Z}(y|z) =
    \begin{cases}
    \frac{1}{d_{Z,z}} & \text{if } y \mapsto z \text{ under channel }\mathcal{B}, \\
    0 & \text{otherwise}.
    \end{cases}
\end{align}
This uniform distribution over the preimage $\{y : z(y) = z\}$ induces the joint distribution $P_{Y,Z}(y,z) = q_Z(z) \cdot P_{Y|Z}(y|z)$. The corresponding input distribution is $P_X(x) = P_{Y,Z}(y(x), z(x))$.

For this distribution, the entropies evaluate to
\begin{subequations}
    \begin{equation}
    \begin{aligned}
    H(Z)
    &= -\sum_{z \in \mathcal{Z}} q_Z(z) \ln q_Z(z) \\
    &= H(q_Z),
    \end{aligned}
    \end{equation}
    \begin{equation}
    \begin{aligned}
    H(Y|Z)
    &= \sum_{z \in \mathcal{Z}} q_Z(z) H(Y|Z=z) \\
    &= \sum_{z \in \mathcal{Z}} q_Z(z) \ln d_{Z,z}.
    \end{aligned}
    \end{equation}
\end{subequations}
By direct substitution of the definition of $q_Z(z)$, these expressions match exactly the parametric equations as in Eq.~\eqref{eq:parametric_simplified}:
\begin{align}
    R_1(t)
    &= t^{-1}\left(\ln\sum_{z \in \mathcal{Z}} d_{Z,z}^t - H(q_Z)\right)\notag\\
    &= H(Y|Z),\\
    R_2(t)
    &= H(q_Z)\notag\\
    &= H(Z),
\end{align}
where the first equality follows from the identity
\begin{equation}
\begin{aligned}
    \ln\sum_{z \in \mathcal{Z}} d_{Z,z}^t - H(q_Z)
    &= \sum_{z \in \mathcal{Z}} q_Z(z) \ln d_{Z,z}^t \\
    &= t \sum_z q_Z(z) \ln d_{Z,z} \\
    &= t H(Y|Z).
\end{aligned}
\end{equation}

By Corollary~\ref{cor:corner_points}, the rate pair $(H(Y|Z), H(Z))$ is achievable for this input distribution. As $t$ varies over $[0,1]$, we trace out the entire curved segment $A$--$B$. The line segments $A$--$A'$ and $B$--$C$ are achieved by time-sharing and operating at the corner points $(0, \ln|\mathcal{Z}|)$ and $(\ln|\mathcal{X}|, 0)$ respectively, completing the boundary of $\tilde{\mathcal{C}}(\mathcal{B})$.
\end{IEEEproof}

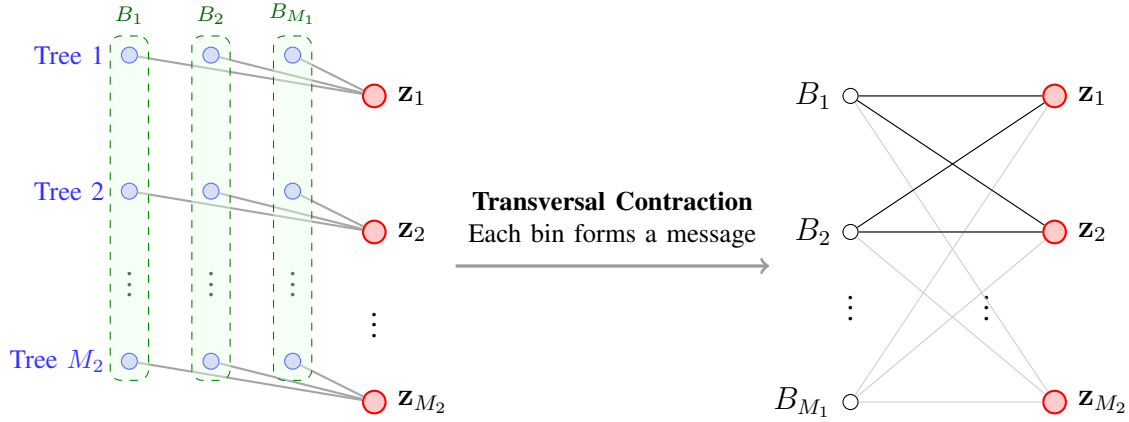
\begin{figure}[t]
    \centering
    \begin{tikzpicture}[
        scale=0.9,
        y_node/.style={circle, fill=blue!20, draw=blue, inner sep=2pt, minimum size=6pt},
        z_node/.style={circle, fill=red!20, draw=red, inner sep=3pt, thick, minimum size=6pt},
        contraction_arrow/.style={->, shorten >=5pt, shorten <=5pt, very thick, draw=gray!80},
        group_box/.style={draw=green!50!black, dashed, rounded corners, inner sep=4pt, fill=green!10, fill opacity=0.4}
    ]

    
    \node[z_node, label=right:$\mathbf{z}_1$] (z1) at (3.6, 3) {};
    \node[y_node] (y1_1) at (0, 3.6) {}; 
    \node[y_node] (y1_2) at (1.2, 3.6) {};
    \node[y_node] (y1_3) at (2.4, 3.6) {};
    \foreach \node in {y1_1, y1_2, y1_3} \draw[thick, gray!70] (\node) -- (z1);
    \node[left, font=\small, color=blue!80] at (-0.2, 3.6) {Tree 1};

    \node[z_node, label=right:$\mathbf{z}_2$] (z2) at (3.6, 1) {};
    \node[y_node] (y2_1) at (0, 1.6) {};
    \node[y_node] (y2_2) at (1.2, 1.6) {};
    \node[y_node] (y2_3) at (2.4, 1.6) {};
    \foreach \node in {y2_1, y2_2, y2_3} \draw[thick, gray!70] (\node) -- (z2);
    \node[left, font=\small, color=blue!80] at (-0.2, 1.6) {Tree 2};

    \node at (0, 0.35) {$\vdots$};
    \node at (1.2, 0.35) {$\vdots$};
    \node at (2.4, 0.35) {$\vdots$};
    \node at (3.6, -0.25) {$\vdots$};

    \node[z_node, label=right:$\mathbf{z}_{M_2}$] (zM) at (3.6, -1.5) {};
    \node[y_node] (yM_1) at (0, -0.9) {};
    \node[y_node] (yM_2) at (1.2, -0.9) {};
    \node[y_node] (yM_3) at (2.4, -0.9) {};
    \foreach \node in {yM_1, yM_2, yM_3} \draw[thick, gray!70] (\node) -- (zM);
    \node[left, font=\small, color=blue!80] at (-0.2, -0.9) {Tree $M_2$};

    \node[group_box, fit=(y1_1) (yM_1), label=above:\textcolor{green!40!black}{\scriptsize $B_1$}] (bin1) {};
    \node[group_box, fit=(y1_2) (yM_2), label=above:\textcolor{green!40!black}{\scriptsize $B_2$}] (bin2) {};
    \node[group_box, fit=(y1_3) (yM_3), label=above:\textcolor{green!40!black}{\scriptsize $B_{M_1}$}] (bin3) {};

    \draw[contraction_arrow] (4.6, 0.5) -- (9.6, 0.5);
    \node[align=center, font=\small, above] at (7.1, 0.6) {
        \textbf{Transversal Contraction}\\
        Each bin forms a message
    };

    \begin{scope}[shift={(10.6,0)}]
        \node[draw, circle, inner sep=2pt, label=left:$B_1$] (m1) at (0, 3) {};
        \node[draw, circle, inner sep=2pt, label=left:$B_2$] (m2) at (0, 1) {};
        \node at (0, 0) {$\vdots$};
        \node[draw, circle, inner sep=2pt, label=left:$B_{M_1}$] (mM) at (0, -1.5) {};

        \node[z_node, label=right:$\mathbf{z}_1$] (rz1) at (3, 3) {};
        \node[z_node, label=right:$\mathbf{z}_2$] (rz2) at (3, 1) {};
        \node at (2, 0) {$\vdots$};
        \node[z_node, label=right:$\mathbf{z}_{M_2}$] (rzM) at (3, -1.5) {};

        \foreach \m in {m1, m2, mM} {
            \foreach \z in {rz1, rz2, rzM} {
                \draw[thin, gray!40] (\m) -- (\z);
            }
        }
        \draw[thin, black] (m1) -- (rz1); \draw[thin, black] (m1) -- (rz2);
        \draw[thin, black] (m2) -- (rz1); \draw[thin, black] (m2) -- (rz2);
    \end{scope}

    \end{tikzpicture}
    \caption{Transversal contraction for a degraded det-BC. The type-induced subgraph (left) forms a forest of $M_2$ disjoint trees. Grouping corresponding leaves across trees (green boxes) yields bins that are each connected to all column vertices, producing the complete bipartite graph $K_{M_1, M_2}$ (right).}
    \label{fig:degraded_subgraph}
\end{figure}

\begin{remark}[Geometric Interpretation]
The degraded channel structure provides clear geometric intuition. The type-induced subgraph $G(\mathcal{B}^n, P_{\mathbf{x}})$ is a regular forest of $M_2 = |T(P_{\mathbf{z}})|$ disjoint star graphs, where each root $\mathbf{z} \in T(P_{\mathbf{z}})$ connects to exactly $d_Z(P_{\mathbf{x}}) = \exp\left[nH(Y|Z) + O(\ln n)\right]$ leaves $\mathbf{y} \in T(P_{\mathbf{y}})$. As illustrated in Fig.~\ref{fig:degraded_subgraph}, the transversal contraction groups the $k$-th leaf from each tree into bin $B_k$ for $k = 1, \ldots, M_1$. Since each tree contributes exactly one leaf to each bin, every bin $B_k$ contains precisely one vertex connected to each column vertex $\mathbf{z}_j$. This deterministic construction guarantees $K_{M_1, M_2} \prec_c G(\mathcal{B}^n, P_{\mathbf{x}})$ without requiring randomization, in contrast to the general case analyzed in Lemma~\ref{lem:subgraph_covering}.
\end{remark}

\begin{lemma}[Converse]
\label{lem:degraded_converse}
If a rate pair $(R_1, R_2)$ is achievable for the degraded det-BC $\mathcal{B}$, then $(R_1, R_2) \in \tilde{\mathcal{C}}(\mathcal{B})$.
\end{lemma}

\begin{IEEEproof}
By the convexity of the capacity region, it suffices to prove that for any achievable rate pair with $R_1 = R_1(t)$ for some $t \in [0,1]$, the rate $R_2$ satisfies $R_2 \leq R_2(t)$, where $R_1(t)$ and $R_2(t)$ are given by the parametric equations \eqref{eq:parametric_R1_simplified}--\eqref{eq:parametric_R2_simplified}. Since any rate achievable with finite blocklength is achievable asymptotically, we consider the limit $n \to \infty$.

By Theorem~\ref{thm:bipartite_blocklength}, any achievable rate pair $(R_1, R_2)$ requires the existence of a contraction $K_{M_1, M_2} \prec_c G(\mathcal{B}^n)$ with $\frac{1}{n}\ln M_i \to R_i$. Our goal is to establish an upper bound on the maximum number $M_2$ of receiver~2 vertices achievable for a given number $M_1$ of receiver~1 vertices through contraction.

\textbf{Structure of the full product graph.}
Unlike in the achievability proof where we analyzed type-induced subgraphs, the converse requires examining the complete graph $G(\mathcal{B}^n)$ since extracting subgraphs discards edges and would yield a loose bound. For degraded det-BCs, $G(\mathcal{B}^n)$ inherits the forest structure: it is a union of disjoint star components, where each sequence $\mathbf{z} \in \mathcal{Z}^n$ serves as the root of a star with degree $d_\mathbf{z}$ and leaf set $\{\mathbf{y} : z(\mathbf{y}) = \mathbf{z}\}$. For a given output type $P_{\mathbf{z}}$ at receiver~2, all sequences $\mathbf{z} \in T(P_{\mathbf{z}})$ have identical degree:
\begin{align}
    d_\mathbf{z} = \prod_{z\in\mathcal{Z}} d_{Z,z}^{n P_{\mathbf{z}}(z)},
\end{align}
and the number of such sequences is:
\begin{align}
    |T(P_{\mathbf{z}})| = \frac{n!}{\prod_{z\in\mathcal{Z}}(n P_{\mathbf{z}}(z))!}.
\end{align}
We retain these exact expressions rather than asymptotic approximations to track the precise combinatorial structure.

\textbf{Contraction strategy.}
To obtain the complete bipartite graph $K_{M_1, M_2}$, we perform contraction in two conceptual steps: (i) merge receiver~2 vertices within and across types until each column vertex has degree at least $M_1$; (ii) apply transversal binning (as illustrated in Fig.~\ref{fig:degraded_subgraph}) to group row vertices into $M_1$ bins, ensuring complete connectivity. Our bound focuses on step (i), which limits the achievable $M_2$.

\textbf{Bounding the number of vertices after contraction.}
For column vertices with degree at least $M_1$, no contraction is needed, since each can independently serve as a column vertex in $K_{M_1, M_2}$. For vertices with degree less than $M_1$, multiple vertices (possibly from different types) must be merged to reach the target degree. The key observation is that the total number of merged vertices achieving degree at least $M_1$ is bounded by the total degree budget divided by $M_1$. Summing over all column vertices $\mathbf{z} \in \mathcal{Z}^n$:
\begin{align}
    M_2
    &\leq \left(\sum_{\mathbf{z} : d_\mathbf{z} \geq M_1} 1\right) + \frac{\sum_{\mathbf{z} : d_\mathbf{z} < M_1} d_\mathbf{z}}{M_1}.
\end{align}
This formulation makes clear that contraction can occur across type boundaries, i.e., the total degree from all low-degree vertices (regardless of type) is pooled together to form new vertices of degree $M_1$. For analytical convenience, we aggregate vertices by their type $P_{\mathbf{z}}$. Since all sequences in $T(P_{\mathbf{z}})$ have the same degree $d_\mathbf{z}$:
\begin{align}
    M_2
    &\leq \sum_{P_\mathbf{z} : d_\mathbf{z} \geq M_1} |T(P_{\mathbf{z}})| + \frac{1}{M_1}\sum_{P_\mathbf{z} : d_\mathbf{z} < M_1} |T(P_{\mathbf{z}})| \cdot d_\mathbf{z} \notag\\
    &= \sum_{P_\mathbf{z}} |T(P_{\mathbf{z}})| \min\left\{1, \frac{d_\mathbf{z}}{M_1}\right\}.
\end{align}
Define $M_{2,P_\mathbf{z}} = |T(P_{\mathbf{z}})| \min\{1, d_\mathbf{z}/M_1\}$, then $M_2 = \sum_{P_\mathbf{z}} M_{2,P_\mathbf{z}}$. Since the number of types grows polynomially as $(n+1)^{|\mathcal{Z}|}$, the exponential order of $M_2$ is determined by the dominant type:
\begin{equation}
\begin{aligned}
    R_2
    &= \lim_{n \to \infty} \frac{1}{n}\ln M_2 \\
    &\leq \max_{P_{\mathbf{z}}} \lim_{n \to \infty} \frac{1}{n}\ln M_{2,P_\mathbf{z}}.
\end{aligned}
\end{equation}

\textbf{Optimizing over types.}
For a given $t \in [0,1]$, let $q_Z$ be the distribution defined by \eqref{eq:type q}. Set the target degree:
\begin{equation}
\begin{aligned}
    M_1
    &= \exp\left(n \sum_{z \in \mathcal{Z}} q_Z(z) \ln d_{Z,z}\right)\\
    &= \exp(n R_1(t)),
\end{aligned}
\end{equation}
where we used the identity $R_1(t) = t^{-1}(\ln\sum_z d_{Z,z}^t - H(q_Z)) = \sum_z q_Z(z) \ln d_{Z,z}$ from the achievability proof. Then
\begin{equation}
\begin{aligned}
    &\quad\,\frac{1}{n}\ln M_{2,P_\mathbf{z}} \\
    &= \frac{1}{n}\ln |T(P_{\mathbf{z}})| + \frac{1}{n}\ln\min\left\{1, \frac{d_\mathbf{z}}{M_1}\right\} \\
    &= H(P_{\mathbf{z}}) + o(1) + \frac{1}{n}\min\left\{0, \sum_{z} n P_{\mathbf{z}}(z) \ln d_{Z,z} - n\sum_z q_Z(z) \ln d_{Z,z}\right\} \\
    &= H(P_{\mathbf{z}}) + \min\left\{0, \sum_z (P_{\mathbf{z}}(z) - q_Z(z)) \ln d_{Z,z}\right\} + o(1),
\end{aligned}
\end{equation}
where we applied Stirling's approximation to the polynomial coefficient. Using the inequality $\min\{0, a\} \leq ta$ for $t \in [0,1]$ and $a \in \mathbb{R}$:
\begin{align}
    &\quad\,\frac{1}{n}\ln M_{2,P_{\mathbf{z}}} \notag\\
    &\leq H(P_{\mathbf{z}}) + t \sum_{z \in \mathcal{Z}} (P_{\mathbf{z}}(z) - q_Z(z)) \ln d_{Z,z} + o(1) \notag \\
    &\overset{(a)}{=} H(P_{\mathbf{z}}) + \sum_{z \in \mathcal{Z}} (P_{\mathbf{z}}(z) - q_Z(z)) \ln \left(q_Z(z) \sum_{z' \in \mathcal{Z}} d_{Z,z'}^t\right) + o(1) \notag \\
    &\overset{(b)}{=} H(P_{\mathbf{z}}) + \sum_{z \in \mathcal{Z}} (P_{\mathbf{z}}(z) - q_Z(z)) \ln q_Z(z) + o(1) \notag\\
    &= -\sum_{z \in \mathcal{Z}} q_Z(z) \ln q_Z(z) - \sum_{z \in \mathcal{Z}} P_{\mathbf{z}}(z) \ln \frac{P_{\mathbf{z}}(z)}{q_Z(z)} + o(1) \notag \\
    &= H(q) - D_{\mathrm{KL}}(P_{\mathbf{z}} \parallel q_Z) + o(1),
\end{align}
where (a) substitutes $d_{Z,z}^t = q_Z(z) \sum_{z'} d_{Z,z'}^t$ from \eqref{eq:type q}, and (b) uses the normalization $\sum_z (P_{\mathbf{z}}(z) - q_Z(z)) = 0$. Since $D_{\mathrm{KL}}(P_{\mathbf{z}} \parallel q_Z) \geq 0$ with equality if and only if $P_{\mathbf{z}} = q_Z$, we have
\begin{equation}
\begin{aligned}
    R_2
    &\leq \max_{P_{\mathbf{z}}} \lim_{n \to \infty} \frac{1}{n}\ln M_{2,P_{\mathbf{z}}}\\
    &\leq H(q_Z)\\
    &= R_2(t).
\end{aligned}
\end{equation}
The maximum is achieved when $P_{\mathbf{z}} = q_Z$, which corresponds precisely to the input distribution constructed in Lemma~\ref{lem:degraded_achievability}. This completes the converse.
\end{IEEEproof}

The capacity region of degraded det-BCs follows directly from the achievability and converse results established above.

\begin{theorem}
\label{thm:degraded_capacity}
The capacity region of a degraded det-BC $\mathcal{B}$ is $\mathcal{C}(\mathcal{B})=\tilde{\mathcal{C}}(\mathcal{B})$.
\end{theorem}

\begin{IEEEproof}
The theorem follows immediately from Lemma~\ref{lem:degraded_achievability} and Lemma~\ref{lem:degraded_converse}, which establish $\tilde{\mathcal{C}}(\mathcal{B}) \subseteq \mathcal{C}(\mathcal{B})$ and $\mathcal{C}(\mathcal{B}) \subseteq \tilde{\mathcal{C}}(\mathcal{B})$, respectively.
\end{IEEEproof}


\subsection{General Det-BCs}
\label{subsec:general_proof}

We now extend the capacity characterization to arbitrary deterministic broadcast channels. The achievability of $\tilde{\mathcal{C}}(\mathcal{B})$ follows directly from Corollary~\ref{cor:corner_points}, which establishes the achievability of corner points for any input distribution. The converse proof exploits the relationship between general and degraded det-BCs through vertex splitting operations. By relating the general case to the degraded case analyzed in Section~\ref{subsec:degraded_proof}, we establish that $\mathcal{C}(\mathcal{B}) = \tilde{\mathcal{C}}(\mathcal{B})$ for all det-BCs.

\begin{lemma}[Achievability for General Det-BCs]
\label{lem:general_achievability}
Any rate pair $(R_1, R_2) \in \tilde{\mathcal{C}}(\mathcal{B})$ is achievable for any det-BC $\mathcal{B}$.
\end{lemma}

\begin{IEEEproof}
By the convexity of the capacity region and symmetry between receivers, it suffices to prove the achievability of the curved segment $A$--$B$.

The proof follows the same construction as Lemma~\ref{lem:degraded_achievability}. For any $t \in [0,1]$, define the distribution $q_Z$ by Eq.~\eqref{eq:type q} and the uniform conditional distribution $P_{Y|Z}$ over preimages of each $z \in \mathcal{Z}$ as in Eq.~\eqref{eq:uniform_conditional}. This construction yields the rate pair $(R_1(t), R_2(t)) = (H(Y|Z), H(Z))$, which is achievable by Corollary~\ref{cor:corner_points}. As $t$ varies over $[0,1]$, this traces out the curved segment $A$--$B$. The remaining boundary follows by symmetry and time-sharing.
\end{IEEEproof}

\begin{remark}[Invariance Under Vertex Splitting]
\label{rem:invariance under vertex splitting}
General and degraded det-BCs share the same curved boundary segments because vertex splitting preserves conditional entropy. Consider splitting receiver~1 vertices to obtain a degraded channel $\mathcal{B}_{Y \to Z}$ from $\mathcal{B}$, as illustrated in Fig.~\ref{fig:vertex contraction}. For any joint distribution $P_{Y,Z}$ on $\mathcal{B}_{Y \to Z}$, merging the split vertices back to their original labels in $\mathcal{B}$ preserves both $H(Y|Z)$ and $H(Z)$.

Taking Fig.~\ref{fig:contraction, Y2Z} and Fig.~\ref{fig:contraction, blackwell} as an example, splitting vertex $y^{(2)}$ and then merging it back gives
\begin{equation*}
\begin{aligned}
P_{Y,Z}(\mathcal{B}_{Y\to Z})
&=
\tikzmarknode{A}{
\begin{pmatrix}
p_1 & 0 \\
p_2 & 0 \\
0   & p_3
\end{pmatrix}}
\\[5mm]
P_{Y,Z}(\mathcal{B})
&=
\tikzmarknode{B}{
\begin{pmatrix}
p_1 & 0 \\
p_2 & p_3
\end{pmatrix}} .
\end{aligned}
\hspace{1cm}   
\end{equation*}%
\begin{tikzpicture}[remember picture,overlay]
\draw[->,thick]
    ([xshift=5mm]A.east)
    --
    node[right,align=left] {\small merge\\[-1mm]\small rows 2,3}
    ([xshift=5mm]A.east |- B.east);
\end{tikzpicture}%
The conditional entropy $H(Y|Z=z)$ is preserved because it depends only on the probability distribution within each column $z$, while $H(Z)$ is preserved because the marginal column sums remain unchanged. Consequently, the curved segments parametrized by $(H(Y|Z), H(Z))$ are identical for both general and degraded configurations.
\end{remark}

\begin{remark}[Why degree sequences determine the capacity region]
\label{rem:degree_seq_sufficiency}
Corollary~\ref{cor:degree_sufficiency} states that the capacity region depends on the channel only through the degree sequences $D_Y\left(\mathcal{B}\right)$ and $D_Z\left(\mathcal{B}\right)$. For degraded det-BCs, this dependence is unsurprising, since the degree sequence already captures essentially the entire channel structure. The more remarkable fact is that the same conclusion continues to hold for general det-BCs, even when the detailed edge arrangement of the bipartite graph can be highly nontrivial.

Remark~\ref{rem:invariance under vertex splitting} showed that $H(Y|Z)$ and $H(Z)$ depend only on the column structure of the bipartite graph, not on which specific receiver-1 outputs occupy each column. We now show how the boundary-achieving input distributions in Lemma~\ref{lem:general_achievability} exploit this invariance, leading to a capacity characterization that depends only on the degree sequence.

For segment~$A$--$B$, the boundary-achieving distribution at parameter $t\in[0,1]$ assigns probability to input symbol $x$ proportional to $d_{Z,z(x)}^{t-1}$. Consequently, all input symbols mapping to the same receiver-2 output $z$ receive equal probability mass. We refer to this property as \emph{column uniformity}.
Under column uniformity, the induced output distribution satisfies $P_Z(z)\propto d_{Z,z}^{t}$ and therefore depends only on the degree of $z$. Moreover, conditioned on $Z=z$, $H(Y|Z=z)=\ln d_{Z,z}$ depends only on the column's size rather than the identity of its members. Hence both $H(Z)$ and $H(Y|Z)$ depend only on the degree sequence $D_Z(\mathcal{B})$.

As $t$ varies over $[0,1]$, these boundary-achieving distributions sweep out the entire segment~$A$--$B$. The symmetric construction for segment~$C$--$D$ depends only on $D_Y(\mathcal{B})$. Therefore two channels with identical degree sequences are indistinguishable to every boundary-achieving distribution and consequently have the same capacity region, even if their bipartite graphs have completely different edge arrangements.
\end{remark}

We next turn to the converse proof. The following lemma establishes a supporting-hyperplane representation of $\tilde{\mathcal{C}}(\mathcal{B})$. Specifically, it shows that the parametric characterization in Definition~\ref{def:explicit_capacity_region} is equivalent to the intersection of two one-parameter families of half-planes, indexed by $t\in[0,1]$ and determined respectively by the degree sequences $D_Z(\mathcal{B})$ and $D_Y(\mathcal{B})$. This representation makes the role of the degree sequences explicit and will be used in the sequel to characterize the capacity region of a general det-BC as the intersection of two degraded capacity regions.

\begin{lemma}[Supporting-Hyperplane Representation]
\label{lem:SHR}
Define
\begin{subequations}
\begin{align}
    c_Z(t) &= \ln \sum_{z \in \mathcal{Z}} d_{Z,z}^{t},\\
    c_Y(t) &= \ln \sum_{y \in \mathcal{Y}} d_{Y,y}^{t},
\end{align}
\label{eq:log_partition}
\end{subequations}
for all $t \in \left[ 0, 1 \right]$. Then
\begin{equation}
    \label{eq:SHR_identity}
    \tilde{\mathcal{C}}\left(\mathcal{B}\right)
    =
    \left\{
    \left(R_1,R_2\right)\in\mathbb{R}_{\geq 0}^2
    \;\middle|\;
    \begin{aligned}
    &tR_1+R_2 \leq c_Z(t),\\
    &R_1+tR_2 \leq c_Y(t),
    \end{aligned}
    \quad\forall\,t\in\left[0,1\right]
    \right\}.
\end{equation}
\end{lemma}

\begin{IEEEproof}
Denote the right-hand side of \eqref{eq:SHR_identity} by $\mathcal H(\mathcal B)$.  The proof proceeds by identifying $c_Z(t_0)$ and $c_Y(t_0)$, for each $t_0 \in \left[0,1\right]$, as the support function of $\tilde{\mathcal{C}}\left(\mathcal{B}\right)$ in the directions $\left(t_0, 1\right)$ and $\left(1, t_0\right)$, respectively;
that is,
\begin{subequations}
\label{eq:support_functions_intro}
\begin{align}
    c_Z(t_0)
    =\max_{\left(R_1,R_2\right)\in\tilde{\mathcal{C}}\left(\mathcal{B}\right)}
    \left(t_0 R_1 + R_2\right),
    \label{eq:support_functions_intro_z}\\
    \qquad
    c_Y(t_0)
    =\max_{\left(R_1,R_2\right)\in\tilde{\mathcal{C}}\left(\mathcal{B}\right)}
    \left(R_1 + t_0 R_2\right).
    \label{eq:support_functions_intro_y}
\end{align}
\end{subequations}
These identities are established in Step~1. Step~2 then shows that $\tilde{\mathcal{C}}\left(\mathcal{B}\right)\subseteq\mathcal{H}\left(\mathcal{B}\right)$ follows immediately from the definition of the support function, while Step~3 establishes the reverse inclusion via the supporting hyperplane theorem applied to the closed convex set $\tilde{\mathcal{C}}\left(\mathcal{B}\right)$.

\noindent
\textbf{Preliminaries.}
By~\eqref{eq:type q} and~\eqref{eq:parametric_R1}, the derivative of $c_Z$ satisfies
\begin{equation}
    c_Z'(t)
    =\frac{\sum_{z\in\mathcal{Z}} d_{Z,z}^{t} \ln d_{Z,z}}{\sum_{z\in\mathcal{Z}} d_{Z,z}^{t}}
    = R_1(t),
    \label{eq:cZ_prime}
\end{equation}
where $R_1(t)$ denotes the first coordinate of the boundary
point of segment $A$--$B$ at parameter $t$.
Differentiating again,
\begin{equation}
    c_Z''(t)
    = \mathrm{Var}_{q_Z}\left[\ln d_{Z,z}\right]
    \geq 0,
    \label{eq:cZ_convex}
\end{equation}
so $c_Z$ is convex on $\left[0,1\right]$, and non-decreasing since $c_Z'(t) = R_1(t) \geq 0$. In particular, $c_Z(0) = \ln\left|\mathcal{Z}\right|$ and $c_Z(1) = \ln\left|\mathcal{X}\right|$. By the symmetry of Definition~\ref{def:explicit_capacity_region} between the two receivers, the analogous identities hold for $c_Y$: $c_Y'(s) = R_2(s)$, where $R_2(s)$ is the second coordinate of the boundary point of segment $C$--$D$ at parameter $s$, $c_Y$ is convex and non-decreasing, and $c_Y(0) = \ln\left|\mathcal{Y}\right|$, $c_Y(1) = \ln\left|\mathcal{X}\right|$. Since $\sum_z d_{Z,z} = \sum_y d_{Y,y} = \left|\mathcal{X}\right|$, we have $c_Z(1) = c_Y(1) = \ln\left|\mathcal{X}\right|$.

\noindent
\textbf{Step 1 (Support functions).}
Fix $t_0 \in \left[0,1\right]$. We first establish \eqref{eq:support_functions_intro_z} in this step, and then \eqref{eq:support_functions_intro_y} follows by symmetry. Geometrically, $c_Z(t_0)$ is the value of the support function of $\tilde{\mathcal{C}}\left(\mathcal{B}\right)$ in the direction $\left(t_0,1\right)$: the line $t_0 R_1 + R_2 = c_Z(t_0)$ is a supporting hyperplane of $\tilde{\mathcal{C}}\left(\mathcal{B}\right)$, touching the region at the boundary point $\left(R_1(t_0), R_2(t_0)\right)$ on segment $A$--$B$, where by Corollary~\ref{cor:tangent_slope} the tangent slope equals $-t_0$, which is precisely the slope of this hyperplane.

\textit{Lower bound.}
The point $\left(R_1(t_0), R_2(t_0)\right) \in\tilde{\mathcal{C}}\left(\mathcal{B}\right)$ achieves
\begin{equation}
\begin{aligned}
    t_0 R_1(t_0) + R_2(t_0)
    &= t_0 c_Z'(t_0) + \left(c_Z(t_0) - t_0 c_Z'(t_0)\right)\\
    &= c_Z(t_0),
\end{aligned}
\label{eq:lower_bound}
\end{equation}
using~\eqref{eq:cZ_prime} and the identity $R_2(t) = c_Z(t) - t c_Z'(t)$, which follows from~\eqref{eq:parametric_R2} together with~\eqref{eq:cZ_prime}. Hence the maximum in~\eqref{eq:support_functions_intro_z} is at least $c_Z(t_0)$.

\textit{Upper bound.}
It remains to show $t_0 R_1 + R_2 \leq c_Z(t_0)$ for every $\left(R_1,R_2\right) \in \tilde{\mathcal{C}}\left(\mathcal{B}\right)$. Since $\tilde{\mathcal{C}}\left(\mathcal{B}\right)$ is the downward-left closure of the five boundary segments of Definition~\ref{def:explicit_capacity_region}, and the linear functional $\left(R_1,R_2\right) \mapsto t_0 R_1 + R_2$ is non-decreasing in each coordinate (as $t_0 \geq 0$), it suffices to verify the bound on each of these five segments. We treat $A$--$A'$ and $A$--$B$ first, and then $B$--$C$--$D$--$D'$ as a group.

On $A$--$A'$, since $0\leq R_1\leq R_1^A$ and $R_2=R_2(A)=\ln\left|\mathcal{Z}\right|$), it suffices to consider point $A$. On $A$--$B$, at parameter $t \in \left[0,1\right]$,
\begin{equation}
\begin{aligned}
    t_0 R_1(t) + R_2(t)
    &=c_Z(t) + \left(t_0 - t\right) c_Z'(t)\\
    &\leq c_Z(t_0),
\end{aligned}
\label{eq:AB_bound}
\end{equation}
where the inequality is the convexity of $c_Z$ established in~\eqref{eq:cZ_convex}.

For $B$--$C$--$D$--$D'$, the bound $t_0 R_1 + R_2 \leq c_Z(t_0)$ follows from the following two properties, which we establish below:
\begin{subequations}
\begin{align}
    R_1 + R_2
    &\leq c_Z(1),
    \label{eq:property_i}
    \\
    R_1
    &\geq c_Z'(1).
    \label{eq:property_ii}
\end{align}
\end{subequations}
Given~\eqref{eq:property_i} and~\eqref{eq:property_ii},
\begin{equation}
\begin{aligned}
    t_0 R_1 + R_2
    &= \left( R_1 + R_2 \right) + \left( t_0 - 1 \right) R_1 \\
    &\leq c_Z(1) + \left( t_0 - 1 \right) R_1 \\
    &\leq c_Z(1) + \left( t_0 - 1 \right) c_Z'(1) \\
    &\leq c_Z(t_0),
    \label{eq:BCD_bound}
\end{aligned}
\end{equation}
where the first inequality is~\eqref{eq:property_i}, the second uses $t_0-1 \leq 0$ together with~\eqref{eq:property_ii}, and the third is the convexity of $c_Z$ as in~\eqref{eq:AB_bound}.

\textit{Proof of property~\eqref{eq:property_i}.}
On $B$--$C$, $R_1+R_2 = \ln\left|\mathcal{X}\right|$ by Definition~\ref{def:explicit_capacity_region}, since this segment has slope $-1$. On $C$--$D$, parametrized by $s \in \left[0,1\right]$ with $R_1(s) = c_Y(s) - sc_Y'(s)$ and $R_2(s) = c_Y'(s)$ ($s=1$ at $C$, $s=0$ at $D$),
\begin{equation}
    R_1(s) + R_2(s) = c_Y(s) + \left(1-s\right)c_Y'(s),
\end{equation}
whose derivative $\left(1-s\right)c_Y''(s) \geq 0$ shows it is non-decreasing in $s$. Hence $R_1+R_2 \leq R_1(1)+R_2(1) = c_Y(1)=\ln\left|\mathcal{X}\right| = c_Z(1)$ throughout $C$--$D$. On $D$--$D'$, $0\leq R_2\leq R_2(D)$ and $R_1=R_1^D=\ln\left|\mathcal{Y}\right|$, so it suffices to consider point $D$.

\textit{Proof of property~\eqref{eq:property_ii}.}
By~\eqref{eq:cZ_prime} at $t=1$, $R_1^B = c_Z'(1)$. By Corollary~\ref{cor:BC_length}, $R_1^C - R_1^B = I\left(Y;Z\right) \geq 0$, so $R_1^C\geq R_1^B = c_Z'(1)$; since $B$--$C$ has slope $-1$ with $R_1$ increasing from $B$ to $C$, $R_1 \geq R_1^B = c_Z'(1)$ throughout $B$--$C$. On $C$--$D$, $\frac{\mathrm{d}}{\mathrm{d}s}R_1(s)=-sc_Y''(s) \leq 0$, so $R_1(s) \geq R_1(1) = R_1^C\geq c_Z'(1)$ for $s \in \left[0,1\right]$. On $D$--$D'$, $R_1$ is constant and equal to $R_1^D = R_1(0)\geq R_1(1) = R_1^C \geq c_Z'(1)$, by the same monotonicity.

This establishes $t_0 R_1+R_2 \leq c_Z(t_0)$ on $B$--$C$--$D$--$D'$, completing the upper bound and establishing~\eqref{eq:support_functions_intro_z}.

\noindent
\textbf{Step 2 ($\tilde{\mathcal{C}}\left(\mathcal{B}\right) \subseteq\mathcal{H}\left(\mathcal{B}\right)$).}
Let $\left(R_1^{*},R_2^{*}\right)\in\tilde{\mathcal{C}}\left(\mathcal{B}\right)$ and $t_0\in\left[0,1\right]$. By~\eqref{eq:support_functions_intro},
\begin{subequations}
\begin{align}
    t_0 R_1^{*}+R_2^{*} &\leq c_Z(t_0),\\
    R_1^{*}+t_0 R_2^{*} &\leq c_Y(t_0).
\end{align}
\end{subequations}
Since $t_0 \in \left[0,1\right]$ was arbitrary, $\left(R_1^{*},R_2^{*}\right)\in \mathcal{H}\left(\mathcal{B}\right)$.

\noindent
\textbf{Step 3 ($\mathcal{H}\left(\mathcal{B}\right)\subseteq \tilde{\mathcal{C}}\left(\mathcal{B}\right)$).}
We prove the contrapositive. Suppose $\left(R_1^{*},R_2^{*}\right)\notin \tilde{\mathcal{C}}\left(\mathcal{B}\right)$. Since $\tilde{\mathcal{C}}\left(\mathcal{B}\right)$ is a non-empty closed convex set, the supporting hyperplane theorem provides $\alpha,\beta \geq 0$, not both zero, such that
\begin{equation}
    \alpha R_1^{*}+\beta R_2^{*}
    >\max_{\left(R_1,R_2\right)\in\tilde{\mathcal{C}}\left(\mathcal{B}\right)}
    \left(\alpha R_1+\beta R_2\right).
    \label{eq:separation}
\end{equation}
We show $\left(R_1^{*},R_2^{*}\right)\notin \mathcal{H}\left(\mathcal{B}\right)$ by considering the ratio $t^{*}=\alpha/\beta$ (with $t^{*}=\infty$ if $\beta=0$).

If $\beta>0$ and $t^{*} \in \left[0,1\right]$, dividing \eqref{eq:separation} by $\beta$ and applying~\eqref{eq:support_functions_intro_z} gives $t^{*}R_1^{*}+R_2^{*} > c_Z(t^{*})$, violating the constraint $tR_1 + R_2 \leq c_Z(t)$ in $\mathcal{H}\left(\mathcal{B}\right)$.

If $\beta>0$ and $t^{*}>1$, set $s^{*}=1/t^{*}\in \left(0,1\right)$; dividing~\eqref{eq:separation} by $\alpha>0$ and applying~\eqref{eq:support_functions_intro_y} gives $R_1^{*}+s^{*}R_2^{*} > c_Y(s^{*})$, violating the constraint $R_1 + tR_2 \leq c_Y(t)$.

If $\beta=0$ (so $\alpha>0$),~\eqref{eq:separation} gives $R_1^{*} > c_Y(0) = \ln\left|\mathcal{Y}\right|$, violating the constraint $R_1 + tR_2 \leq c_Y(t)$ at $t=0$; symmetrically, if $\alpha=0$, $R_2^{*} > c_Z(0) = \ln\left|\mathcal{Z}\right|$ violates the constraint $tR_1 + R_2 \leq c_Z(t)$ at $t_0=0$.

In every case $\left(R_1^{*},R_2^{*}\right)\notin \mathcal{H}\left(\mathcal{B}\right)$, establishing the contrapositive and completing the proof.
\end{IEEEproof}

\begin{lemma}[Intersection of Degraded Capacity Regions]
\label{lem:intersection_identity}
Let $\mathcal{B}$ be a det-BC with bipartite graph $G\left(\mathcal{B}\right) = \left(\mathcal{Y}, \mathcal{Z}, E\right)$, and let $\mathcal{B}_{Y \to Z}$ and $\mathcal{B}_{Z \to Y}$ be the degraded channels obtained by splitting all vertices of degree greater than one into degree-$1$ vertices at receiver~1 and receiver~2, respectively. Then
\begin{equation}
    \tilde{\mathcal{C}}\left(\mathcal{B}_{Y \to Z}\right)\cap \tilde{\mathcal{C}}\left(\mathcal{B}_{Z \to Y}\right)
    = \tilde{\mathcal{C}}\left(\mathcal{B}\right).
    \label{eq:intersection of degraded capacity region}
\end{equation}
\end{lemma}

\begin{IEEEproof}
We establish the identity through three steps: a degree-sequence preservation property of vertex splitting, a supporting-hyperplane representation of each region, and the intersection.

\textit{Degree-sequence preservation.}
Splitting all degree-$\geq 2$ vertices of receiver~1 in $\mathcal{B}$ to obtain $\mathcal{B}_{Y \to Z}$ modifies $\mathcal{Y}$ but leaves every vertex in $\mathcal{Z}$ and its incident edges unchanged. Therefore
\begin{subequations}
\begin{align}
    D_Z\left(\mathcal{B}_{Y \to Z}\right)
    &= D_Z\left(\mathcal{B}\right),\\
    \left| E\left(\mathcal{B}_{Y \to Z}\right) \right|
    &=\left| \mathcal{X} \right|.
\end{align}
\label{eq:DZ_preserved}
\end{subequations}
By the symmetric argument,
\begin{subequations}
\begin{align}
    D_Y\left(\mathcal{B}_{Z \to Y}\right)
    &= D_Y\left(\mathcal{B}\right),\\
    \left| E\left(\mathcal{B}_{Z \to Y}\right) \right|
    &=\left| \mathcal{X} \right|.
\end{align}
\label{eq:DY_preserved}
\end{subequations}

\textit{Half-plane characterizations of the degraded regions.}
Apply Lemma~\ref{lem:SHR} to $\mathcal{B}_{Y \to Z}$. Since $D_Z\left(\mathcal{B}_{Y \to Z}\right) = D_Z\left(\mathcal{B}\right)$ by~\eqref{eq:DZ_preserved}, the function $c_Z$ in~\eqref{eq:log_partition} is the same for both channels. Since $D_Y\left(\mathcal{B}_{Y \to Z}\right) = \left\{ 1, 1, \ldots, 1 \right\}$ together with $|E(\mathcal{B}_{Y\to Z})|=|\mathcal{X}|$ from \eqref{eq:DZ_preserved} gives $|\mathcal{Y}(\mathcal{B}_{Y\to Z})|=|\mathcal{X}|$, we have $c_Y^{Y\to Z}(t)=\ln|\mathcal{X}|$ for all $t\geq 0$. The constraint $R_1 + tR_2 \leq c_Y(t)$ of Lemma~\ref{lem:SHR} reduces to $R_1 + tR_2 \leq \ln \left| \mathcal{X} \right|$, which is implied by the other constraint $tR_1 + R_2 \leq c_Z(t)$ at $t = 1$ together with $R_2 \geq 0$. Therefore
\begin{equation}
    \tilde{\mathcal{C}}\left(\mathcal{B}_{Y \to Z}\right)
    =\left\{\left(R_1,R_2\right)\in\mathbb{R}_{\geq 0}^2
    \;\middle|\;
    tR_1+R_2\leq c_Z(t),
    \quad\forall\, t\in\left[0,1\right]
    \right\}.
    \label{eq:C_YtoZ}
\end{equation}
By the symmetric argument using~\eqref{eq:DY_preserved},
\begin{equation}
    \tilde{\mathcal{C}}\left(\mathcal{B}_{Z \to Y}\right)
    =\left\{\left(R_1,R_2\right)\in\mathbb{R}_{\geq 0}^2
    \;\middle|\;
    R_1+tR_2\leq c_Y(t),
    \quad
    \forall\, t\in\left[0,1\right]
    \right\}.
    \label{eq:C_ZtoY}
\end{equation}

\textit{Intersection.}
Combining~\eqref{eq:C_YtoZ} and~\eqref{eq:C_ZtoY}:
\begin{equation}
\begin{aligned}
    &\quad\,\tilde{\mathcal{C}}\left(\mathcal{B}_{Y \to Z}\right)\cap\tilde{\mathcal{C}}\left(\mathcal{B}_{Z \to Y}\right) \\
    &=\left\{\left(R_1,R_2\right)\in\mathbb{R}_{\geq 0}^2
    \;\middle|\;
    \begin{gathered}
    tR_1+R_2\leq c_Z(t),\\
    R_1+tR_2\leq c_Y(t),
    \end{gathered}
    \quad\forall\, t\in\left[0,1\right]
    \right\}\\
    &= \tilde{\mathcal{C}}\left(\mathcal{B}\right),
\end{aligned}
\end{equation}
where the second equality is Lemma~\ref{lem:SHR} applied to $\mathcal{B}$.
\end{IEEEproof}

\begin{lemma}[Converse for General Det-BCs]
\label{lem:general_converse}
If a rate pair $(R_1, R_2)$ is achievable for a det-BC $\mathcal{B}$, then $(R_1, R_2) \in \tilde{\mathcal{C}}(\mathcal{B})$.
\end{lemma}

\begin{IEEEproof}
We establish an outer bound on $\mathcal{C}(\mathcal{B})$ by relating it to the capacity regions of two degraded channels obtained through vertex splitting operations. Consider the degraded channels $\mathcal{B}_{Y \to Z}$ and $\mathcal{B}_{Z \to Y}$ constructed by splitting all vertices of degree greater than 1 at receiver~1 and receiver~2, respectively, until each resulting graph becomes a forest (Fig.~\ref{fig:vertex contraction}). Since vertex splitting is the inverse operation of contraction, we have $\mathcal{B} \prec_c \mathcal{B}_{Y \to Z}$ and $\mathcal{B} \prec_c \mathcal{B}_{Z \to Y}$. By Lemma~\ref{lem:contraction_capacity}, this implies
\begin{subequations}
\begin{align}
    \mathcal{C}(\mathcal{B}) &\subset \mathcal{C}(\mathcal{B}_{Y \to Z}),\\
    \mathcal{C}(\mathcal{B}) &\subset \mathcal{C}(\mathcal{B}_{Z \to Y}).
\end{align}
\end{subequations}
Since both $\mathcal{B}_{Y \to Z}$ and $\mathcal{B}_{Z \to Y}$ are degraded det-BCs, their capacity regions are characterized by Lemmas~\ref{lem:degraded_achievability} and~\ref{lem:degraded_converse}.
Therefore, we have $\mathcal{C}(\mathcal{B}_{Y \to Z})=\tilde{\mathcal{C}}(\mathcal{B}_{Y \to Z})$, $\mathcal{C}(\mathcal{B}_{Z \to Y})=\tilde{\mathcal{C}}(\mathcal{B}_{Z \to Y})$ and thus
\begin{align}
    \mathcal{C}(\mathcal{B}) \subseteq \tilde{\mathcal{C}}(\mathcal{B}_{Y \to Z}) \cap \tilde{\mathcal{C}}(\mathcal{B}_{Z \to Y}).
\end{align}
By Lemma~\ref{lem:intersection_identity}, $\tilde{\mathcal{C}}\left(\mathcal{B}_{Y \to Z}\right)\cap\tilde{\mathcal{C}}\left(\mathcal{B}_{Z \to Y}\right) =\tilde{\mathcal{C}}\left(\mathcal{B}\right)$. This establishes $\mathcal{C}(\mathcal{B}) \subseteq \tilde{\mathcal{C}}(\mathcal{B})$, completing the converse.
\end{IEEEproof}

The capacity region characterization for general det-BCs now follows immediately by combining the achievability and converse results.

\begin{IEEEproof}[Proof of Theorem~\ref{thm:main_capacity}]
Lemma~\ref{lem:general_achievability} establishes $\tilde{\mathcal{C}}(\mathcal{B}) \subseteq \mathcal{C}(\mathcal{B})$, while Lemma~\ref{lem:general_converse} establishes $\mathcal{C}(\mathcal{B}) \subseteq \tilde{\mathcal{C}}(\mathcal{B})$. Therefore, $\mathcal{C}(\mathcal{B}) = \tilde{\mathcal{C}}(\mathcal{B})$.
\end{IEEEproof}

\begin{remark}[Connection to Zero-Error Capacity]
\label{rem:zero_error}
Let $\mathcal{C}_{\mathrm{ZE}}\left(\mathcal{B}\right)$ denote the zero-error capacity region of the det-BC $\mathcal{B}$, defined as the closure of all rate pairs $\left(R_1, R_2\right)$ achievable by sequences of deterministic codes with zero probability of error. We claim
\begin{equation}
    \mathcal{C}_{\mathrm{ZE}}\left(\mathcal{B}\right)
    = \mathcal{C}\left(\mathcal{B}\right)
    = \tilde{\mathcal{C}}\left(\mathcal{B}\right).
    \label{eq:zero_error_equality}
\end{equation}

The inclusion $\mathcal{C}_{\mathrm{ZE}}\left(\mathcal{B}\right) \subseteq \mathcal{C}\left(\mathcal{B}\right)$ is immediate, since every zero-error code is also achievable under the standard vanishing-error criterion. The converse $\mathcal{C}\left(\mathcal{B}\right) \subseteq \tilde{\mathcal{C}}\left(\mathcal{B}\right)$ has already been established in Lemma~\ref{lem:general_converse}. It only remains to show $\tilde{\mathcal{C}}\left(\mathcal{B}\right) \subseteq \mathcal{C}_{\mathrm{ZE}}\left(\mathcal{B}\right)$.

The achievability proof of Lemma~\ref{lem:general_achievability} in fact constructs zero-error codes. To see this, observe that a valid contraction $K_{M_1,M_2}\prec_c G\left(\mathcal{B}^n\right)$ assigns each message pair $\left(m_1,m_2\right)$ to a unique input sequence $\mathbf{x}\in\mathcal{X}^n$. Since the channel is deterministic, the corresponding output sequences are produced with probability one, and Definition~\ref{def:contraction_order} guarantees that receiver~1 can uniquely recover $m_1$ from $\mathbf{y}$ and receiver~2 can uniquely recover $m_2$ from $\mathbf{z}$. Consequently, the resulting code has zero probability of error.

The only use of randomization in the achievability proof occurs in Lemma~\ref{lem:subgraph_covering}, where random partitioning is employed. Specifically, it is shown that the probability that a randomly chosen partition fails to induce the required contraction tends to zero as $n\to\infty$. Hence, there must exist at least one deterministic partition that produces the desired complete bipartite contraction. Once this partition is fixed, the encoder and both decoders are entirely deterministic, and the resulting code operates with zero error. The randomization is therefore confined to the existence proof and is not part of the encoding or decoding procedure.

Theorem~\ref{thm:main_capacity} therefore characterizes both the classical capacity region and the zero-error capacity region of a deterministic broadcast channel.
\end{remark}


\section{Common Message and Finite-Blocklength Analysis}
\label{sec:common}

This section extends our graph-theoretic framework in two directions. First, we characterize the capacity region with common message, establishing achievable rate triples $(R_0, R_1, R_2)$ where $R_0$ represents the common message rate. A key insight is that the graph-based approach naturally associates common message with the connected components of the channel graph $G(\mathcal{B})$. This reveals that coding schemes achieving the curved boundaries of the private-message capacity region $\mathcal{C}(\mathcal{B})$ inherently support a non-trivial amount of \emph{publicly decodable message}—information that can be decoded by both receivers. Such information can be interpreted operationally as \emph{cross-leakage}: private information intended for one receiver that becomes decodable by the other due to the channel's topological structure. Second, we leverage the discrete nature of our graph formulation to address the finite-blocklength regime, formulating the search for optimal rate pairs as an integer linear programming (ILP) problem.


\subsection{Common Message and Publicly Decodable Rates}
\label{subsec:common}

Our analysis thus far has focused on the private-message capacity region, where independent messages are sent to each receiver. We now extend the graph-theoretic characterization from Theorem~\ref{thm:bipartite_blocklength} to accommodate common messages—information intended for both receivers. The extension naturally arises from the observation that disconnected channel graphs can support disjoint coding structures.

\subsubsection{Graph Structure for Common Messages}

To analyze channels whose graph $G(\mathcal{B})$ contains multiple connected components, we introduce the following characterization:

\begin{figure}[t]
    \centering
    \begin{tikzpicture}[
        scale=0.9,
        y_node/.style={circle, fill=blue!20, draw=blue, inner sep=3pt, minimum size=6pt},
        z_node/.style={circle, fill=red!20, draw=red, inner sep=3pt, thick, minimum size=6pt},
        contraction_arrow/.style={->, shorten >=5pt, shorten <=5pt, very thick, draw=gray!80},
        group_box/.style={draw=green!50!black, thick, dashed, rounded corners, inner sep=4pt}
    ]

    \node[y_node, label={[name=y1_1_label]left:$\mathbf{y}_1^{(1)}$}] (y1_1) {};
    \node[y_node, below=of y1_1, label=left:$\mathbf{y}_2^{(1)}$] (y2_1) {};
    \node[below=0.3 of y2_1] (ydots_1) {$\vdots$};
    \node[y_node, below=0.3 of ydots_1, label=left:$\mathbf{y}_{M_1}^{(1)}$] (yM1_1) {};

    \node[z_node, right=1.5 of y1_1, label=right:$\mathbf{z}_1^{(1)}$] (z1_1) {};
    \node[z_node, below=of z1_1, label=right:$\mathbf{z}_2^{(1)}$] (z2_1) {};
    \node[below=0.3 of z2_1] (zdots_1) {$\vdots$};
    \node[z_node, below=0.3 of zdots_1, label={[name=zM2_1_label]right:$\mathbf{z}_{M_2}^{(1)}$}] (zM2_1) {};

    \begin{scope}[on background layer]
        \foreach \y in {y1_1, y2_1, yM1_1}
            \foreach \z in {z1_1, z2_1, zM2_1}
                \draw[black!40] (\y) -- (\z);
    \end{scope}

    \node[group_box, draw=black!40, fit=(y1_1_label) (zM2_1_label), label=above:\textbf{\small $K_{M_1, M_2}$ (Copy 1)}] (comp1) {};

    \node[right=0.1 of comp1, font=\huge] (dots_comp) {$\cdots$};

    \node[y_node, right=1 of dots_comp, yshift=2cm, label={[name=y1_k_label]left:$\mathbf{y}_1^{(k)}$}] (y1_k) {};
    \node[y_node, below=of y1_k, label=left:$\mathbf{y}_2^{(k)}$] (y2_k) {};
    \node[below=0.3cm of y2_k] (ydots_k) {$\vdots$};
    \node[y_node, below=0.3 of ydots_k, label=left:$\mathbf{y}_i^{(k)}$] (yi_k) {};
    \node[below=0.3 of yi_k] (ydots2_k) {$\vdots$};
    \node[y_node, below=0.3 of ydots2_k, label=left:$\mathbf{y}_{M_1}^{(k)}$] (yM1_k) {};

    \node[z_node, right=1.5 of y1_k, label=right:$\mathbf{z}_1^{(k)}$] (z1_k) {};
    \node[z_node, below=of z1_k, label=right:$\mathbf{z}_2^{(k)}$] (z2_k) {};
    \node[below=0.3 of z2_k] (zdots_k) {$\vdots$};
    \node[z_node, below=0.3 of zdots_k, label=right:$\mathbf{z}_j^{(k)}$] (zj_k) {};
    \node[below=0.3 of zj_k] (zdots2_k) {$\vdots$};
    \node[z_node, below=0.3 of zdots2_k, label={[name=zM2_k_label]right:$\mathbf{z}_{M_2}^{(k)}$}] (zM2_k) {};

    \draw[red, ultra thick] (yi_k) -- (zj_k) node[midway, above=0.2, font=\bfseries\small, sloped] {Codeword $\mathbf{x}(k,i,j)$};

    \begin{scope}[on background layer]
        \foreach \y in {y1_k, y2_k, yi_k, yM1_k}
            \foreach \z in {z1_k, z2_k, zj_k, zM2_k}
                \draw[black!40] (\y) -- (\z);
    \end{scope}

    \node[group_box, fit=(y1_k_label) (zM2_k_label), label=above:\textbf{\small $K_{M_1, M_2}$ (Copy k)}] (compk) {};

    \node[right=5.5 of comp1, font=\huge] (dots2_comp) {$\cdots$};

    \node[y_node, right=1.2 of dots2_comp, yshift=1.5cm, label={[name=y1_M0_label]left:$\mathbf{y}_1^{(M_0)}$}] (y1_M0) {};
    \node[y_node, below=of y1_M0, label=left:$\mathbf{y}_2^{(M_0)}$] (y2_M0) {};
    \node[below=0.3 of y2_M0] (ydots_M0) {$\vdots$};
    \node[y_node, below=0.3 of ydots_M0, label=left:$\mathbf{y}_{M_1}^{(M_0)}$] (yM1_M0) {};

    \node[z_node, right=1.5 of y1_M0, label=right:$\mathbf{z}_1^{(M_0)}$] (z1_M0) {};
    \node[z_node, below=of z1_M0, label=right:$\mathbf{z}_2^{(M0)}$] (z2_M0) {};
    \node[below=0.3 of z2_M0] (zdots_M0) {$\vdots$};
    \node[z_node, below=0.3 of zdots_M0, label={[name=zM2_M0_label]right:$\mathbf{z}_{M_2}^{(M_0)}$}] (zM2_M0) {};

    \begin{scope}[on background layer]
        \foreach \y in {y1_M0, y2_M0, yM1_M0}
            \foreach \z in {z1_M0, z2_M0, zM2_M0}
                \draw[black!20] (\y) -- (\z);
    \end{scope}

    \node[group_box, draw=black!40, fit=(y1_M0_label) (zM2_M0_label), label=above:\textbf{\small $K_{M_1, M_2}$ (Copy $M_0$)}] (compM0) {};

    \end{tikzpicture}
    \caption{Encoding scheme based on the disjoint union graph $\bigsqcup_{k=1}^{M_0} K_{M_1, M_2}$. The common message $k$ selects the $k$-th copy of the complete bipartite graph. The private message pair $(i, j)$ then selects a specific edge $\left(\mathbf{y}_i^{(k)}, \mathbf{z}_j^{(k)}\right)$ within that component, which corresponds to the transmitted codeword.}
    \label{fig:encoding_common}
\end{figure}
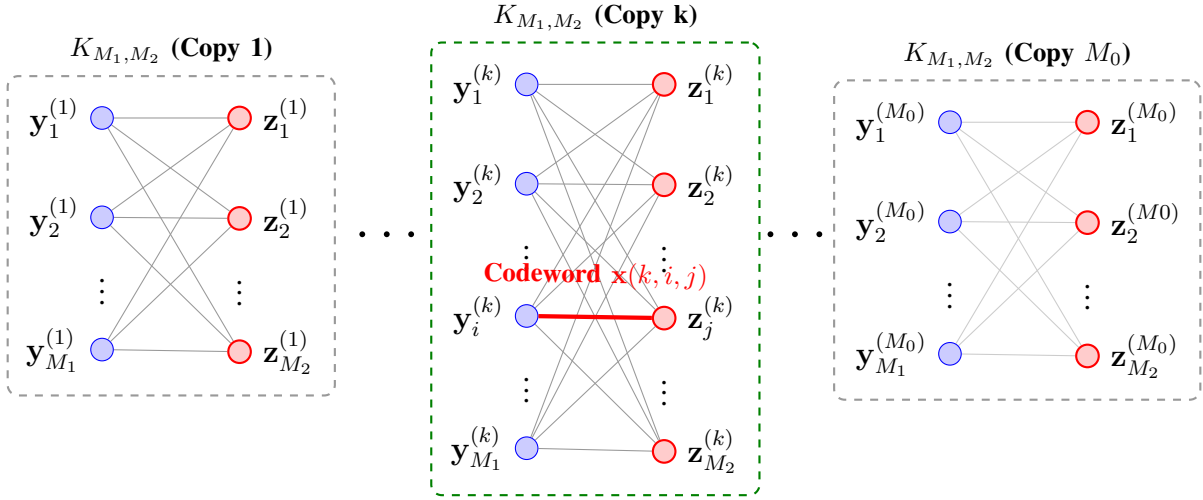

\begin{theorem}[Achievability with Common Messages]
\label{thm:graph_achievability_triple}
For a det-BC $\mathcal{B}$ with blocklength $n$, a rate triple $(R_0, R_1, R_2)$ (in nats) is achievable if there exists a bipartite graph consisting of $M_0$ disjoint copies of the complete bipartite graph $K_{M_1, M_2}$, denoted by $\bigsqcup_{k=1}^{M_0} K_{M_1, M_2}$, such that 
\begin{align}
    \bigsqcup_{k=1}^{M_0} K_{M_1, M_2} \prec_c G(\mathcal{B}^n),
\end{align}
where
\begin{subequations}
\begin{align}
    R_0 &= \frac{1}{n}\ln M_0, \\
    R_1 &= \frac{1}{n}\ln M_1, \\
    R_2 &= \frac{1}{n}\ln M_2.
\end{align}
\end{subequations}
Here, $R_0$ represents the common message rate decoded by both receivers, while $R_1$ and $R_2$ represent the private message rates for receivers~1 and~2, respectively.
\end{theorem}

\begin{IEEEproof}
The construction parallels Lemma~\ref{lem:bipartite_single_use}. Given the contraction $\bigsqcup_{k=1}^{M_0} K_{M_1, M_2} \prec_c G(\mathcal{B}^n)$, we construct an $(n, M_0, M_1, M_2)$ code as follows.

\textbf{Message structure.}
The message set is $\{1, \ldots, M_0\} \times \{1, \ldots, M_1\} \times \{1, \ldots, M_2\}$, where the first component $k$ indexes the common message for both receivers, while components $i$ and $j$ index the private messages for receivers~1 and~2, respectively.

\textbf{Encoding.}
As illustrated in Fig.~\ref{fig:encoding_common}, the encoder uses the structure of the disjoint union $\bigsqcup_{k=1}^{M_0} K_{M_1, M_2}$ to partition the message space. Consider the $k$-th copy of $K_{M_1, M_2}$ in the disjoint union. Its $M_1$ vertices on the receiver~1 side correspond to distinct output sequences $\{\mathbf{y}_1^{(k)}, \ldots, \mathbf{y}_{M_1}^{(k)}\}$, and its $M_2$ vertices on the receiver~2 side correspond to distinct output sequences $\{\mathbf{z}_1^{(k)}, \ldots, \mathbf{z}_{M_2}^{(k)}\}$. For each valid message triple $(k, i, j) \in [M_0] \times [M_1] \times [M_2]$, the selected edge connects vertex $\mathbf{y}_i^{(k)}$ and $\mathbf{z}_j^{(k)}$. This edge directly represents the codeword $\mathbf{x}(k,i,j)$ to be transmitted.

Since $\bigsqcup_{k=1}^{M_0} K_{M_1, M_2} \prec_c G(\mathcal{B}^n)$, there exists a contraction mapping from $G(\mathcal{B}^n)$ to the disjoint union. For each message triple $(k,i,j)$, the encoder transmits a pre-image input sequence $\mathbf{x}_{k,i,j} \in \mathcal{X}^n$ under this contraction that maps to the edge connecting $\mathbf{y}_i^{(k)}$ and $\mathbf{z}_j^{(k)}$.

\textbf{Decoding.}
Upon observing output sequence $\mathbf{y}$, receiver~1 identifies the unique component index $k$ and private message index $i$ such that $\mathbf{y}$ contracts to $\mathbf{y}_i^{(k)}$, thereby recovering $(k, i)$. Similarly, receiver~2 observes $\mathbf{z}$ and recovers $(k, j)$ by identifying the component and index such that $\mathbf{z}$ contracts to $\mathbf{z}_j^{(k)}$.

The disjointness of the $M_0$ components ensures unique determination of the common message $k$ at both receivers, since output sequences from different components contract to disjoint vertex sets. Within each component, the complete bipartite structure $K_{M_1, M_2}$ guarantees reliable decoding of private messages, as established in Lemma~\ref{lem:bipartite_single_use}.
\end{IEEEproof}

\begin{remark}
This theorem provides a sufficient graph-theoretic condition for achievability with common messages. The characterization is not tight in general, as contraction operations preserve connectivity and cannot create disjoint components from a connected graph. The disjoint union structure must already be present in $G(\mathcal{B}^n)$ or arise through subgraph selection (edge deletion), which is not captured by contraction alone. Nevertheless, this framework proves particularly useful for deriving explicit achievable rate triples, as demonstrated below.
\end{remark}

\subsubsection{Component Index and Type Decomposition}

To apply Theorem~\ref{thm:graph_achievability_triple}, we formalize the notion of connected components in the channel graph. Let the vertex sets $\mathcal{Y}$ and $\mathcal{Z}$ of the bipartite graph $G(\mathcal{B})$ be partitioned into $|\mathcal{U}|$ disjoint connected components. We define a discrete random variable $U$ taking values in $\mathcal{U} = \{1, \ldots, |\mathcal{U}|\}$ as the component index. Since components are disjoint, $U$ is uniquely determined by either output: there exist deterministic functions $u_Y: \mathcal{Y} \to \mathcal{U}$ and $u_Z: \mathcal{Z} \to \mathcal{U}$ such that $U = u_Y(Y) = u_Z(Z)$ whenever $(Y,Z)$ correspond to a channel use. This structural property implies $H(U|Y) = H(U|Z) = 0$.

Extending the type-based analysis from Subsection~\ref{subsec:type}, we define the empirical distribution of component sequences:

\begin{definition}[Component Type]
\label{def:component_type}
For a sequence $\mathbf{u}=(u_1,\ldots,u_n)\in\mathcal{U}^n$ of component indices, let $n_U(k)$ denote the number of occurrences of component $k \in \mathcal{U}$. The type $P_\mathbf{u}$ is the empirical distribution
\begin{align}
    P_\mathbf{u}(u^{(k)}) = \frac{n_U(k)}{n}, \quad k=1,\ldots,|\mathcal{U}|.
\end{align}
The type class $T(P_\mathbf{u})$ denotes the set of all sequences with type $P_\mathbf{u}$.
\end{definition}

Conditioning on the component sequence decomposes the product channel graph into component-specific subgraphs, whose asymptotic properties we now characterize:

\begin{lemma}[Conditional Type-Induced Subgraphs]
\label{lem:conditional_properties}
For a given input distribution $P_X$ inducing output distribution $P_{Y,Z}$ and component distribution $P_U$, the following hold for sufficiently large $n$:

(i) Component type class size:
\begin{align}
    |T(P_\mathbf{u})| = \exp\left[nH(U) + O(\ln n)\right].
\end{align}

(ii) Conditional subgraph structure:
For a typical component sequence $\mathbf{u} \in T(P_\mathbf{u})$, the conditional type-induced subgraph $G(\mathcal{B}^n, P_{\mathbf{x}} | \mathbf{u})$ has vertex set cardinalities
\begin{subequations}
\begin{align}
    |T(P_{\mathbf{y}}|\mathbf{u})| &= \exp\left[ n H(Y|U) + O(\ln n) \right], \\
    |T(P_{\mathbf{z}}|\mathbf{u})| &= \exp\left[ n H(Z|U) + O(\ln n) \right],
\end{align}
\end{subequations}
and vertex degrees
\begin{subequations}
\begin{align}
    d_Y(P_{\mathbf{x}}|\mathbf{u}) &= \exp\left[ n H(Z|Y) + O(\ln n) \right], \\
    d_Z(P_{\mathbf{x}}|\mathbf{u}) &= \exp\left[ n H(Y|Z) + O(\ln n) \right],
\end{align}
\end{subequations}
where all entropies are evaluated with respect to $P_{X,Y,Z,U}$.
\end{lemma}

\begin{IEEEproof}
Part (i) follows from standard type-counting arguments: $|T(P_\mathbf{u})| = n!/\prod_k (n P_\mathbf{u}(k))! = \exp\left[nH(P_\mathbf{u}) + O(\ln n)\right]$ by Stirling's approximation.

For part (ii), fixing $\mathbf{u}$ restricts the output sequences to those consistent with the component allocation. The conditional type classes $T(P_{\mathbf{y}}|\mathbf{u})$ and $T(P_{\mathbf{z}}|\mathbf{u})$ have sizes determined by the conditional entropy $H(Y|U)$ and $H(Z|U)$ through the same multinomial coefficient argument applied in Lemma~\ref{lem:submatrix_asymptotics}. The degree calculations are the same as in Lemma~\ref{lem:submatrix_asymptotics}.
\end{IEEEproof}

\subsubsection{Achievable Rate Triples}

Building on the component decomposition, we establish the existence of the disjoint union structure required by Theorem~\ref{thm:graph_achievability_triple}:

\begin{lemma}[Random Coding for Disjoint Union]
\label{lem:subgraph_covering_with_common}
For any det-BC $\mathcal{B}$ and input distribution $P_X$, the product channel graph $G(\mathcal{B}^n)$ satisfies
\begin{align}
    \label{eq:subgraph_contraction_prob_with_common}
    \lim_{n \to \infty} \Pr\left\{ \bigsqcup_{k=1}^{M_0} K_{M_1, M_2} \prec_c G(\mathcal{B}^n) \right\} = 1
\end{align}
if, for any $\epsilon > 0$, the message sizes satisfy
\begin{align}
    \frac{1}{n} \ln M_0 &= H(U) + O\left(\frac{\ln n}{n}\right), \\
    \frac{1}{n} \ln M_1 &= H(Y|Z) - \epsilon, \\
    \frac{1}{n} \ln M_2 &= H(Z|U) + O\left(\frac{\ln n}{n}\right),
\end{align}
where entropies are evaluated under $P_{Y,Z,U}$ induced by $P_X$.
\end{lemma}

\begin{IEEEproof}
The proof extends the random binning argument from Lemma~\ref{lem:subgraph_covering} to the component-wise structure. We construct the graph homomorphism by treating each typical component sequence independently.

Let $\mathcal{T}_n = T(P_\mathbf{u})$ denote the set of typical component sequences with cardinality $M_0 = |\mathcal{T}_n| = \exp\left[nH(U) + O(\ln n)\right]$. For each $\mathbf{u} \in \mathcal{T}_n$, consider the conditional type-induced subgraph $G(\mathcal{B}^n, P_{\mathbf{x}}|\mathbf{u})$ with
\begin{itemize}
    \item Column vertices: $T(P_{\mathbf{z}}|\mathbf{u})$ with $|T(P_{\mathbf{z}}|\mathbf{u})| = M_2 = \exp\left[nH(Z|U) + O(\ln n)\right]$
    \item Row vertices: $T(P_{\mathbf{y}}|\mathbf{u})$ randomly partitioned into $M_1 = \lfloor\exp(n(H(Y|Z)-\epsilon))\rfloor$ bins
\end{itemize}

We set $M_2 = |T(P_{\mathbf{z}}|\mathbf{u})|$ to include all column vertices within each component. Each row bin has size
\begin{align}
    |B_i| &= \frac{|T(P_{\mathbf{y}}|\mathbf{u})|}{M_1}\notag\\
    &= \exp\left[n(H(Y|U) - H(Y|Z) + \epsilon) + O(\ln n)\right]\notag\\
    &= \exp\left[n(I(Y;Z|U) + \epsilon) + O(\ln n)\right],
\end{align}
where we used $H(Y|U) - H(Y|Z) = I(Y;Z|U)$.

Let $E_{i,j,k}$ denote the event that, in component $k$, bin $B_i$ contains no neighbor of column vertex $\mathbf{z}_j$. By Lemma~\ref{lem:conditional_properties}, column vertices have degree $d_Z = \exp\left[nH(Y|Z) + O(\ln n)\right]$. Following the hypergeometric analysis from Lemma~\ref{lem:subgraph_covering}, we have
\begin{align}
    \Pr\{E_{i,j,k}\} &\leq \left(1 - \frac{d_Z}{|T(P_{\mathbf{y}}|\mathbf{u})|}\right)^{|B_i|} \notag\\
    &\leq \exp\left(-|B_i| \frac{d_Z}{|T(P_{\mathbf{y}}|\mathbf{u})|}\right) \notag\\
    &= \exp\Biggl\{-\exp\left[n(I(Y;Z|U) + \epsilon) + O(\ln n)\right] \frac{\exp\left[nH(Y|Z) + O(\ln n)\right]}{\exp\left[nH(Y|U) + O(\ln n)\right]}\Biggr\} \notag\\
    &= \exp\left\{-\exp\left[n\epsilon + O(\ln n)\right]\right\}.
\end{align}

Applying the union bound over all $M_0 \times M_1 \times M_2$ component-bin-column triples yields
\begin{align}
    &\quad\,\Pr\left\{\bigsqcup_{k=1}^{M_0}K_{M_1, M_2} \nprec_c G(\mathcal{B}^n) \right\} \notag\\
    &\leq \sum_{k=1}^{M_0}\sum_{i=1}^{M_1} \sum_{j=1}^{M_2} \Pr\{E_{i,j,k}\} \notag\\
    &\leq M_0 M_1 M_2 \, \exp\left\{-\exp\left[n\epsilon + O(\ln n)\right]\right\} \notag\\
    &= \exp\left[n(H(U) + H(Y|Z) + H(Z|U)) + O(\ln n)\right] \exp\left\{-\exp\left[n\epsilon + O(\ln n)\right]\right\} \notag\\
    \label{eq:probability, subgraph_covering_with_common}
    &= \exp\left[nH(Y,Z) + O(\ln n)\right] \, \exp\left\{-\exp\left[n\epsilon + O(\ln n)\right]\right\}.
\end{align}
Since $M_{0} M_{1} M_{2}=\exp\left[nH(Y,Z)+O(\ln n)\right]$ grows \emph{exponentially} in $n$, while the factor $\exp\left\{-\exp\left[n\epsilon+O(\ln n)\right]\right\}$ decays \emph{doubly exponentially} in $n$, the right-hand side of \eqref{eq:probability, subgraph_covering_with_common} vanishes as $n\to\infty$ by the same argument as in the proof of Lemma~\ref{lem:subgraph_covering} (cf.~\eqref{eq:probability, subgraph covering}--\eqref{eq:double exponential dominates}), yielding the desired convergence \eqref{eq:subgraph_contraction_prob_with_common}.
\end{IEEEproof}

The immediate consequence establishes achievable rate triples:

\begin{corollary}[Corner Points with Common Messages]
\label{cor:achievable_triple}
For any det-BC $\mathcal{B}$ and input distribution $P_X$, the rate triple
\begin{align}
    (R_0, R_1, R_2) = (H(U), H(Y|Z), H(Z|U))
\end{align}
is achievable, where entropies are evaluated under $P_{Y,Z,U}$ induced by $P_X$. By symmetry, $(H(U), H(Y|U), H(Z|Y))$ is also achievable.
\end{corollary}

\begin{IEEEproof}
This follows immediately from Lemma~\ref{lem:subgraph_covering_with_common} and Theorem~\ref{thm:graph_achievability_triple} by taking $\epsilon \to 0$ and $n \to \infty$.
\end{IEEEproof}

\begin{remark}[Connection to Marton's Inner Bound]
\label{rem:marton_connection}
Corollary~\ref{cor:achievable_triple} represents a special instance within the framework of Marton's inner bound \cite{marton_coding_1979} for broadcast channels with common messages. In the general (stochastic) setting, Marton's bound involves auxiliary variables $(U, V_1, V_2)$ forming the Markov chain $U \to (V_1, V_2) \to X \to (Y, Z)$, with achievable common rate $R_0 \leq \min\{I(U;Y), I(U;Z)\}$.

In our deterministic framework, the component index variable $U$, which is defined as the index of connected components in $G(\mathcal{B})$, plays the role of Marton's common message auxiliary variable. A key simplification arises from the fact that $U$ is a deterministic function of both outputs: $U = u_Y(Y) = u_Z(Z)$. This implies $I(U;Y) = H(U) - H(U|Y) = H(U)$ and similarly $I(U;Z) = H(U)$, so the common rate constraint reduces to $R_0 \leq H(U)$. 

The graph-theoretic construction provides an intuitive physical interpretation: the common message corresponds to identifying which connected component of the channel graph is being used. However, we emphasize that this choice of $U$ as the component index is not exhaustive. The complete common-and-private capacity region $\mathcal{R}(\mathcal{B})$ may admit other choices of auxiliary variables with more complex structural interpretations, which remain to be characterized. The achievable rate triples in Theorem~\ref{thm:achievability_with_common} thus constitute a subset of the full capacity region boundary, corresponding specifically to the component-based decomposition of the channel graph.
\end{remark}

\subsubsection{Parametric Characterization of Achievable Triples}

We now derive explicit parametric forms for achievable rate triples along the curved boundary segments:

\begin{theorem}[Parametric Achievable Triples]
\label{thm:achievability_with_common}
For a det-BC $\mathcal{B}$ with connected components indexed by $\mathcal{U} = \{1, \ldots, |\mathcal{U}|\}$, the rate triple $(R_0, R_1, R_2)$ is achievable if
\begin{subequations}
\label{eq:achievable_triple}
\begin{align}
    R_0 &= -\sum_{u\in\mathcal{U}} q_U(u) \ln q_U(u) = H(q_U), \label{eq:R0_explicit}\\
    R_1 &= \sum_{z\in\mathcal{Z}} q_Z(z) \ln d_{Z,z} = H(Y|Z), \label{eq:R1_explicit}\\
    R_2 &= -\sum_{u\in\mathcal{U}} q_U(u) \sum_{z\in u} q_{Z|U}(z|u) \ln q_{Z|U}(z|u) = H(Z|U), \label{eq:R2_explicit}
\end{align}
\end{subequations}
where the probability distributions are parametrized by $t \in [0,\infty)$ as
\begin{align}
    q_U(u) &= \frac{\sum_{z\in u} d_{Z,z}^t}{\sum_{z'\in\mathcal{Z}} d_{Z,z'}^t}, \label{eq:type_u}\\
    q_{Z|U}(z|u) &= \frac{d_{Z,z}^t}{\sum_{z'\in u} d_{Z,z'}^t}, \label{eq:type_z_given_u}\\
    q_Z(z) &= \frac{d_{Z,z}^t}{\sum_{z'\in\mathcal{Z}} d_{Z,z'}^t}. \label{eq:type_z_marginal}
\end{align}
Here, the notation ``$z \in u$'' denotes that output symbol $z$ belongs to the $u$-th connected component of the bipartite graph $G(\mathcal{B})$.
\end{theorem}

\begin{IEEEproof}
We construct the input distribution using the same parameterization that achieves the curved boundary of the private capacity region. Specifically, we adopt the output distribution $q_Z$ defined by \eqref{eq:type_z_marginal}, which is proportional to $d_{Z,z}^t$, and the uniform conditional distribution $P_{Y|Z}$ over preimages as in Lemma~\ref{lem:degraded_achievability}.

Since the component index $U$ is a deterministic function of $Z$ (specifically, $U = u$ if and only if $z \in u$), the induced distribution on $\mathcal{U}$ is obtained by summing $q_Z(z)$ over all $z$ in each component as follows:
\begin{equation}
\begin{aligned}
    q_U(u)
    &= P_U(u)\\
    &= \sum_{z \in u} P_Z(z)\\
    &= \sum_{z \in u} q_Z(z)\\
    &= \frac{\sum_{z\in u} d_{Z,z}^t}{\sum_{z'\in\mathcal{Z}} d_{Z,z'}^t},
\end{aligned}
\end{equation}
establishing \eqref{eq:type_u}. By the definition of conditional probability we have
\begin{equation}
\begin{aligned}
    q_{Z|U}(z|u)
    &= P_{Z|U}(z|u)\\
    &= \frac{P_Z(z)}{P_U(u)}\\
    &= \frac{q_Z(z)}{q_U(u)}\\
    &= \frac{d_{Z,z}^t}{\sum_{z'\in u} d_{Z,z'}^t},
\end{aligned}
\end{equation}
yielding \eqref{eq:type_z_given_u}. Substituting these distributions into the achievable rates from Corollary~\ref{cor:achievable_triple} completes the proof.
\end{IEEEproof}

\subsubsection{Geometric Interpretation and Publicly Decodable Message}

The parametric characterization reveals a fundamental connection between the private and common-message capacity regions:

\begin{remark}[Rate Reallocation]
\label{rem:rate_reallocation}
Let $\mathcal{R}(\mathcal{B})$ denote the capacity region with common messages. Since common messages are decoded by both receivers, any portion of it can be arbitrarily relabeled as private message for either receiver. If $(R_0, R_1, R_2) \in \mathcal{R}(\mathcal{B})$ is achievable, then for all $\delta_1, \delta_2 \geq 0$ with $\delta_1 + \delta_2 \leq R_0$, we have
\begin{align}
    (R_0 - \delta_1 - \delta_2, R_1 + \delta_1, R_2 + \delta_2) \in \mathcal{R}(\mathcal{B}).
\end{align}
\end{remark}

\begin{corollary}[Boundary Optimality]
\label{cor:boundary_optimality}
The achievable rate triples $(R_0, R_1, R_2)$ characterized in Theorem~\ref{thm:achievability_with_common} lie on the boundary of the capacity region $\mathcal{R}(\mathcal{B})$.
\end{corollary}

\begin{IEEEproof}
We first observe that for any triple $(R_0, R_1, R_2)$ from Theorem~\ref{thm:achievability_with_common}, the projected rate pair onto the private capacity region is given by
\begin{equation}
\begin{aligned}
    (R_1, R_0 + R_2)
    &= (H(Y|Z), H(U) + H(Z|U)) \\
    &= (H(Y|Z), H(Z)).
\end{aligned}
\end{equation}
This pair precisely parametrizes the curved boundary segment $A$--$B$ of $\mathcal{C}(\mathcal{B})$ as established in Definition~\ref{def:explicit_capacity_region}.

Now suppose, for contradiction, that some triple $(R_0, R_1, R_2)$ lies in the interior of $\mathcal{R}(\mathcal{B})$. Then there exists $\delta > 0$ such that $(R_0 + \delta, R_1, R_2) \in \mathcal{R}(\mathcal{B})$. By Remark~\ref{rem:rate_reallocation}, reallocating the entire common rate to receiver~2 yields the following achievable private-message pair:
\begin{align}
    (R_1, R_2 + R_0 + \delta) = (H(Y|Z), H(Z) + \delta) \in \mathcal{C}(\mathcal{B}).
\end{align}
However, we have shown that $(R_1, R_0 + R_2) = (H(Y|Z), H(Z))$ lies on the boundary curve $A$--$B$ of $\mathcal{C}(\mathcal{B})$. The achievability of $(H(Y|Z), H(Z) + \delta)$ would violate the converse bound (Lemma~\ref{lem:general_converse}), yielding a contradiction. Therefore, the original triple must lie on the boundary of $\mathcal{R}(\mathcal{B})$.
\end{IEEEproof}

\begin{remark}[Publicly Decodable Message and Cross-Leakage]
\label{rem:geometric_interpretation}
Corollary~\ref{cor:boundary_optimality} reveals that coding schemes achieving the curved boundary segments of $\mathcal{C}(\mathcal{B})$ inherently support a non-trivial common message rate $R_0$ when the channel graph contains multiple connected components. We term this a \emph{publicly decodable message}.
While the common message $R_0$ is a code-design choice, a publicly decodable message is content that both receivers can decode, regardless of whether it was originally designated as common. Remark~\ref{rem:rate_reallocation} shows that any such publicly decodable rate can always be relabeled as part of $R_0$ without loss of total rate. Thus the two notions coincide at the level of the achievable region, but they remain conceptually distinct.

This phenomenon admits an alternative interpretation as \emph{cross-leakage}: private information intended for one receiver that becomes decodable by the other due to the channel's topological structure. Specifically, $R_0$ quantifies the portion of private information that can be ``lifted'' to a common message without reducing the total achievable rate. In this sense, $R_0$ measures the structural information leakage inherent in the channel graph $G(\mathcal{B})$.

The parametrization \eqref{eq:type_u}--\eqref{eq:type_z_given_u} shows that $R_0$ depends on the number and relative sizes of connected components in $G(\mathcal{B})$. For a connected graph ($|\mathcal{U}| = 1$), we have $H(U) = 0$ for all $t$, yielding $R_0 = 0$ (no leakage). As the graph becomes more fragmented, $R_0$ can increase, reflecting greater potential for information sharing across receivers.
\end{remark}

\begin{remark}[Characterization of Segment $B$--$C$]
\label{rem:BC_segment_origin}
The presence of publicly decodable message provides a structural mechanism for achieving the linear segment $B$--$C$ with slope $-1$ in the capacity region $\mathcal{C}(\mathcal{B})$. Point $B$ corresponds to the parameter $t=1$. If the associated common rate satisfies $R_0(t=1) = H(U|t=1) > 0$, then part of segment $B$--$C$ can be achieved by continuously reallocating this common rate between the two receivers via Remark~\ref{rem:rate_reallocation}, i.e.,
\begin{align}
    (R_1, R_2) = (H(Y|Z) + \alpha R_0, H(Z|U) + (1-\alpha) R_0)
\end{align}
for $\alpha \in [0,1]$. This reallocation maintains constant sum rate $R_1 + R_2 = H(Y,Z)$ and generates a line segment of length $\sqrt{2} R_0(t=1)$. However, this does not fully explain the full segment $B$--$C$. The full segment $B$--$C$ generally has length $|BC| = \sqrt{2} I(Y;Z)$ (Corollary~\ref{cor:BC_length}), where $I(Y;Z) \geq R_0(t=1)$ with strict inequality in general. Detailed numerical results are presented in Section~\ref{sec:numerical}.
\end{remark}

\begin{remark}[Connection to Classical Degraded Channel Results]
\label{rem:degraded_consistency}
This perspective aligns with classical results for degraded broadcast channels \cite[Sect.~5.7]{el_gamal_network_2011}. For a physically degraded channel where receiver~2 is degraded with respect to receiver~1, the private-message capacity region is achievable even when receiver~1 is required to decode receiver~2's message.
In our framework, this classical result corresponds to the scenario where the entire private message for the weaker receiver constitutes publicly decodable message. Such a message can be relabeled as a common message (representing complete leakage to the stronger receiver) without loss in the achievable rate region.
\end{remark}

\subsubsection{Bounds on the Capacity Region with Common Messages}
\label{sec:common_bounds}

We begin with the degraded case, for which the capacity region with common messages admits a complete characterization.

\begin{theorem}[Complete Characterization of $\mathcal{R}\left(\mathcal{B}\right)$ for Degraded Det-BCs]
\label{thm:degraded_3D_exact}
Let $\mathcal{B}_{Y\to Z}$ be the degraded channel obtained from $\mathcal{B}$ by splitting all receiver-1 vertices of degree greater than one, so that $D_Y\left(\mathcal{B}_{Y\to Z}\right) = \{1,1,\ldots,1\}$. Then the capacity region with common messages of $\mathcal{B}_{Y\to Z}$ is given by
\begin{equation}
\begin{aligned}
    \mathcal{R}\left(\mathcal{B}_{Y\to Z}\right)
    =\left\{
    \left(R_0,R_1,R_2\right)\in\mathbb{R}_{\geq 0}^{3}
    :
    \left(R_1,R_0+R_2\right)
    \in\tilde{\mathcal{C}}\left(\mathcal{B}_{Y\to Z}\right)
    \right\}.
\end{aligned}
\label{eq:degraded_3D_exact}
\end{equation}
By symmetry, the capacity region of the degraded channel $\mathcal{B}_{Z\to Y}$ (obtained by splitting receiver-2 vertices) satisfies
\begin{equation}
\begin{aligned}
    \mathcal{R}\left(\mathcal{B}_{Z\to Y}\right)
    =\left\{
    \left(R_0,R_1,R_2\right)\in\mathbb{R}_{\geq 0}^{3}
    :
    \left(R_0+R_1,R_2\right)
    \in\tilde{\mathcal{C}}\left(\mathcal{B}_{Z\to Y}\right)
    \right\}.
\end{aligned}
\label{eq:degraded_3D_exact_sym}
\end{equation}
\end{theorem}

\begin{IEEEproof}
We establish both inclusions for~\eqref{eq:degraded_3D_exact}; \eqref{eq:degraded_3D_exact_sym} follows by exchanging the roles of the two receivers.

\textit{Achievability.}
Let $\left(R_1, R_2'\right) \in \tilde{\mathcal{C}} \left(\mathcal{B}_{Y\to Z}\right)$ and set $R_2' = R_0 + R_2$ for any $R_0, R_2 \geq 0$. By Theorem~\ref{thm:degraded_capacity}, there exists a sequence of codes achieving private rates $\left(R_1, R_2'\right)$ for the degraded channel $\mathcal{B}_{Y\to Z}$. By Remark~\ref{rem:degraded_consistency}, receiver~1 (the stronger receiver) can decode receiver~2's message as well as its own; in particular, the code achieving $\left(R_1, R_2'\right)$ allows both receivers to decode the component of rate $R_2'$ within receiver~2's message. Splitting receiver~2's message into a common part at rate $R_0$ (decodable by both receivers) and a private part at rate $R_2 = R_2' - R_0$ (intended only for receiver~2) yields an achievable rate triple $\left(R_0, R_1, R_2\right)$.

\textit{Converse.}
Suppose $\left(R_0, R_1, R_2\right) \in\mathcal{R}\left(\mathcal{B}_{Y\to Z}\right)$. By Remark~\ref{rem:rate_reallocation} with $\delta_1 = 0$ and $\delta_2 = R_0$, the rate triple $\left(0, R_1, R_0 + R_2\right)$ is also achievable. Since the resulting rate triple has no common message, it follows from Theorem~\ref{thm:main_capacity} that $(R_1,R_0+R_2)\in\tilde{\mathcal C}(\mathcal B_{Y\to Z})$, which proves the converse.
\end{IEEEproof}

\begin{remark}
Theorem~\ref{thm:degraded_3D_exact} admits a simple geometric interpretation. The three-dimensional region $\mathcal{R}\left(\mathcal{B}_{Y\to Z}\right)$ is obtained by taking the two-dimensional region $\tilde{\mathcal{C}}\left(\mathcal{B}_{Y\to Z}\right)$ in the $\left(R_1, R_2\right)$-plane and ``expanding'' it into three dimensions by allowing the original $R_2$ to be freely split between $R_0$ and $R_2$. Concretely, $\mathcal{R}\left(\mathcal{B}_{Y\to Z}\right)$ is the convex hull of the union of two boundary copies of $\tilde{\mathcal{C}}\left(\mathcal{B}_{Y\to Z}\right)$: the $R_0 = 0$ face, $\left\{\left(0, R_1, R_2\right) :\left(R_1, R_2\right) \in\tilde{\mathcal{C}}\left(\mathcal{B}_{Y\to Z}\right)\right\}$, and the $R_2 = 0$ face, $\left\{\left(R_0, R_1, 0\right) :\left(R_1, R_0\right) \in\tilde{\mathcal{C}}\left(\mathcal{B}_{Y\to Z}\right)\right\}$.
\end{remark}

\begin{corollary}[Explicit Outer Bound on $\mathcal{R}\left(\mathcal{B}\right)$ for General Det-BCs]
\label{cor:outer_bound_3D}
For any det-BC $\mathcal{B}$, the capacity region with common messages satisfies
\begin{equation}
    \mathcal{R}\left(\mathcal{B}\right)
    \subseteq\mathcal{O}\left(\mathcal{B}\right)
    :=\mathcal{O}_1\left(\mathcal{B}\right)
    \cap\mathcal{O}_2\left(\mathcal{B}\right),
    \label{eq:outer_bound_3D}
\end{equation}
where
\begin{align}
    \mathcal{O}_1\left(\mathcal{B}\right)
    &:= \left\{\left(R_0, R_1, R_2\right) \in \mathbb{R}_{\geq 0}^3:
    \left(R_0 + R_1, R_2\right)\in \tilde{\mathcal{C}}\left(\mathcal{B}\right)\right\},
    \label{eq:O1}
    \\
    \mathcal{O}_2\left(\mathcal{B}\right)
    &:= \left\{
    \left(R_0, R_1, R_2\right) \in \mathbb{R}_{\geq 0}^3:
    \left(R_1, R_0 + R_2\right)\in \tilde{\mathcal{C}}\left(\mathcal{B}\right)\right\}.
    \label{eq:O2}
\end{align}
\end{corollary}

\begin{IEEEproof}
Let $(R_0,R_1,R_2)\in\mathcal R(\mathcal B)$. Applying Remark~\ref{rem:rate_reallocation} with $\delta_1 = R_0$ and $\delta_2 = 0$, the triple $\left(R_0, R_1, R_2\right)$ is achievable, so $\left(R_0+R_1, R_2\right) \in \tilde{\mathcal{C}}\left(\mathcal{B}\right)$.
Similarly, for $\mathcal{O}_2$, applying Remark~\ref{rem:rate_reallocation} with $\delta_1 = 0$ and $\delta_2 = R_0$, the triple $\left(R_0, R_1, R_2\right)$ is achievable, so $\left(R_1, R_0+R_2\right) \in \tilde{\mathcal{C}}\left(\mathcal{B}\right)$.
\end{IEEEproof}

\begin{remark}
The outer bound in Corollary~\ref{cor:outer_bound_3D} is expressed directly in terms of the private-message capacity region $\tilde{\mathcal C}(\mathcal B)$, and is therefore computable once $\tilde{\mathcal C}(\mathcal B)$ is known. For degraded channels, Theorem~\ref{thm:degraded_3D_exact} shows that this outer bound is tight (after replacing $\tilde{\mathcal C}(\mathcal B)$ by the corresponding degraded private-message capacity region). For general det-BCs, however, the bound is generally not tight, as illustrated numerically in Section~\ref{subsec:region_with_common}.
\end{remark}

\begin{corollary}[Constructive Inner Bound on $\mathcal{R}\left(\mathcal{B}\right)$]
\label{cor:inner_bound_3D}
Define the private-message face of $\mathcal{R}\left(\mathcal{B}\right)$ as
\begin{equation}
    \tilde{\mathcal{C}}_0\left(\mathcal{B}\right)
    :=\left\{\left(0, R_1, R_2\right):
    \left(R_1, R_2\right) \in \tilde{\mathcal{C}}\left(\mathcal{B}\right)\right\},
    \label{eq:C0_def}
\end{equation}
let $\mathcal{T}\left(\mathcal{B}\right)$ denote the set of achievable rate triples from Theorem~\ref{thm:achievability_with_common}, and let
\begin{equation}
    R_{0,\max}\left(\mathcal{B}\right)
    :=\max_{P_X}\min\left\{I\left(X; Y\right), I\left(X; Z\right)\right\}
    \label{eq:R0max}
\end{equation}
denote the common-message capacity, where the maximization is over all input distributions $P_X$. Then
\begin{align}
    &\mathcal{R}\left(\mathcal{B}\right)
    \supseteq\mathcal{I}\left(\mathcal{B}\right)
    := \mathrm{conv}\left(\tilde{\mathcal{C}}_0\left(\mathcal{B}\right)
    \cup\mathcal{T}\left(\mathcal{B}\right)
    \cup\left\{\left(R_{0,\max}\left(\mathcal{B}\right), 0, 0\right)\right\}\right).
    \label{eq:inner_bound_3D}
\end{align}
\end{corollary}

\begin{IEEEproof}
Since $\mathcal R(\mathcal B)$ is convex, it suffices to verify that each set inside the convex hull is achievable. The set $\tilde{\mathcal{C}}_0\left(\mathcal{B}\right)$ is achievable: by Theorem~\ref{thm:main_capacity}, every rate pair $\left(R_1, R_2\right)\in \tilde{\mathcal{C}}\left(\mathcal{B}\right)$ is achievable with $R_0 = 0$. The set $\mathcal{T}\left(\mathcal{B}\right)$ is achievable by Theorem~\ref{thm:achievability_with_common}. The point $\left(R_{0,\max}\left(\mathcal{B}\right), 0, 0\right)$ is achievable by its definition.
\end{IEEEproof}

\begin{remark}
For channels whose graph $G(\mathcal B)$ is connected, Theorem~\ref{thm:achievability_with_common} always yields $R_0=0$. Consequently, $\mathcal T(\mathcal B)\subseteq\tilde{\mathcal C}_0(\mathcal B)$, and the inner bound simplifies to $\mathcal I(\mathcal B)=\operatorname{conv}\left(\tilde{\mathcal C}_0(\mathcal B)\cup\{(R_{0,\max},0,0)\}\right)$. Numerical results in Section~\ref{subsec:region_with_common} show that this bound is generally not tight, indicating that additional achievable rate triples remain to be characterized.
\end{remark}


\subsection{Integer Linear Programming for Finite-Blocklength Analysis}
\label{subsec:ILP}

The asymptotic analysis in previous sections provides closed-form characterizations of the capacity region. However, practical communication systems operate at finite blocklengths, where determining achievable rate pairs requires verifying whether a complete bipartite graph $K_{M_1, M_2}$ can be obtained through contraction from the channel graph $G(\mathcal{B}^n)$. 

We formulate this graph contraction problem as an integer linear program (ILP). The key idea is to interpret contraction as a partitioning problem: vertices of $G(\mathcal{B}^n)$ are grouped into super-nodes such that the resulting quotient graph admits a graph homomorphism to $K_{M_1, M_2}$. This reformulation enables the use of modern ILP solvers to compute optimal achievable rate pairs at finite blocklengths.

\subsubsection{Problem Setup}

Consider the $n$-fold product channel graph $G(\mathcal{B}^n) = (\mathcal{Y}^n, \mathcal{Z}^n, E(\mathcal{B}^n))$ with edge set
\begin{align}
    E(\mathcal{B}^n) = \{(\mathbf{y}, \mathbf{z}) : \exists\, \mathbf{x} \in \mathcal{X}^n \text{ s.t. } \mathbf{y} = y(\mathbf{x}), \mathbf{z} = z(\mathbf{x})\}.
\end{align}
For a fixed row message size $M_1$, we seek the maximum column message size $M_2$ such that $K_{M_1, M_2} \prec_c G(\mathcal{B}^n)$. By Theorem~\ref{thm:bipartite_blocklength}, this characterizes the finite-blocklength achievable rate pair $(R_1, R_2) = (\frac{1}{n}\ln M_1, \frac{1}{n}\ln M_2)$.

\subsubsection{Decision Variables}

We introduce binary decision variables to represent the vertex partitioning and induced edge structure.

\textbf{Vertex assignment variables:} $u_{\mathbf{y}, i} \in \{0,1\}$ for $\mathbf{y} \in \mathcal{Y}^n$, $i \in [M_1]$, indicates whether sequence $\mathbf{y}$ is assigned to row super-node $i$ (representing message $i$ for receiver~1). Similarly, $v_{\mathbf{z}, j} \in \{0,1\}$ for $\mathbf{z} \in \mathcal{Z}^n$, $j \in [|\mathcal{Z}^n|]$, indicates whether sequence $\mathbf{z}$ is assigned to column super-node $j$. Here $[M_1] = \{1, \ldots, M_1\}$ and $[|\mathcal{Z}^n|] = \{1, \ldots, |\mathcal{Z}^n|\}$ denote the index sets. We allow up to $|\mathcal{Z}^n|$ potential column super-nodes to avoid presupposing the value of $M_2$.

\textbf{Column activation variables:} $\beta_j \in \{0,1\}$ for $j \in [|\mathcal{Z}^n|]$ indicates whether column super-node $j$ is active (i.e., corresponds to a valid message for receiver~2).

\textbf{Edge mapping variables:} $w_{\mathbf{y}, \mathbf{z}, i, j} \in \{0,1\}$ for $(\mathbf{y}, \mathbf{z}) \in E(\mathcal{B}^n)$, $i \in [M_1]$, $j \in [|\mathcal{Z}^n|]$, indicates whether edge $(\mathbf{y}, \mathbf{z})$ contributes to the connection between row super-node $i$ and column super-node $j$. These auxiliary variables enable linearization of the logical condition ``super-nodes $i$ and $j$ are connected''.

\subsubsection{ILP Formulation}

For fixed blocklength $n$ and row message size $M_1$, we solve:
\begin{align}
    \text{maximize} \quad & M_2 = \sum_{j=1}^{|\mathcal{Z}^n|} \beta_j \label{eq:ilp_objective}
\end{align}
subject to the following constraints:

\textbf{Constraint 1 (Vertex partition):} Each vertex must be assigned to exactly one super-node, i.e.,
\begin{subequations}
\label{eq:partition_constraints}
\begin{align}
    \sum_{i=1}^{M_1} u_{\mathbf{y}, i} &= 1, \quad \forall \mathbf{y} \in \mathcal{Y}^n, \label{eq:partition_row}\\
    \sum_{j=1}^{|\mathcal{Z}^n|} v_{\mathbf{z}, j} &= 1, \quad \forall \mathbf{z} \in \mathcal{Z}^n. \label{eq:partition_col}
\end{align}
\end{subequations}

\textbf{Constraint 2 (Complete bipartite structure):} Every active column super-node $j$ must be connected to every row super-node $i$, i.e.,
\begin{align}
    \sum_{(\mathbf{y}, \mathbf{z}) \in E(\mathcal{B}^n)} w_{\mathbf{y}, \mathbf{z}, i, j} \geq \beta_j, \quad \forall i \in [M_1], \, j \in [|\mathcal{Z}^n|]. \label{eq:connectivity}
\end{align}
This constraint ensures that if column super-node $j$ is active ($\beta_j = 1$), then for each row super-node $i$, there exists at least one edge $(\mathbf{y}, \mathbf{z})$ with $\mathbf{y}$ assigned to $i$ and $\mathbf{z}$ assigned to $j$.

\textbf{Constraint 3 (Edge mapping logic):} The auxiliary variables $w_{\mathbf{y}, \mathbf{z}, i, j}$ must satisfy
\begin{subequations}
\label{eq:edge_logic}
\begin{align}
    w_{\mathbf{y}, \mathbf{z}, i, j} &\leq u_{\mathbf{y}, i}, \label{eq:edge_logic_a}\\
    w_{\mathbf{y}, \mathbf{z}, i, j} &\leq v_{\mathbf{z}, j}, \label{eq:edge_logic_b}\\
    w_{\mathbf{y}, \mathbf{z}, i, j} &\geq u_{\mathbf{y}, i} + v_{\mathbf{z}, j} - 1, \label{eq:edge_logic_c}
\end{align}
\end{subequations}
for all $(\mathbf{y}, \mathbf{z}) \in E(\mathcal{B}^n)$, $i \in [M_1]$, $j \in [|\mathcal{Z}^n|]$. These constraints linearize the logical AND operation: $w_{\mathbf{y}, \mathbf{z}, i, j} = 1$ if and only if both $u_{\mathbf{y}, i} = 1$ and $v_{\mathbf{z}, j} = 1$.

\textbf{Constraint 4 (Variable domains):}
\begin{align}
    u_{\mathbf{y}, i}, v_{\mathbf{z}, j}, \beta_j, w_{\mathbf{y}, \mathbf{z}, i, j} \in \{0, 1\}.
\end{align}

\subsubsection{Valid Inequalities for Solver Acceleration}

The formulation \eqref{eq:ilp_objective}--\eqref{eq:edge_logic} is minimal in the sense that removing any constraint changes the feasible region. However, adding redundant \emph{valid inequalities} can significantly accelerate branch-and-bound solvers by tightening the linear programming relaxation:

\textbf{Non-empty super-node constraints:}
\begin{align}
    \sum_{\mathbf{y} \in \mathcal{Y}^n} u_{\mathbf{y}, i} \geq 1, \quad \forall i \in [M_1]. \label{eq:nonempty_row}
\end{align}
This ensures each row super-node contains at least one vertex, a condition implicitly required for any feasible solution with $M_2 > 0$.

\textbf{Column activation coupling:}
\begin{align}
    \sum_{\mathbf{z} \in \mathcal{Z}^n} v_{\mathbf{z}, j} \leq |\mathcal{Z}^n| \beta_j, \quad \forall j \in [|\mathcal{Z}^n|]. \label{eq:activation_coupling}
\end{align}
This forces all assignment variables $v_{\mathbf{z}, j}$ to zero whenever column super-node $j$ is inactive ($\beta_j = 0$), eliminating fractional solutions in the LP relaxation where vertices are assigned to inactive super-nodes.

\textbf{Symmetry breaking:}
\begin{align}
    \beta_1 \geq \beta_2 \geq \cdots \geq \beta_{|\mathcal{Z}^n|}. \label{eq:symmetry_breaking}
\end{align}

Since column super-nodes are interchangeable, we can impose an ordering on their activation status without loss of generality. This constraint ensures that active super-nodes ($\beta_j = 1$) are indexed consecutively from $j=1$, while inactive ones ($\beta_j = 0$) have higher indices. This partially reduces the search space.

These valid inequalities do not change the optimal value $M_2^*$ but can reduce solver runtime, particularly for larger blocklengths.

\begin{remark}[Encoding and Decoding Scheme at Finite Blocklength]
\label{rem:ILP coding scheme}
The optimal ILP solution not only determines the maximum achievable $M_{2}$ for a given blocklength $n$ and row message size $M_{1}$, but also directly yields an optimal encoding and decoding scheme for that specific blocklength. Concretely, the solution specifies the vertex contraction mapping from $G\left(\mathcal{B}^{n}\right)$ to $K_{M_{1},M_{2}}$, which identifies the partition of input sequences $\mathcal{X}^{n}$ into $M_{1} \times M_{2}$ codeword classes. The encoder maps each message pair $\left(m_{1}, m_{2}\right)$ to an input sequence in the corresponding class, and each receiver decodes its intended message by identifying the contraction class of its received sequence. We stress that this scheme is optimal at the given finite blocklength $n$; it does not constitute an infinite-blocklength capacity-achieving scheme. The asymptotic capacity region $\mathcal{C}\left(\mathcal{B}\right)$ is characterized independently by Theorem~\ref{thm:main_capacity} through the graph-theoretic framework and the contraction-order analysis of Sections~\ref{sec:graph} and~\ref{sec:derivation}, without relying on the ILP.
\end{remark}

\begin{remark}[Computational Complexity]
\label{rem:ilp_complexity}
The formulation involves $O\left(|\mathcal{Y}^n| M_1 + |\mathcal{Z}^n|^2\right)$ assignment and activation variables, and $O(|\mathcal{X}^n| M_1 |\mathcal{Z}^n|)$ edge mapping variables, where we used $|E(\mathcal{B}^n)| = |\mathcal{X}^n|$ for deterministic channels. The exponential growth of $|\mathcal{X}^n|$ with blocklength $n$ limits practical computations to small blocklengths. Nevertheless, even moderate values of $n$ suffice to observe the transition from finite-blocklength behavior toward asymptotic capacity, providing valuable insights into convergence rates and informing practical system design.
\end{remark}

\begin{remark}[Extension to Common Messages]
\label{rem:ilp_common}
The ILP framework can be extended to optimize rate triples $(R_0, R_1, R_2)$ for the common-and-private message scenario. Following Theorem~\ref{thm:graph_achievability_triple}, the objective becomes maximizing the number of disjoint $K_{M_1, M_2}$ components. This requires introducing component assignment variables to partition vertices into groups corresponding to different common message values. The formulation becomes more complex but remains structurally similar, with additional constraints to enforce that each component forms a complete bipartite subgraph $K_{M_1, M_2}$.
\end{remark}

\subsubsection{Simplification for Degraded Channels}

For degraded det-BCs, the finite-blocklength formulation admits significant simplification by exploiting the forest structure of the channel graph. Consider a degraded channel $\mathcal{B}_{Y\to Z}$ with $D_Y(\mathcal{B}_{Y\to Z}) = \{1, 1, \ldots, 1\}$ and degree sequence $D_Z(\mathcal{B}_{Y\to Z})$ at receiver~2. The problem of maximizing $M_2$ for a fixed $M_1$ can be reformulated as a standard \emph{bin covering problem}: given items with sizes specified by the degree multiset $D_Z^{\otimes n}(\mathcal{B}_{Y\to Z})$ and bin capacity $M_1$, partition the items into the maximum number of disjoint bins such that each bin contains items with total size at least $M_1$.

The key observation is that vertices on the receiver~2 side correspond to items, with vertex degree $d_{\mathbf{z}}$ serving as item size. Vertices with $d_{\mathbf{z}} \geq M_1$ can each form a separate bin, while those with $d_{\mathbf{z}} < M_1$ must be grouped to meet the capacity threshold. This interpretation eliminates the need for:
\begin{itemize}
    \item Row assignment variables $u_{\mathbf{y}, i}$ (since each row vertex has degree 1, the assignment is implicit);
    \item Edge mapping variables $w_{\mathbf{y}, \mathbf{z}, i, j}$ (connectivity is determined by item sizes);
    \item Connectivity constraints \eqref{eq:connectivity} (replaced by bin capacity constraints).
\end{itemize}

The reduced ILP formulation becomes
\begin{alignat}{2}
    \text{Maximize:}\quad
    && M_2
    &= \sum_{j=1}^{|\mathcal{Z}^n|} \beta_j
    \label{eq:ilp_degraded_obj}\\
    \text{Subject to:}\quad
    && \sum_{j=1}^{|\mathcal{Z}^n|} v_{\mathbf{z}, j}
    &= 1,
    \quad \forall \mathbf{z} \in \mathcal{Z}^n,
    \label{eq:ilp_degraded_partition}\\
    &&
    \sum_{\mathbf{z}\in\mathcal{Z}^n}
    v_{\mathbf{z}, j}\, d_{\mathbf{z}}
    &\geq M_1\,\beta_j,
    \quad \forall j \in [|\mathcal{Z}^n|],
    \label{eq:ilp_degraded_capacity}\\
    &&
    v_{\mathbf{z}, j}, \beta_j
    &\in \{0,1\}.
    \label{eq:ilp_degraded_domain}
\end{alignat}
Here, $v_{\mathbf{z}, j}$ indicates whether item (vertex) $\mathbf{z}$ is assigned to bin $j$, and $\beta_j$ indicates whether bin $j$ is active. Constraint \eqref{eq:ilp_degraded_partition} ensures each item is assigned to exactly one bin, while \eqref{eq:ilp_degraded_capacity} enforces that active bins have total size at least $M_1$.

This formulation reduces the number of variables from $O(|\mathcal{X}^n| M_1 |\mathcal{Z}^n|)$ in the general ILP to $O(|\mathcal Z^n|^2)$ for degraded channels, resulting in a substantial reduction in computational complexity. The reduced problem can be solved much more efficiently using specialized algorithms for the classical bin covering problem or its LP relaxations. Section~\ref{subsec:FBR} reports the corresponding problem sizes and solver runtimes for both the general Blackwell channel and its degraded counterpart. Moreover, as discussed in Remark~\ref{rem:bin_cover_approx}, the reduced bin covering problem admits classical approximation algorithms whose running time is polynomial in the number of items $m=|\mathcal{Z}^n|$, while the induced rate loss is asymptotically negligible.

\begin{remark}[Approximation Algorithms for Bin Covering]
\label{rem:bin_cover_approx}
The degraded-channel formulation in~\eqref{eq:ilp_degraded_obj} admits several classical approximation algorithms for the bin covering problem, including the Dual Next-Fit algorithm (approximation ratio $1/2$, runtime $O(m)$) \cite{assmann_dual_1984}, the Bidirectional Bin-Filling algorithm (ratio $2/3$, runtime $O(m\log m)$), and the Three-Classes Bin-Filling algorithm (ratio $3/4$, runtime $O(m\log^2 m)$) \cite{csirik_two_1999}, where $m=|\mathcal Z^n|$ denotes the number of items in the bin covering instance. Although these algorithms do not always attain the optimal number of covered bins, they are asymptotically sufficient for our purpose.

Indeed, let $M_2^*(n)$ denote the optimal number of covered bins and suppose an approximation algorithm achieves $M_2^{A}(n)\ge \alpha M_2^*(n)$ for some constant $\alpha\in(0,1)$. Then
\begin{align}
    \frac1n\ln M_2^{A}(n)
    \geq\frac1n\ln M_2^*(n)+\frac{\ln\alpha}{n}.
\end{align}
Since $\alpha$ is independent of $n$, the approximation incurs only an $O(1/n)$ rate loss, which is asymptotically negligible compared with the capacity characterization.

It should be emphasized that approximation does not alter the exponential dependence on the channel blocklength, since the problem itself contains $|\mathcal Z^n|=\exp\left[nH(Z)+O(\ln n)\right]$ items. Its practical benefit instead lies in avoiding the substantially higher computational cost of solving the exact ILP by general-purpose branch-and-bound algorithms, thereby significantly extending the range of tractable blocklengths.
\end{remark}


\section{Numerical Results}
\label{sec:numerical}

In this section, we present numerical results that illustrate and validate our theoretical characterizations. The visualizations provide intuitive insights into the geometric structure of the capacity region and demonstrate the role of graph-theoretic properties in determining achievable rates.


\subsection{Private-Message Capacity Region}
\label{subsec:region w/o common}

We begin by examining the private-message capacity region characterized in Section~\ref{sec:derivation}, using the Blackwell channel as a canonical example.

\begin{figure}[t]
    \centering
    \subfloat[]{\includegraphics[width=.47\linewidth]{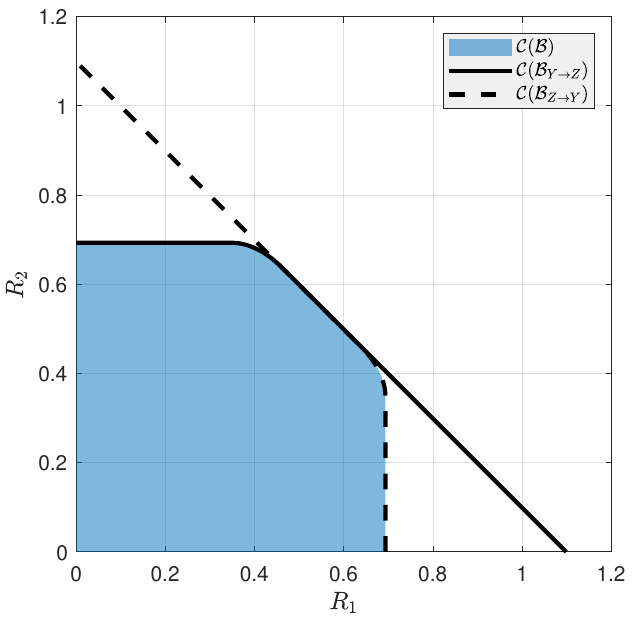}
    \label{fig:blackwell_num1}}
    \hfil
    \subfloat[]{\includegraphics[width=.5\linewidth]{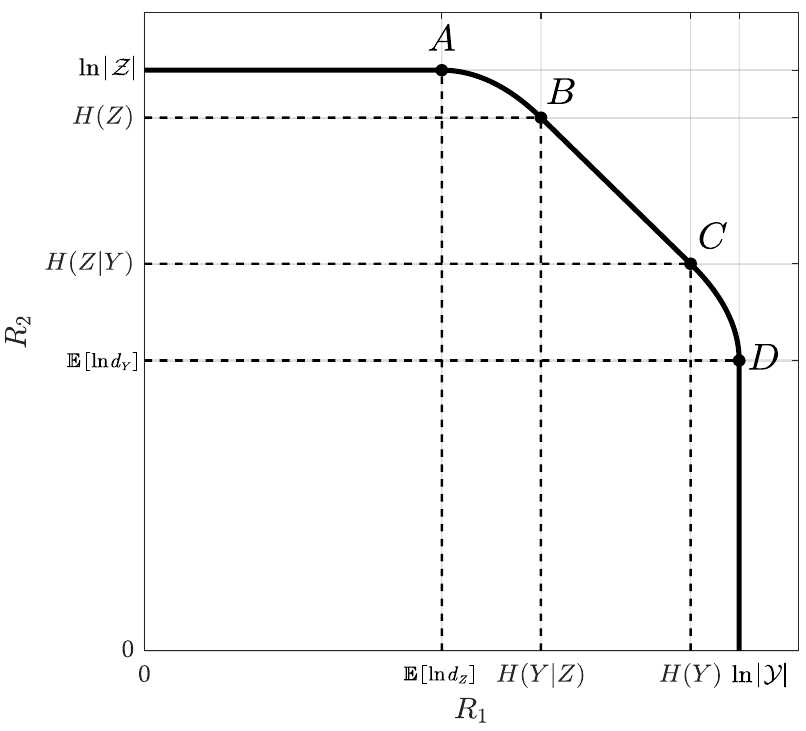}
    \label{fig:blackwell_num2}}
    \caption{Private-message capacity region of the Blackwell channel.
    (a) Capacity regions of the Blackwell channel $\mathcal{C}(\mathcal{B})$ and its two induced degraded channels $\mathcal{C}(\mathcal{B}_{Y\to Z})$ and $\mathcal{C}(\mathcal{B}_{Z\to Y})$ obtained via vertex splitting (see Fig.~\ref{fig:vertex contraction} for channel structures).
    (b) Detailed view of $\mathcal{C}(\mathcal{B})$ with corner points $A$, $B$, $C$, $D$ and their coordinate projections. All entropies are evaluated under a uniform input distribution $P_X(x) = 1/|\mathcal{X}|$.}
    \label{fig:blackwell_num}
\end{figure}

\textbf{Illustrative example: Blackwell channel.}
Fig.~\ref{fig:blackwell_num} illustrates the explicit capacity region structure for the Blackwell channel. Fig.~\ref{fig:blackwell_num1} displays the capacity region $\mathcal{C}(\mathcal{B})$ alongside those of the two degraded channels $\mathcal{B}_{Y\to Z}$ and $\mathcal{B}_{Z\to Y}$ obtained through vertex splitting (the channel structures are shown in Fig.~\ref{fig:vertex contraction}). As established by Lemma~\ref{lem:general_converse}, the capacity region $\mathcal{C}(\mathcal{B})$ is precisely the intersection $\mathcal{C}(\mathcal{B}_{Y\to Z}) \cap \mathcal{C}(\mathcal{B}_{Z\to Y})$. Each degraded channel contributes one curved boundary segment to $\mathcal{C}(\mathcal{B})$: segment $A$--$B$ from $\mathcal{B}_{Y\to Z}$ and segment $C$--$D$ from $\mathcal{B}_{Z\to Y}$. The overlapping portion of the two degraded regions forms the linear segment $B$--$C$ with slope $-1$ on the boundary of $\mathcal{C}(\mathcal{B})$.

Fig.~\ref{fig:blackwell_num2} provides a detailed view of $\mathcal{C}(\mathcal{B})$ with the corner points $A$, $B$, $C$, $D$ explicitly marked and their coordinates indicated by dashed projection lines. The $R_1$-coordinates of these points are $\mathbb{E}[\ln d_Z]$, $H(Y|Z)$, $H(Y)$, and $\ln|\mathcal{Y}|$ respectively, where $\mathbb{E}[\ln d_Z] = \frac{1}{|\mathcal{Z}|}\sum_{z\in\mathcal{Z}} \ln d_{Z,z}$ denotes the average log-degree at receiver~2. By symmetry, the $R_2$-coordinates follow analogous expressions in terms of receiver~1's parameters. 
The geometric structure directly reveals key information-theoretic quantities. For instance, the horizontal projection of segment $B$--$C$ equals $R_1^C - R_1^B = H(Y) - H(Y|Z) = I(Y;Z)$, consistent with Corollary~\ref{cor:BC_length}, where the mutual information quantifies the length of the sum-rate-optimal linear segment.

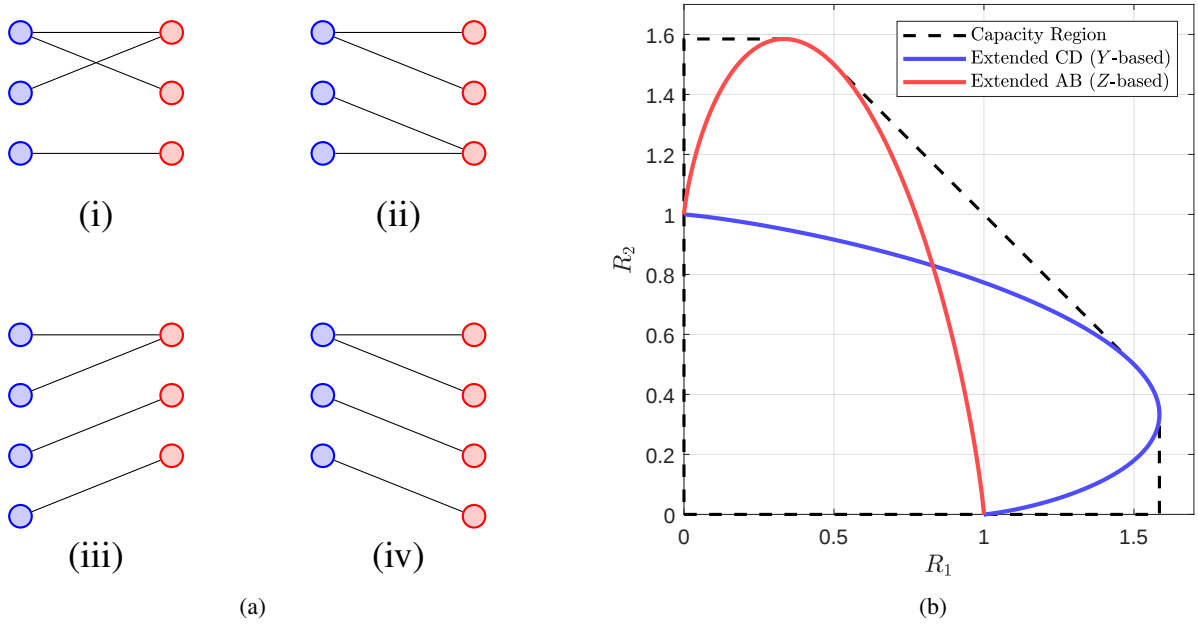
\begin{figure}[t]
    \centering
    \subfloat[]{
    \makebox[0.48\linewidth][c]{%
    \begin{tikzpicture}[
        scale=1,
        y_node/.style={circle, fill=blue!20, draw=blue, inner sep=3pt, thick, minimum size=6pt},
        z_node/.style={circle, fill=red!20, draw=red, inner sep=3pt, thick, minimum size=6pt}
    ]
    \begin{scope}
        \node[y_node] (y1) at (0, .8) {}; 
        \node[y_node] (y2) at (0, 0) {};
        \node[y_node] (y3) at (0, -.8) {}; 
        
        \node[z_node] (z1) at (2, .8) {};
        \node[z_node] (z2) at (2, 0) {};
        \node[z_node] (z3) at (2, -.8) {};

        \draw[thin, black] (y1) -- (z1);
        \draw[thin, black] (y1) -- (z2);
        \draw[thin, black] (y2) -- (z1);
        \draw[thin, black] (y3) -- (z3);
        
        \node[align=center, font=\large, above] at (1, -2) {(i)};
    \end{scope}

    \begin{scope}[shift={(4,0)}]
        \node[y_node] (y1) at (0, .8) {}; 
        \node[y_node] (y2) at (0, 0) {};
        \node[y_node] (y3) at (0, -.8) {}; 
        
        \node[z_node] (z1) at (2, .8) {};
        \node[z_node] (z2) at (2, 0) {};
        \node[z_node] (z3) at (2, -.8) {};

        \draw[thin, black] (y1) -- (z1);
        \draw[thin, black] (y1) -- (z2);
        \draw[thin, black] (y2) -- (z3);
        \draw[thin, black] (y3) -- (z3);
        
        \node[align=center, font=\large, above] at (1, -2) {(ii)};
    \end{scope}

    \begin{scope}[shift={(0,-4)}]
        \node[y_node] (y1) at (0, .8) {}; 
        \node[y_node] (y2) at (0, 0) {};
        \node[y_node] (y3) at (0, -.8) {}; 
        \node[y_node] (y4) at (0, -1.6) {}; 
        
        \node[z_node] (z1) at (2, .8) {};
        \node[z_node] (z2) at (2, 0) {};
        \node[z_node] (z3) at (2, -.8) {};

        \draw[thin, black] (y1) -- (z1);
        \draw[thin, black] (y2) -- (z1);
        \draw[thin, black] (y3) -- (z2);
        \draw[thin, black] (y4) -- (z3);
        
        \node[align=center, font=\large, above] at (1, -2.5) {(iii)};
    \end{scope}

    \begin{scope}[shift={(4,-4)}]
        \node[y_node] (y1) at (0, .8) {}; 
        \node[y_node] (y2) at (0, 0) {};
        \node[y_node] (y3) at (0, -.8) {}; 
        
        \node[z_node] (z1) at (2, .8) {};
        \node[z_node] (z2) at (2, 0) {};
        \node[z_node] (z3) at (2, -.8) {};
        \node[z_node] (z4) at (2, -1.6) {};

        \draw[thin, black] (y1) -- (z1);
        \draw[thin, black] (y1) -- (z2);
        \draw[thin, black] (y2) -- (z3);
        \draw[thin, black] (y3) -- (z4);
        
        \node[align=center, font=\large, above] at (1, -2.5) {(iv)};
    \end{scope}
    \end{tikzpicture}
    }
    \label{fig:same_degree_graph}}%
    \subfloat[]{
    \makebox[0.48\linewidth][c]{%
    \includegraphics[width=.48\linewidth]{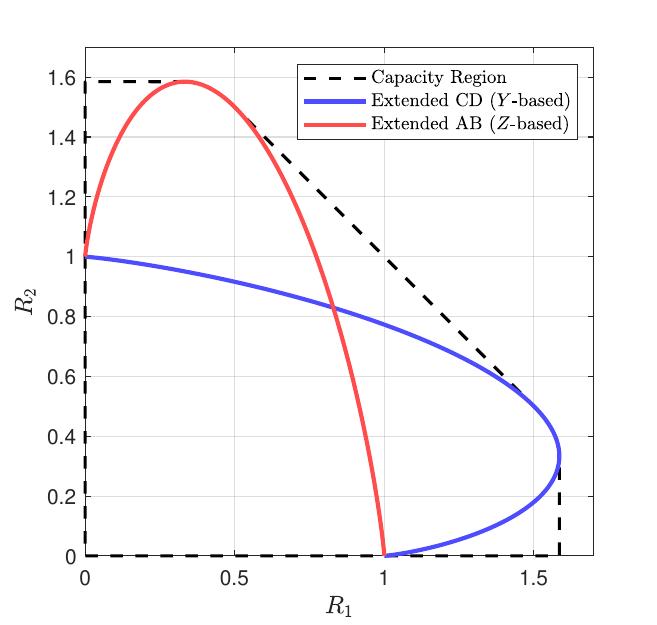}
    \label{fig:same_degree_capacity_region}
    }
    }

    \caption{Two structurally distinct det-BCs with identical degree sequences and their shared private-message capacity region. (a) Bipartite graphs of two different det-BCs (i) and (ii), and their common induced degraded channels (iii) $\mathcal{B}_{Y\to Z}$ and (iv) $\mathcal{B}_{Z\to Y}$ obtained via vertex splitting. (b) The capacity region $\mathcal{C}(\mathcal{B})$, identical for both channels. The red and blue curves represent the full parametric trajectories for $t \in (-\infty, \infty)$, contributed by degree sequences $D_Z$ and $D_Y$, respectively.}
    \label{fig:same_degree}
\end{figure}

\textbf{Degree sequence sufficiency.}
Fig.~\ref{fig:same_degree} demonstrates that structurally distinct det-BCs can share identical private-message capacity regions. The two channels shown in Fig.~\ref{fig:same_degree_graph}(i) and (ii) have different edge configurations but identical degree sequences $D_Y(\mathcal{B}) = D_Z(\mathcal{B}) = \{2, 1\}$. Consequently, they induce the same degraded channels $\mathcal{B}_{Y\to Z}$ and $\mathcal{B}_{Z\to Y}$ through vertex splitting, shown in (iii) and (iv), respectively.

As established in Section~\ref{sec:derivation}, the degree sequence $D_Z(\mathcal{B})$ determines the curved segment $A$--$B$ (red curve in Fig.~\ref{fig:same_degree_capacity_region}), while $D_Y(\mathcal{B})$ determines segment $C$--$D$ (blue curve). The capacity region boundary is completed by time-sharing. This confirms Corollary~\ref{cor:degree_sufficiency}: the capacity region depends on the channel transition probability $p(y,z|x)$ only through the degree sequences $D_Y$ and $D_Z$, independent of the specific edge configuration.
Fig.~\ref{fig:same_degree_capacity_region} also illustrates the extended parametric curves. While the curved segments on $\partial\mathcal{C}(\mathcal{B})$ correspond to $t \in [0,1]$ in the parametric equations \eqref{eq:parametric_R1_simplified}--\eqref{eq:parametric_R2_simplified}, the figure displays the full trajectories for $t \in (-\infty, \infty)$. These extended curves lie entirely within the capacity region and reveal the complete geometric structure of the parametric family.

\begin{figure}[t]
    \centering    
    \subfloat[]{
    \begin{tikzpicture}[
        scale=1,
        y_node/.style={circle, fill=blue!20, draw=blue, inner sep=3pt, thick, minimum size=6pt},
        z_node/.style={circle, fill=red!20, draw=red, inner sep=3pt, thick, minimum size=6pt}
    ]
    \begin{scope}
        \node[y_node] (y1) at (0, 1.5) {}; 
        \node[y_node] (y2) at (0, .5) {};
        \node[y_node] (y3) at (0, -.5) {}; 
        \node[y_node] (y4) at (0, -1.5) {};
        
        \node[z_node] (z1) at (2, .5) {};
        \node[z_node] (z2) at (2, -.5) {};

        \node at (1,-2.5) {};

        \foreach \y in {y1,y2,y3,y4} {
            \foreach \z in {z1, z2} {
                \draw[thin, black] (\y) -- (\z);
            }
        }
    \end{scope}
    \end{tikzpicture}
    \includegraphics[width=.3\linewidth]{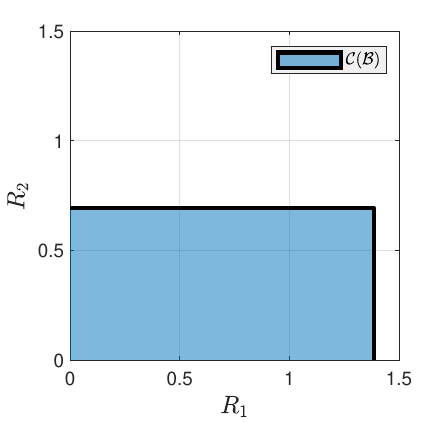}
    \label{fig:general_1_complete}}
    \hfil
    \subfloat[]{
    \begin{tikzpicture}[
        scale=1,
        y_node/.style={circle, fill=blue!20, draw=blue, inner sep=3pt, thick, minimum size=6pt},
        z_node/.style={circle, fill=red!20, draw=red, inner sep=3pt, thick, minimum size=6pt}
    ]
    \begin{scope}
        \node[y_node] (y1) at (0, .9) {}; 
        \node[y_node] (y2) at (0, .3) {};
        \node[y_node] (y3) at (0, -.3) {}; 
        \node[y_node] (y4) at (0, -.9) {};
        
        \node[z_node] (z1) at (2, .9) {}; 
        \node[z_node] (z2) at (2, .3) {};
        \node[z_node] (z3) at (2, -.3) {}; 
        \node[z_node] (z4) at (2, -.9) {};

        \node at (1,-2.5) {};

        \draw[thin, black] (y1) -- (z1);
        \draw[thin, black] (y2) -- (z2);
        \draw[thin, black] (y3) -- (z3);
        \draw[thin, black] (y4) -- (z4);
    \end{scope}
    \end{tikzpicture}
    \includegraphics[width=.3\linewidth]{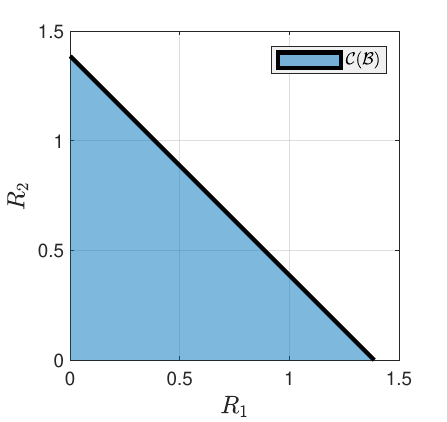}
    \label{fig:general_2_identity}}
    \hfil
    \subfloat[]{
    \begin{tikzpicture}[
        scale=1,
        y_node/.style={circle, fill=blue!20, draw=blue, inner sep=3pt, thick, minimum size=6pt},
        z_node/.style={circle, fill=red!20, draw=red, inner sep=3pt, thick, minimum size=6pt}
    ]
    \begin{scope}
        \node[y_node] (y1) at (0, 1.5) {}; 
        \node[y_node] (y2) at (0, .5) {};
        \node[y_node] (y3) at (0, -.5) {}; 
        \node[y_node] (y4) at (0, -1.5) {};
        
        \node[z_node] (z1) at (2, .5) {};
        \node[z_node] (z2) at (2, -.5) {};

        \node at (1,-2.5) {};

        \foreach \y in {y1,y2} {
            \draw[thin, black] (\y) -- (z1);
        }
        \foreach \y in {y3,y4} {
            \draw[thin, black] (\y) -- (z2);
        }
    \end{scope}
    \end{tikzpicture}
    \includegraphics[width=.3\linewidth]{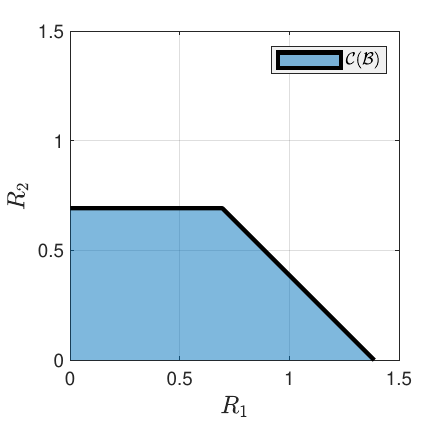}
    \label{fig:general_3_degraded_biregular}}
    \hfil
    \subfloat[]{
    \begin{tikzpicture}[
        scale=1,
        y_node/.style={circle, fill=blue!20, draw=blue, inner sep=3pt, thick, minimum size=6pt},
        z_node/.style={circle, fill=red!20, draw=red, inner sep=3pt, thick, minimum size=6pt}
    ]
    \begin{scope}
        \node[y_node] (y1) at (0, 1.5) {}; 
        \node[y_node] (y2) at (0, .5) {};
        \node[y_node] (y3) at (0, -.5) {}; 
        \node[y_node] (y4) at (0, -1.5) {};
        
        \node[z_node] (z1) at (2, 1.5) {};
        \node[z_node] (z2) at (2, .5) {};
        \node[z_node] (z3) at (2, -.5) {};
        \node[z_node] (z4) at (2, -1.5) {};

        \node at (1,-2.5) {};

        \foreach \y in {y1,y2} {
            \foreach \z in {z1, z2} {
                \draw[thin, black] (\y) -- (\z);
            }
        }
        \foreach \y in {y3,y4} {
            \foreach \z in {z3, z4} {
                \draw[thin, black] (\y) -- (\z);
            }
        }
    \end{scope}
    \end{tikzpicture}
    \includegraphics[width=.3\linewidth]{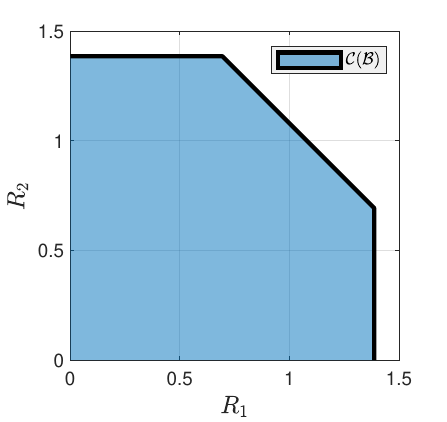}
    \label{fig:general_4_biregular}}
    \hfil
    \subfloat[]{
    \begin{tikzpicture}[
        scale=1,
        y_node/.style={circle, fill=blue!20, draw=blue, inner sep=3pt, thick, minimum size=6pt},
        z_node/.style={circle, fill=red!20, draw=red, inner sep=3pt, thick, minimum size=6pt}
    ]
    \begin{scope}
        \node[y_node] (y1) at (0, 1.5) {}; 
        \node[y_node] (y2) at (0, .5) {};
        \node[y_node] (y3) at (0, -.5) {}; 
        \node[y_node] (y4) at (0, -1.5) {};
        
        \node[z_node] (z1) at (2, 1) {};
        \node[z_node] (z2) at (2, 0) {};
        \node[z_node] (z3) at (2, -1) {};

        \node at (1,-2.5) {};

        \foreach \z in {z1,z2,z3} {
            \draw[thin, black] (y1) -- (\z);
        }
        \draw[thin, black] (y2) -- (z1);
        \draw[thin, black] (y3) -- (z2);
        \draw[thin, black] (y4) -- (z3);
    \end{scope}
    \end{tikzpicture}
    \includegraphics[width=.3\linewidth]{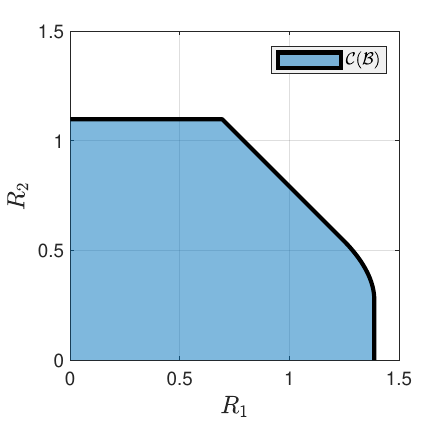}
    \label{fig:general_5_right_regular}}
    \hfil
    \subfloat[]{
    \begin{tikzpicture}[
        scale=1,
        y_node/.style={circle, fill=blue!20, draw=blue, inner sep=3pt, thick, minimum size=6pt},
        z_node/.style={circle, fill=red!20, draw=red, inner sep=3pt, thick, minimum size=6pt}
    ]
    \begin{scope}
        \node[y_node] (y1) at (0, 1.5) {}; 
        \node[y_node] (y2) at (0, .5) {};
        \node[y_node] (y3) at (0, -.5) {}; 
        \node[y_node] (y4) at (0, -1.5) {};
        
        \node[z_node] (z1) at (2, 1.5) {};
        \node[z_node] (z2) at (2, .5) {};
        \node[z_node] (z3) at (2, -.5) {};
        \node[z_node] (z4) at (2, -1.5) {};

        \node at (1,-2.5) {};

        \foreach \y in {y1,y2,y3,y4} {
            \draw[thin, black] (\y) -- (z1);
        }
        \foreach \z in {z2,z3,z4} {
            \draw[thin, black] (y1) -- (\z);
        }
    \end{scope}
    \end{tikzpicture}
    \includegraphics[width=.3\linewidth]{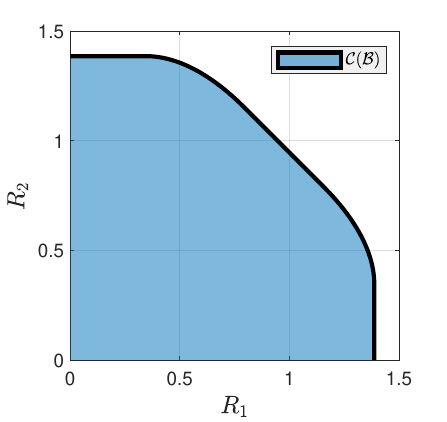}
    \label{fig:general_6_curve}}
    
    \caption{Bipartite graph structures and corresponding private-message capacity regions for representative det-BCs. Each subfigure shows the channel graph (left) and its capacity region (right). (a) Complete bipartite graph. (b) Identity biadjacency matrix (degraded from both perspectives). (c) Degraded channel with constant receiver~2 degrees. (d) Biregular bipartite graph. (e) Constant receiver~2 degrees (non-degraded). (f) General case.}
    \label{fig:general_num}
\end{figure}
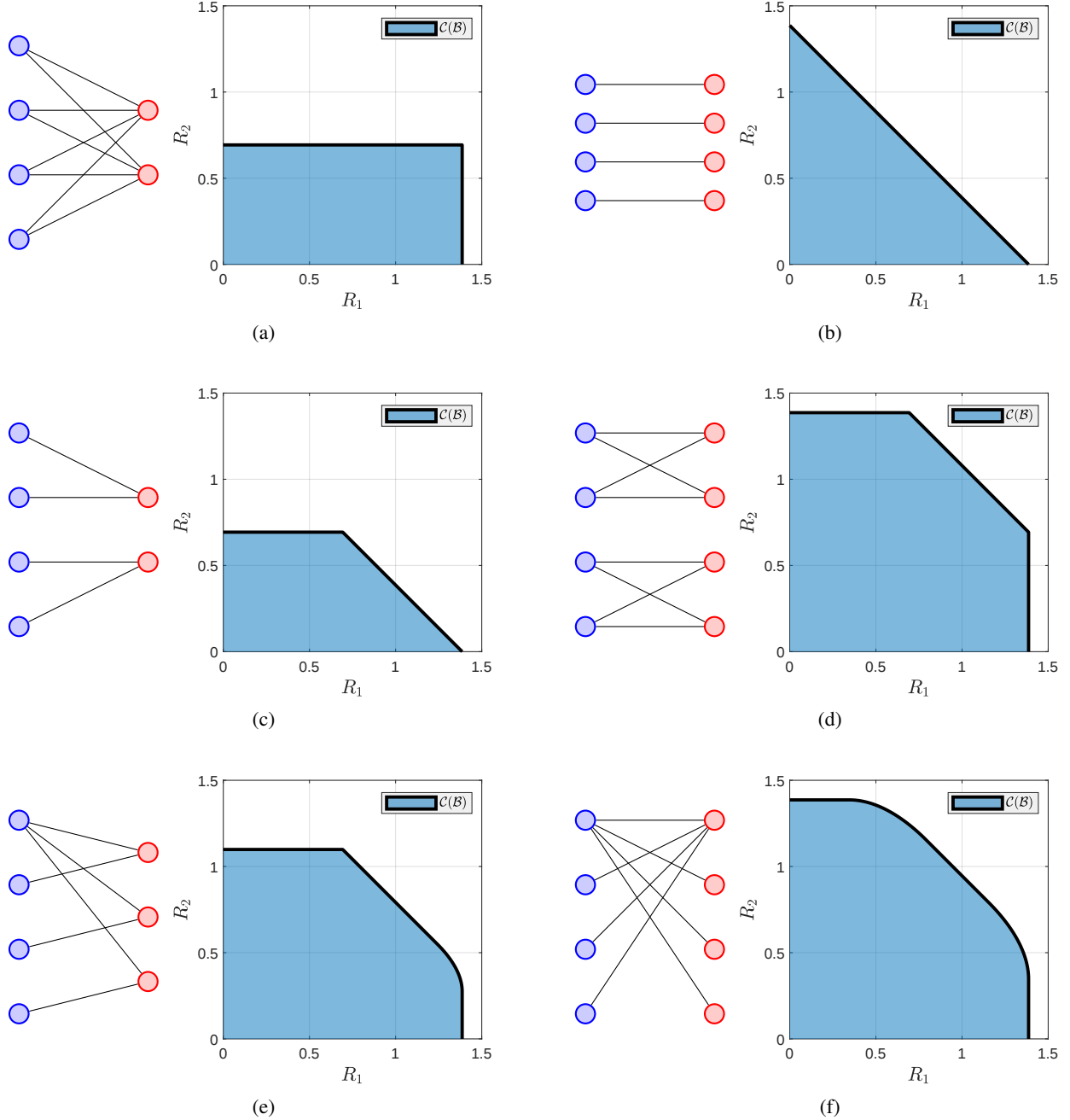

\textbf{Special cases and geometric degeneracies.}
Fig.~\ref{fig:general_num} illustrates how the graph structure determines the geometric properties of the capacity region. These examples reveal the relationship between degree sequence regularity and boundary segment degeneracy.
\begin{enumerate}[(a)]
    \item Complete bipartite graph (Fig.~\ref{fig:general_1_complete}): When $G(\mathcal{B}) = K_{|\mathcal{Y}|, |\mathcal{Z}|}$, all vertices have maximum degree, yielding uniform degree sequences $D_Y = \{|\mathcal{Z}|, \ldots, |\mathcal{Z}|\}$ and $D_Z = \{|\mathcal{Y}|, \ldots, |\mathcal{Y}|\}$. The capacity region degenerates to a rectangle $[0, \ln|\mathcal{Y}|] \times [0, \ln|\mathcal{Z}|]$, as both curved segments and the connecting $B$--$C$ collapse to points.
    \item Identity biadjacency matrix (Fig.~\ref{fig:general_2_identity}): When $A(\mathcal{B}) = I$ (implying $|\mathcal{Y}| = |\mathcal{Z}| = |\mathcal{X}|$), the channel is degraded from both perspectives: $X \to Y \to Z$ and $X \to Z \to Y$. The capacity region becomes the triangle with vertices $(0, 0)$, $(\ln|\mathcal{X}|, 0)$, and $(0, \ln|\mathcal{X}|)$, as segments $A'$--$A$--$B$ and $C$--$D$--$D'$ vanish.
    \item Degraded with constant receiver~2 degrees (Fig.~\ref{fig:general_3_degraded_biregular}): When $\mathcal{B}$ is degraded ($D_Y = \{1, \ldots, 1\}$) and receiver~2 has constant degrees ($D_Z = \{d, \ldots, d\}$), the parametric distribution $q_Z(D_Z, t)$ becomes uniform for all $t$, independent of the parameter. Consequently, segment $A$--$B$ degenerates to a single point, and the capacity region is a trapezoid.
    \item Biregular bipartite graph (Fig.~\ref{fig:general_4_biregular}): When both sides have constant degrees ($D_Y = \{d_Y, \ldots, d_Y\}$ and $D_Z = \{d_Z, \ldots, d_Z\}$), both curved segments $A$--$B$ and $C$--$D$ degenerate to points. The capacity region is a pentagon with only linear boundary segments: $A$--$A'$, $B$--$C$, and $D$--$D'$.
    \item Constant receiver~2 degrees (non-degraded) (Fig.~\ref{fig:general_5_right_regular}): When only receiver~2 has constant degrees ($D_Z = \{d, \ldots, d\}$) but the channel is not degraded, segment $A$--$B$ degenerates while $C$--$D$ remains a proper curve.    
    \item General case (Fig.~\ref{fig:general_6_curve}): When degree sequences contain distinct values, all five boundary segments are present: $A$--$A'$, $A$--$B$, $B$--$C$, $C$--$D$, and $D$--$D'$. This is the most common case, exhibiting the full geometric structure described in Definition~\ref{def:explicit_capacity_region}.
\end{enumerate}
These examples demonstrate that constant degree sequences lead to geometric degeneracies in the capacity region. Specifically, when $D_Z$ has identical elements, segment $A$--$B$ reduces to a point (the parametric family collapses), and similarly for $D_Y$ and segment $C$--$D$.


\subsection{Capacity Region with Common Message}
\label{subsec:region_with_common}

\begin{figure}[t]
    \centering
    \subfloat[]{\includegraphics[width=.48\linewidth]{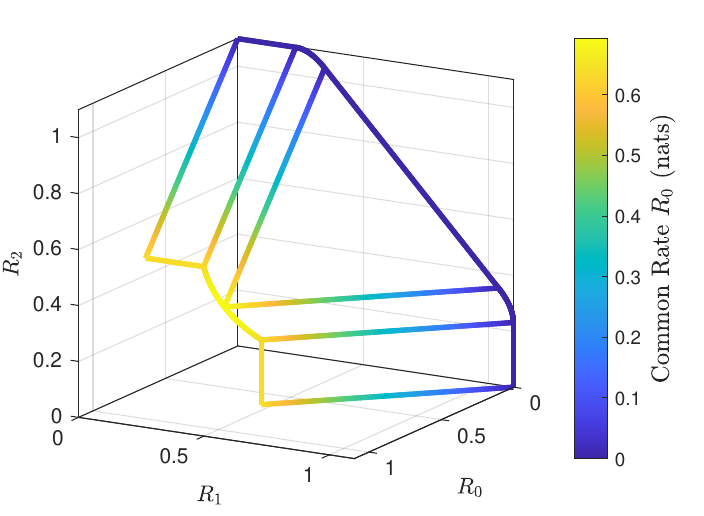}
    \label{fig:with_common_1}}
    \hfil
    \subfloat[]{\includegraphics[width=.48\linewidth]{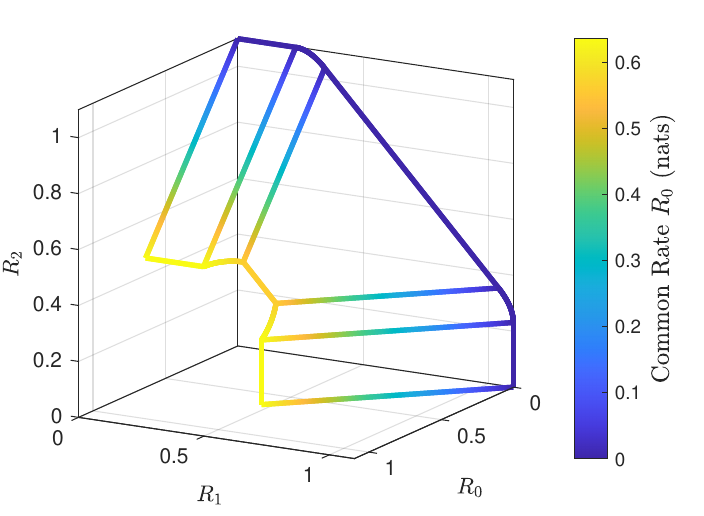}
    \label{fig:with_common_2}}

    \caption{Partial visualization of the three-dimensional capacity region with common messages. The regions are shown for two structurally distinct channels from Fig.~\ref{fig:same_degree_graph}: (a) channel (i) and (b) channel (ii). Colors indicate the common message rate $R_0$ (blue: $R_0 = 0$, yellow: larger $R_0$ characterized by Theorem~\ref{thm:achievability_with_common}).}
    \label{fig:with_common}
\end{figure}

We next turn to the capacity region with common messages. Unlike the private-message capacity region, which is completely determined by the degree sequences, the common-message capacity region depends on finer structural properties of the bipartite graph. The following numerical examples illustrate both this structural dependence and the current state of our analytical characterization.

Fig.~\ref{fig:with_common} illustrates the achievable rate triples characterized in Section~\ref{subsec:common} for the two channels in Fig.~\ref{fig:same_degree_graph}. Although these channels share the same private-message capacity region (Fig.~\ref{fig:same_degree}), their three-dimensional regions differ because the common-message capacity depends on structural properties of the bipartite graph that are not captured by the degree sequences alone.
The blue curves correspond to the private-message capacity region embedded in the plane $R_0=0$, while the yellow curves represent the achievable triples characterized by Theorem~\ref{thm:achievability_with_common}. The intermediate colored curves are obtained by the rate reallocation property (Remark~\ref{rem:rate_reallocation}), which redistributes common-message rate into private-message rates while preserving the total rate $R_0+R_1+R_2$.
The figure therefore visualizes only the portion of $\mathcal R(\mathcal B)$ that is currently characterized. Beyond the yellow curves, the true boundary remains unknown. We conjecture that increasing the common-message rate beyond this point reduces the maximum achievable sum rate, but a rigorous characterization of this tradeoff for the true capacity region $\mathcal{R}(\mathcal{B})$ remains an open problem.

\begin{figure}[t]
    \centering
    \subfloat[]{\includegraphics[width=.48\linewidth]{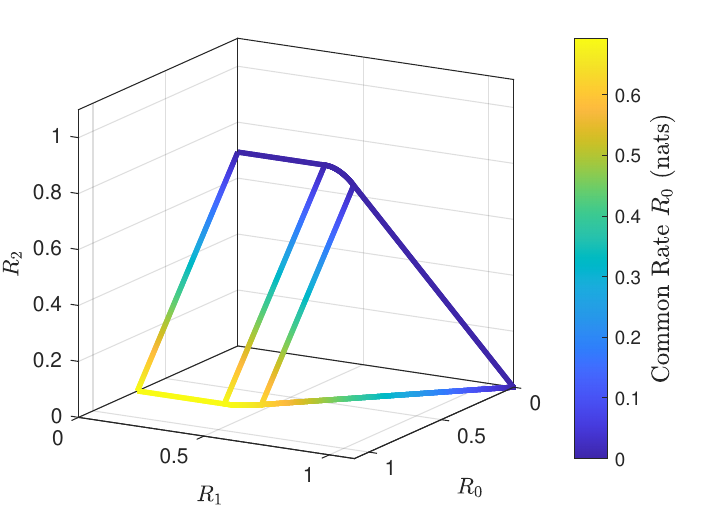}
    \label{fig:with_common_degraded}}
    \hfil
    \subfloat[]{\includegraphics[width=.48\linewidth]{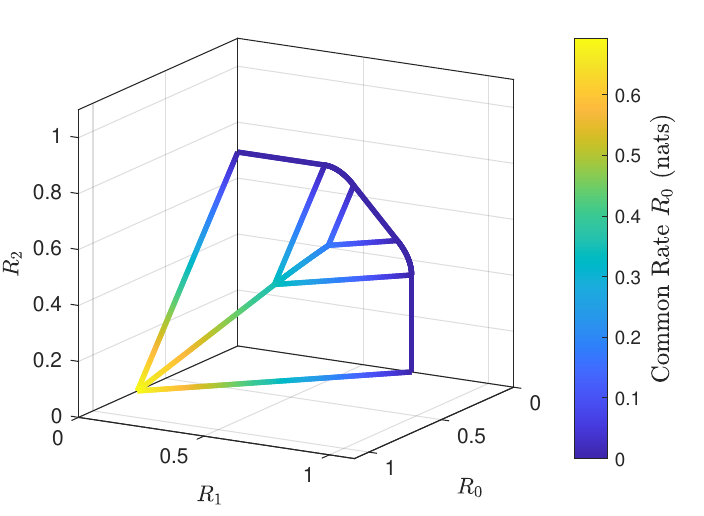}
    \label{fig:with_common_blackwell}}

    \caption{Complete characterization for degraded det-BCs and outer bound for general det-BCs. (a) Complete characterization $\mathcal{R}(\mathcal{B}_{Y\to Z})$ for the degraded channel $\mathcal{B}_{Y\to Z}$ from Fig.~\ref{fig:contraction, Y2Z}. (b) Outer bound $\mathcal{O}\left(\mathcal{B}\right)$ for the Blackwell channel.}
    \label{fig:with_common_outer_bound}
\end{figure}

Fig.~\ref{fig:with_common_outer_bound} summarizes our current analytical understanding. For degraded det-BCs, Theorem~\ref{thm:degraded_3D_exact} completely characterizes the capacity region. As shown in Fig.~\ref{fig:with_common_degraded}, the entire three-dimensional boundary is obtained from the private-message capacity region through the transformation $(R_1,R_0+R_2)\in\tilde{\mathcal C}(\mathcal B_{Y\to Z})$.
For general det-BCs, such a characterization is unavailable. Fig.~\ref{fig:with_common_blackwell} therefore shows the explicit outer bound $\mathcal O(\mathcal B)$ given by Corollary~\ref{cor:outer_bound_3D} for the Blackwell channel. The relationship between this outer bound, our analytical inner bound, and numerically optimized achievable regions is examined next.

\begin{figure}[t]
    \centering
    \subfloat[]{\includegraphics[width=.48\linewidth]{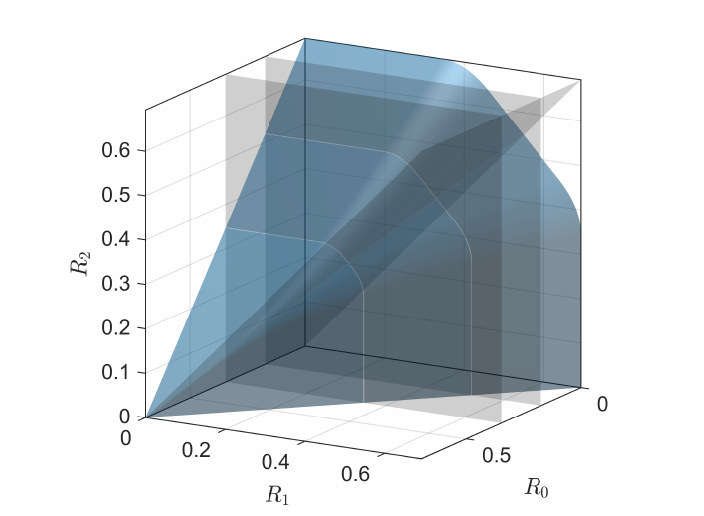}
    \label{fig:with_common_cross_sec}}
    \hfil
    \subfloat[]{\includegraphics[width=.48\linewidth]{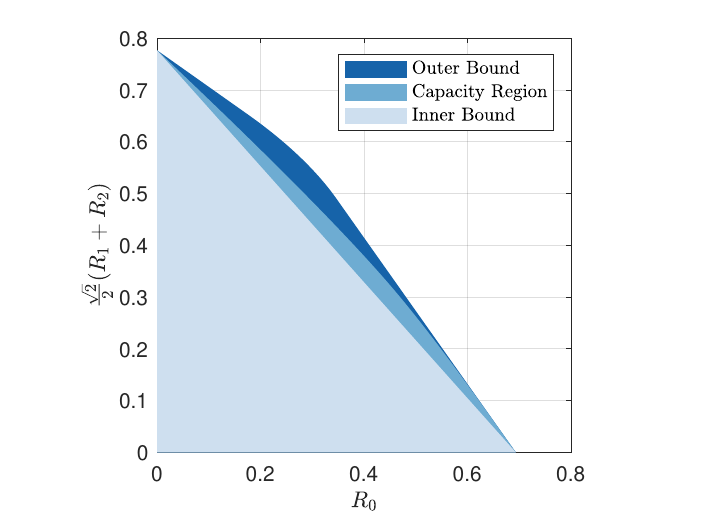}
    \label{fig:with_common_sec_R1R2}}
    \hfil
    \subfloat[]{\includegraphics[width=.48\linewidth]{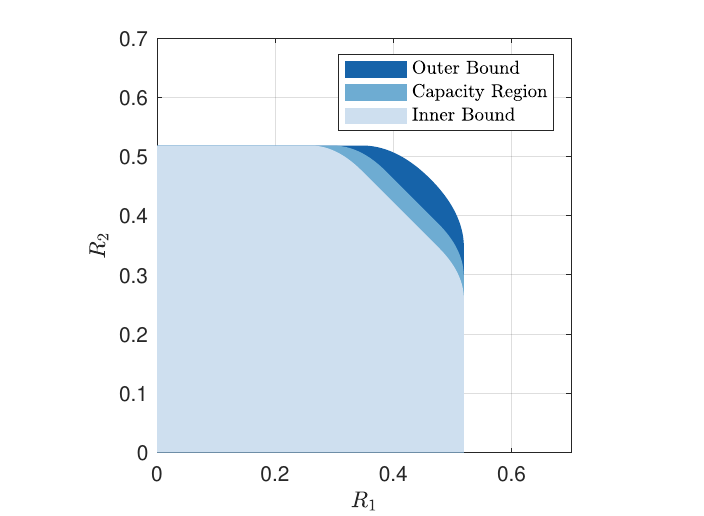}
    \label{fig:with_common_sec_R0c1}}
    \hfil
    \subfloat[]{\includegraphics[width=.48\linewidth]{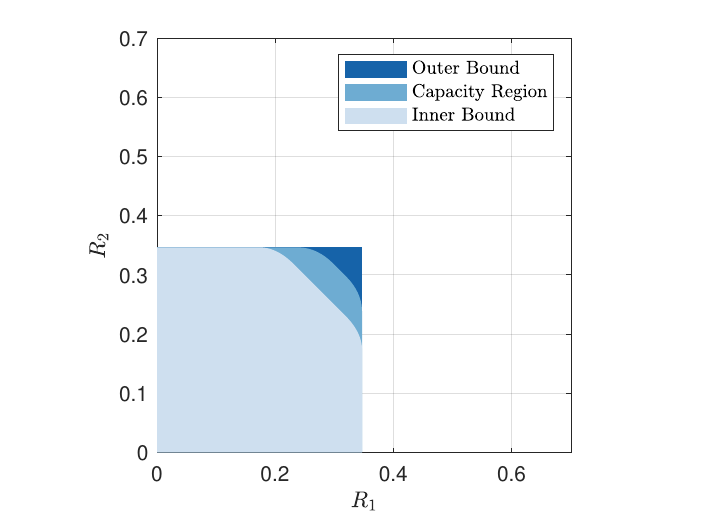}
    \label{fig:with_common_sec_R0c2}}

    \caption{Common-message capacity region for the Blackwell channel. (a) Three-dimensional view of the numerically computed achievable region, with three slicing planes shown. (b) Cross-section at $R_1=R_2$. (c) Cross-section at $R_0=\ln 3-\frac{4}{3}\ln 2$. (d) Cross-section at $R_0=\frac{1}{2}\ln 2$. In each panel, the outer bound $\mathcal{O}\left(\mathcal{B}\right)$ (Corollary~\ref{cor:outer_bound_3D}), the numerically optimized Marton's achievable region, and the analytical inner bound $\mathcal{I}\left(\mathcal{B}\right)$ (Corollary~\ref{cor:inner_bound_3D}) are shown.}
    \label{fig:blackwell_3D}
\end{figure}

Fig.~\ref{fig:blackwell_3D} compares the analytical inner bound $\mathcal I(\mathcal B)$, the explicit outer bound $\mathcal O(\mathcal B)$, and a numerically optimized achievable region for the Blackwell channel. The numerically optimized achievable region is obtained by numerically optimizing Marton's coding scheme over the auxiliary random variable $U$ and the input distribution. Fig.~\ref{fig:with_common_cross_sec} shows the three-dimensional achievable region together with three representative slicing planes, while Figs.~\ref{fig:with_common_sec_R1R2}--\ref{fig:with_common_sec_R0c2} compare the three regions on these cross-sections.
In the symmetric slice $R_1=R_2$ (Fig.~\ref{fig:with_common_sec_R1R2}), the numerically optimized achievable region lies strictly between them, indicating that neither analytical bound is tight. Figs.~\ref{fig:with_common_sec_R0c1} and~\ref{fig:with_common_sec_R0c2} show the achievable $(R_1,R_2)$ region for two representative common-message rates. In both cases, the numerical boundary again lies strictly between the analytical inner and outer bounds.

Taken together, Figs.~\ref{fig:with_common_outer_bound} and~\ref{fig:blackwell_3D} provide a complete picture of the current state of the problem. For degraded det-BCs, the capacity region with common messages admits a complete analytical characterization (Theorem~\ref{thm:degraded_3D_exact}). For general det-BCs, both an explicit analytical inner bound and an explicit outer bound can be constructed from the private-message capacity region $\tilde{\mathcal C}(\mathcal B)$, while numerical optimization confirms that neither is tight in general. Closing this gap and obtaining an explicit characterization of $\mathcal R(\mathcal B)$ remains an important open problem.


\subsection{Finite-Blocklength Regime}
\label{subsec:FBR}

We now present numerical results for the finite-blocklength regime, computed using the ILP formulation from Subsection~\ref{subsec:ILP}.
All computations were performed using MATLAB R2024b \verb|intlinprog| on an Intel Core i9-13900KF CPU with 64 GB RAM.

\begin{table}[t]
\centering
\caption{
Problem size and representative solver runtime of the ILP formulation for the Blackwell channel.
}
\label{tab:ilp_runtime}
\setlength{\tabcolsep}{10pt}
\begin{tabular}{ccccr}
\hline
Blocklength $n$ & Representative $M_1$ & \#variables & \#constraints & Runtime (s) \\
\hline
1 & 2 & 22 & 48 & 0.002 \\
2 & 4 & 180 & 464 & 0.664 \\
3 & 8 & 1864 & 5280 & 16.832 \\
4 & 16 & 21264 & 62528 & 3738.721 \\
\hline
\end{tabular}
\end{table}

\begin{figure}[t]
    \centering
    \subfloat[]{\includegraphics[width=.41\linewidth]{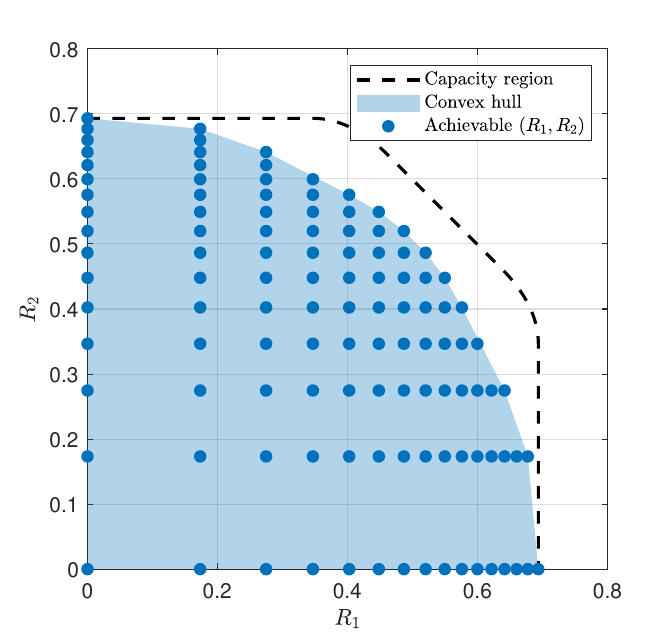}
    \label{fig:fbc_blackwell}}
    \hfil
    \subfloat[]{\includegraphics[width=.57\linewidth]{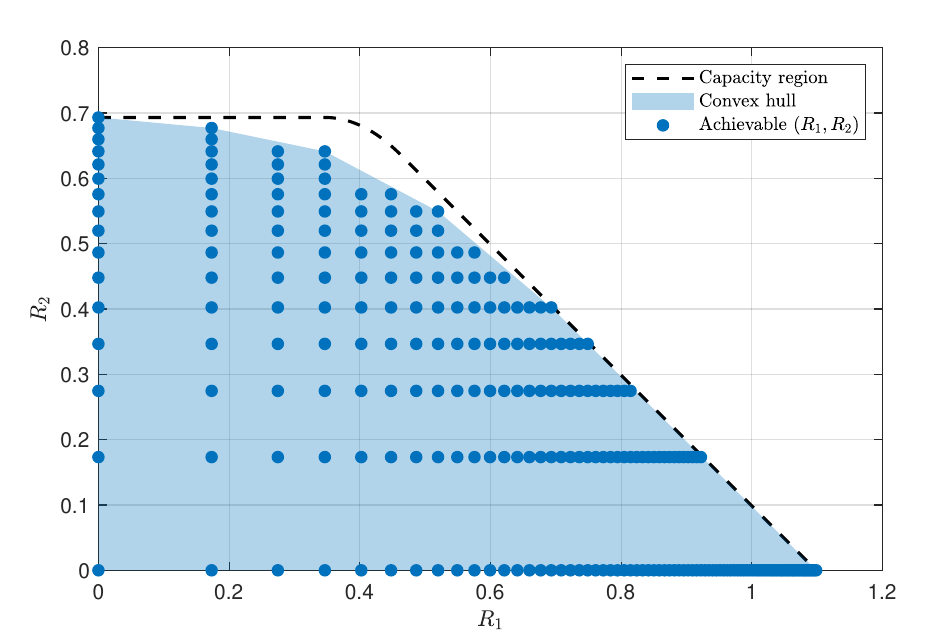}
    \label{fig:fbc_degraded}}

    \caption{Achievable rate pairs at blocklength $n = 4$. (a) Blackwell channel. (b) Degraded channel $\mathcal{B}_{Y\to Z}$ from Fig.~\ref{fig:contraction, Y2Z}. Blue dots represent achievable rate pairs $(R_1, R_2) = (\frac{1}{n}\ln M_1, \frac{1}{n}\ln M_2)$ from ILP solutions. The light blue region shows their convex hull (achievable via time-sharing). The black dashed curve is the asymptotic capacity region boundary $\partial\mathcal{C}(\mathcal{B})$.}
    \label{fig:fbc}
\end{figure}

\begin{figure}[t]
    \centering
    \subfloat[]{\includegraphics[width=.41\linewidth]{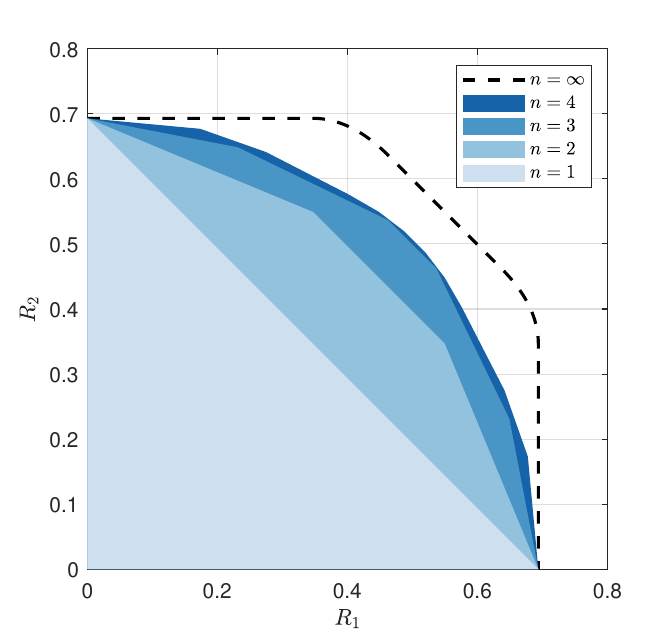}
    \label{fig:fbc_convexhull_blackwell}}
    \hfil
    \subfloat[]{\includegraphics[width=.57\linewidth]{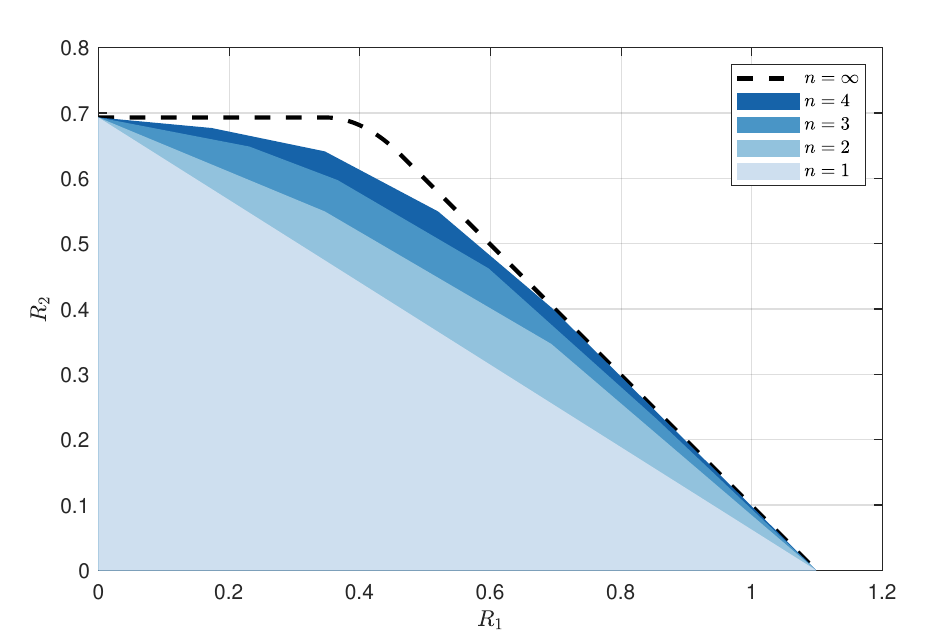}
    \label{fig:fbc_convexhull_degraded}}

    \caption{Evolution of achievable rate regions with blocklength. (a) Blackwell channel. (b) Degraded channel $\mathcal{B}_{Y\to Z}$. Shaded regions show convex hulls of achievable rate pairs for $n = 1, 2, 3, 4$ (darker blue indicates larger $n$). The black dashed curve represents the asymptotic capacity region boundary.}
    \label{fig:fbc_convexhull}
\end{figure}

Table~\ref{tab:ilp_runtime} summarizes the problem sizes and solver runtimes for the Blackwell channel. For each blocklength $n$, all ILP instances corresponding to $M_1=1,2,\ldots,|\mathcal{Y}|^n$ were solved, and the worst-case numbers of decision variables and constraints are reported together with the corresponding solver runtime.
Table~\ref{tab:ilp_runtime} confirms the exponential growth of the ILP problem size predicted in Remark~\ref{rem:ilp_complexity}. Consequently, the solver runtime also increases substantially with the blocklength, limiting exact computation to relatively small values of $n$. Nevertheless, as shown in Figs.~\ref{fig:fbc} and~\ref{fig:fbc_convexhull}, blocklengths $n=3$ and $4$ already capture a substantial fraction of the asymptotic capacity region, especially near the corner points and the linear boundary segments where simple input distributions are optimal.

Fig.~\ref{fig:fbc} displays the achievable rate pairs for the Blackwell channel and its induced degraded channel at blocklength $n = 4$. The blue dots represent discrete achievable pairs obtained by solving the ILP formulation in Subsection~\ref{subsec:ILP} for various values of $M_1$. The light blue shaded region is the convex hull of these points, representing rates achievable through time-sharing between the computed codes. The gap between this convex hull and the asymptotic capacity region $\mathcal{C}(\mathcal{B})$ (black dashed curve) quantifies the rate loss due to finite-blocklength constraints.

Several observations can be made from Fig.~\ref{fig:fbc}: First, the achievable region at $n = 4$ captures a substantial portion of the asymptotic capacity, particularly near the corner points where one receiver has low rate. Second, the remaining gap is more pronounced in the curved boundary regions (segments $A$--$B$ and $C$--$D$), where the optimal input distribution varies continuously with the target rate pair. Third, the degraded channel (Fig.~\ref{fig:fbc_degraded}) exhibits faster convergence toward capacity compared to the general Blackwell channel, consistent with its simpler forest structure and the corresponding reduction to a bin covering problem.

Fig.~\ref{fig:fbc_convexhull} illustrates the convergence of achievable rate regions to the asymptotic capacity as blocklength increases. The convex hulls for $n = 1, 2, 3, 4$ are displayed with progressively darker shades of blue, demonstrating monotonic expansion toward the capacity region boundary (black dashed curve). 

The convergence behavior exhibits notable features:
\begin{itemize}
    \item \textbf{Non-uniform convergence}: The rate of approach to capacity varies across the boundary. Corner points and linear segments (with simple optimal input distributions) converge rapidly, while curved segments require larger blocklengths to achieve near-optimal performance.
    \item \textbf{Blocklength $n = 1$}: Even single-symbol codes capture the extreme corner points of the capacity region, corresponding to scenarios where one receiver operates at maximum rate while the other receives no information.
    \item \textbf{Degraded vs. general channels}: Comparing (a) and (b), the degraded channel exhibits tighter convex hulls at all blocklengths, suggesting that structural simplicity in the channel graph translates to better finite-blocklength performance.
\end{itemize}

These results validate the proposed ILP formulation and provide insights into the finite-blocklength behavior of det-BCs. In particular, they demonstrate that even moderate blocklengths ($n=3$ or $4$) achieve rates close to the asymptotic capacity over a large portion of the boundary, with the largest finite-blocklength loss occurring along the curved boundary segments where the optimal input distribution changes continuously.


\section{Conclusion}
\label{sec:conclusion}

This paper has established an explicit, closed-form characterization of the private-message capacity region for deterministic broadcast channels. By representing det-BCs as bipartite graphs, we have formulated the achievability problem as a graph contraction condition: a rate pair $(R_1, R_2)$ is achievable if and only if the complete bipartite graph $K_{M_1, M_2}$ with $R_i = \frac{1}{n}\ln M_i$ can be obtained by contracting the $n$-fold product channel graph. This graph-theoretic perspective transforms capacity analysis into a combinatorial problem amenable to both asymptotic and finite-blocklength analysis.
Our main result (Theorem~\ref{thm:main_capacity}) provides a parametric characterization of the capacity region boundary through the degree sequences $D_Y(\mathcal{B})$ and $D_Z(\mathcal{B})$ alone. The boundary consists of five segments: two curved segments determined by parametric equations involving weighted degree sums, two vertical/horizontal segments, and one diagonal segment with slope $-1$. Notably, any det-BC's capacity region can be readily computed from its degree sequences, without requiring explicit optimization over input distributions. This degree sequence sufficiency (Corollary~\ref{cor:degree_sufficiency}) reveals that the topological structure of the channel graph, rather than the specific transition probability, determines the fundamental communication limits.
Beyond the private-message setting, we have extended the graph-theoretic framework in two directions. For the common-message problem, we characterized achievable rate triples $\left(R_0,R_1,R_2\right)$ through the connected-component structure of $G\left(\mathcal{B}\right)$ (Theorem~\ref{thm:achievability_with_common}). This framework further yields a complete characterization of the three-dimensional capacity region $\mathcal{R}\left(\mathcal{B}\right)$ for degraded det-BCs (Theorem~\ref{thm:degraded_3D_exact}), together with explicit inner and outer bounds for general det-BCs expressed directly in terms of the private-message capacity region $\tilde{\mathcal{C}}\left(\mathcal{B}\right)$. Numerical results demonstrate that these bounds do not coincide in general, indicating that a complete characterization of $\mathcal{R}\left(\mathcal{B}\right)$ for non-degraded det-BCs remains an open problem. We also formulated the finite-blocklength achievable rate problem as an integer linear program (ILP) that directly yields encoding and decoding schemes, enabling numerical computation of achievable rate pairs at moderate blocklengths. For degraded det-BCs, the ILP further reduces to a bin covering problem with significantly lower computational complexity.

While the present work establishes explicit capacity characterizations for private messages, several important problems remain open. Closing the gap between the inner and outer bounds for $\mathcal{R}\left(\mathcal{B}\right)$ will likely require fundamentally new achievability techniques beyond the current component-based decomposition. On the computational side, although the proposed ILP is tractable for small blocklengths, developing tighter analytical bounds or efficient approximation algorithms for practical blocklengths remains an important direction. More broadly, we expect that the graph-theoretic and combinatorial framework developed here may extend beyond deterministic channels, potentially providing new insights into explicit capacity characterizations for stochastic broadcast channels through weighted or probabilistic graph models.


\bibliographystyle{IEEEtran}
\bibliography{IEEEabrv,references}

\end{document}